\documentclass[aps,prd,preprintnumbers,groupedaddress,nofootinbib,amssymb,notitlepage,eqsecnum]{revtex4-1}
\usepackage{here}
\usepackage{graphicx}
\usepackage{amsmath,amsthm,amssymb}
\usepackage{bm}
\usepackage{color}

\usepackage{amsfonts}
\usepackage{dcolumn}
\usepackage{hyperref}
\allowdisplaybreaks[1]
\usepackage{stackengine}

\newcommand{\fr}[2]{\frac{#1}{#2}}

\newcommand{\bra}[1]{\left( #1 \right)}

\newcommand{\be}{\begin{equation}}  
\newcommand{\ee}{\end{equation}}
\newcommand{\ba}{\begin{eqnarray}}
\newcommand{\ea}{\end{eqnarray}}

\newcommand{\rd}{{\rm d}}
\newcommand{\hr}{\hat{r}}

\newcommand{\bem}{\begin{bmatrix}}
\newcommand{\eem}{\end{bmatrix}}
\newcommand{\Mpl}{M_{\rm Pl}}

\newcommand{\mE}{\mathcal{E}}

\newcommand{\mF}{\mathcal{F}}
\newcommand{\mG}{\mathcal{G}}
\newcommand{\mH}{\mathcal{H}}
\newcommand{\mL}{\mathcal{L}}
\newcommand{\mO}{\mathcal{O}}

\allowdisplaybreaks

\begin{document}

\preprint{YITP-22-41, WUAP-22-03}

\title{Linear stability of black holes with static scalar hair 
in full Horndeski theories: 
generic instabilities and surviving models}

\author{Masato Minamitsuji,$^{1}$ Kazufumi Takahashi,$^{2}$ 
and Shinji Tsujikawa$^{3}$}

\affiliation{
$^{1}$Centro de Astrof\'{\i}sica e Gravita\c c\~ao - CENTRA, Departamento de F\'{\i}sica, Instituto Superior T\'ecnico - IST, Universidade de Lisboa - UL, Av.~Rovisco Pais 1, 1049-001 Lisboa, Portugal\\
$^2$Center for Gravitational Physics and Quantum Information, Yukawa Institute for Theoretical Physics, Kyoto University, 606-8502, Kyoto, Japan\\
$^3$Department of Physics, Waseda University, 3-4-1 Okubo, Shinjuku, Tokyo 169-8555, Japan}

\begin{abstract}
In full Horndeski theories, we show that the static and spherically symmetric black hole (BH) solutions with a static scalar field~$\phi$ whose kinetic term~$X$ is nonvanishing on the BH horizon are generically prone to ghost/Laplacian instabilities. We then search for asymptotically Minkowski hairy BH solutions with a vanishing $X$ on the horizon free from ghost/Laplacian instabilities. We show that models with regular coupling functions of $\phi$ and $X$ result in no-hair Schwarzschild BHs in general. On the other hand, the presence of a coupling between the scalar field and the Gauss-Bonnet (GB) term~$R_{\rm GB}^2$, even with the coexistence of other regular coupling functions, leads to the realization of asymptotically Minkowski hairy BH solutions without ghost/Laplacian instabilities. Finally, we find that hairy BH solutions in power-law $F(R_{\rm GB}^2)$ gravity are plagued by ghost instabilities. These results imply that the GB coupling of the form~$\xi(\phi)R_{\rm GB}^2$ plays a prominent role for the existence of asymptotically Minkowski hairy BH solutions free from ghost/Laplacian instabilities.
\end{abstract}

\maketitle

\section{Introduction}\label{sec:intro}

General Relativity (GR) has been tested by numerous experiments 
in the Solar System. While gravity can be well described by GR 
on the weak gravitational background in our local 
Universe~\cite{Will:2014kxa}, the dawn of 
gravitational-wave (GW) astronomy~\cite{LIGOScientific:2016aoc} 
and black hole (BH) shadow measurements~\cite{EventHorizonTelescope:2019dse}
have started to allow us to probe the physics of extremely 
compact objects like 
BHs and neutron stars~\cite{Berti:2015itd,Barack:2018yly,Berti:2018cxi,Berti:2018vdi}.
On the other hand, we also know that the Universe 
recently entered a phase of accelerated expansion~\cite{SupernovaSearchTeam:1998fmf,SupernovaCosmologyProject:1998vns}. While the cosmological constant  
is the simplest candidate for the source of this acceleration, 
i.e., dark energy, the theoretical value of vacuum energy 
mimicking the cosmological constant is enormously 
larger than the observed dark energy scale~\cite{Weinberg:1988cp}.
The cosmological constant has also been 
plagued by tensions of today's Hubble constant~$H_0$ 
constrained from high- and low-redshift 
measurements~\cite{Riess:2019cxk,DiValentino:2021izs}.
These facts led to the question for the validity of GR on 
large distances relevant to today's cosmic acceleration.
A simple and robust alternative to GR is provided by 
scalar-tensor theories possessing a scalar degree of 
freedom coupled to gravity~\cite{Fujii:2003pa}.

The most general class of scalar-tensor theories with second-order 
Euler-Lagrange equations of motion is called 
Horndeski theories~\cite{Horndeski,Def11,KYY,Charmousis:2011bf}.\footnote{The second-order nature of Euler-Lagrange equations is desirable for avoiding the notorious problem of Ostrogradsky ghost~\cite{Woodard:2015zca}.
Interestingly, there is a larger class of scalar-tensor theories without Ostrogradsky ghost known as DHOST theories~\cite{Langlois:2015cwa,Crisostomi:2016czh,BenAchour:2016fzp,Takahashi:2017pje,Langlois:2018jdg,Takahashi:2021ttd}, which we do not consider in the present paper.}
There have been 
numerous attempts for the theoretical construction of 
dark energy models compatible with observations~\cite{Copeland:2006wr,DeFelice:2010aj,Clifton:2011jh,Joyce:2014kja,Heisenberg:2018vsk,Kase:2018aps}.
Although such a new scalar degree of freedom potentially manifests 
itself in the Solar System, fifth forces mediated by 
the scalar field 
can be screened~\cite{Burrage:2010rs,Kimura:2011dc,Koyama:2013paa,Kase:2013uja} by Vainshtein~\cite{Vainshtein:1972sx} 
or chameleon~\cite{Khoury:2003aq} mechanisms
around a compact body on the weak gravitational background.
In the vicinity of a BH, on the other hand, 
a nonvanishing charge of the scalar field, 
i.e., scalar hair, gives rise to a nontrivial 
field profile affecting the background geometry.
This offers an interesting possibility for probing 
the possible deviation from GR in strong gravity regimes.

In GR, the vacuum, asymptotically flat, static, 
and spherically symmetric solution is uniquely characterized by 
the Schwarzschild geometry containing the mass 
of a compact body.\footnote{Rigorously speaking, an ``asymptotically Minkowski'' metric should be distinguished from an ``asymptotically flat'' metric.
Indeed, there are some cases where the asymptotic form of the metric does not look like Minkowski but all the components of the curvature tensor vanish asymptotically (see, e.g., \cite{Anabalon:2013oea}).
In the present paper, we do not consider such asymptotically locally flat metrics and focus only on asymptotically Minkowski metrics.}
The search for hairy BH solutions endowed with nontrivial 
field profiles has been performed 
for several subclasses of Horndeski theories.
It has been recognized that there 
is no scalar hair for a minimally coupled canonical scalar 
field~\cite{Hawking:1971vc,Bekenstein:1972ny} 
and k-essence~\cite{Graham:2014mda} as well as 
for a nonminimally coupled scalar field with the Ricci 
scalar of the form~
$G_4(\phi)R$~\cite{Hawking:1972qk,Bekenstein:1995un,Sotiriou:2011dz,Faraoni:2017ock}.
If the scalar field is coupled to a 
Gauss-Bonnet (GB) curvature invariant~$R_{\rm GB}^2$
[see Eq.~\eqref{def_gb}] as $\xi(\phi) R_{\rm GB}^2$~\cite{Zwiebach:1985uq,Antoniadis:1993jc,Gasperini:1996fu}, 
where $\xi(\phi)$ is a function of $\phi$, 
there are asymptotically Minkowski hairy BH solutions 
for the dilatonic coupling~$\xi(\phi)  
\sim {\rm e}^{-\phi}$~\cite{Kanti:1995vq,Torii:1996yi,Kanti:1997br,Chen:2006ge,Guo:2008hf,Guo:2008eq,Pani:2009wy,Ayzenberg:2014aka,Maselli:2015tta,Kleihaus:2011tg,Kleihaus:2015aje}, linear coupling~$\xi(\phi) \propto \phi$~\cite{Sotiriou:2013qea,Sotiriou:2014pfa},
and models for spontaneous scalarization of BHs where
$\xi(\phi)\propto \sum_{j\geq 1}c_j\phi^{2j}$ with $c_j$ 
being constant~\cite{Doneva:2017bvd,Silva:2017uqg,Antoniou:2017acq,Blazquez-Salcedo:2018jnn,Minamitsuji:2018xde,Silva:2018qhn,Macedo:2019sem,Doneva:2021tvn}. The nonminimal derivative coupling~$\phi G_{\mu \nu} \nabla^{\mu} \nabla^{\nu}\phi$ to the Einstein tensor $G_{\mu \nu}$, where $\nabla_{\mu}$ is the covariant derivative operator, 
gives rise to non-asymptotically Minkowski hairy BHs~\cite{Rinaldi:2012vy,Anabalon:2013oea,Minamitsuji:2013ura,Cisterna:2014nua,Kolyvaris:2011fk,Minamitsuji:2014hha}
with the static background scalar field, 
but it was recently recognized that they 
are unstable against linear perturbations~\cite{Minamitsuji:2022mlv}.

The purpose of this paper is to elucidate the class of 
Horndeski theories giving rise to asymptotically Minkowski 
BH solutions endowed with scalar 
hair which are free from ghost or Laplacian instabilities.
We will focus on a time-independent background scalar field 
on the static and spherically symmetric background. In this case, 
we can exploit conditions for the absence of ghost/Laplacian instabilities against odd- and even-parity perturbations derived 
in Refs.~\cite{Kobayashi:2012kh,Kobayashi:2014wsa,Kase:2021mix}.
In shift-symmetric Horndeski theories where the coupling 
functions $G_{2,3,4,5}$ depend on 
the canonical kinetic term~$X$ only,
Hui and Nicolis~\cite{Hui:2012qt} showed the 
absence of asymptotically Minkowski hairy BH solutions 
under several assumptions based on the properties of 
a conserved Noether current associated with the shift symmetry 
(see also Ref.~\cite{Minamitsuji:2022mlv} for 
a detailed review).\footnote{We note that
shift-symmetric theories admit asymptotically Minkowski BH solutions with a time-dependent background scalar 
field~$\phi=qt+\Phi(r)$~\cite{Babichev:2013cya,Kobayashi:2014eva,Ogawa:2015pea,Takahashi:2015pad,Takahashi:2016dnv,Babichev:2017lmw,Tretyakova:2017lyg,Takahashi:2019oxz,Motohashi:2019sen,Takahashi:2020hso,Khoury:2020aya,Takahashi:2021bml,Nakashi:2022wdg}, 
which we will not address in this paper.}
Such a no-hair argument based on the  
Noether current cannot be applied to full 
Horndeski theories containing both 
$\phi$- and $X$-dependence in the coupling functions 
and hence breaking the shift symmetry.
A systematic approach to the search for hairy BHs in 
such more general shift-symmetry-breaking Horndeski 
theories would be challenging. 

Our approach is first to show that the BH solutions 
where $X$ is an analytic function 
with a nonvanishing value on the horizon ($X_s \neq 0$) 
can be excluded by linear instability.
This is a generalization of the result
recognized for shift-symmetric Horndeski  
theories~\cite{Minamitsuji:2022mlv}. 
We then search for hairy BHs with $X_s=0$.
Assuming that the deviation from GR with a canonical scalar field is controlled by a single coupling constant, we perform consistent expansions of the metric and scalar field in terms of the coupling constant.
Imposing that the metric is asymptotically Minkowski together 
with a vanishing radial scalar-field derivative 
as the boundary condition at spatial infinity,
we will show that couplings of the 
form~$G_{I} \supset \alpha_I(\phi) F_I(X)$ 
with $I=2,3,4,5$, where $\alpha_I(\phi)$ and $F_I(X)$ 
are analytic functions of $\phi$ and $X$ respectively,
do not generally give rise to hairy asymptotically Minkowski
BH solutions.\footnote{Here, ``$G_{I} \supset \alpha_I(\phi) F_I(X)$'' means that $G_{I}$ contains only the term~$\alpha_I(\phi) F_I(X)$ besides the canonical kinetic term of the scalar field in $G_2$ and the Einstein-Hilbert term in $G_4$.
Likewise, in the present paper, we use the symbol~``$\supset$'' when we incorporate additional terms on top of those of primary interest (i.e., the canonical kinetic term of the scalar field and the Einstein-Hilbert term in Sec.~\ref{nohairsec} and the scalar-GB coupling~$\alpha \xi(\phi) R_{\rm GB}^2$ in Sec.~\ref{GBother}).}

As an extension of the results in shift-symmetric 
Horndeski theories~\cite{Sotiriou:2013qea,Babichev:2017guv}, 
there is a possibility for evading the no-hair property of BHs 
in non-shift-symmetric theories
with a particular choice of the coupling 
functions~$G_2 \supset \alpha_2(\phi) \sqrt{-X}$, 
$G_3 \supset \alpha_3(\phi) \ln |X|$, 
$G_4 \supset \alpha_4(\phi) \sqrt{-X}$, and 
$G_5 \supset \alpha_5(\phi) \ln |X|$,
which are no longer analytic in $X$.
Among them, however, we will see that only the quintic 
coupling~$G_5 \supset \alpha_5(\phi) \ln |X|$ allows us 
for realizing asymptotically Minkowski hairy BHs 
free from ghost or Laplacian instabilities.
In particular, the scalar field coupled to the GB
term of the form~$\alpha\xi(\phi)R_{\rm GB}^2$ 
can be embedded in Horndeski theories, 
where the corresponding action contains 
the quintic coupling of this type. 
For the GB coupling, 
we will derive hairy BH solutions by using expansions 
with respect to a small dimensionless coupling constant.  
Furthermore, we will show that they can satisfy all 
the conditions for the absence of ghost/Laplacian instabilities
against odd- and even-parity perturbations.
We note that, in Refs.~\cite{Torii:1998gm,DeFelice:2011ka,Blazquez-Salcedo:2018jnn,Minamitsuji:2018xde,Blazquez-Salcedo:2020rhf,Blazquez-Salcedo:2020caw,Blazquez-Salcedo:2022omw,Langlois:2022eta}, 
the linear stability of hairy BHs for scalar-GB 
theories has also been investigated.

In the presence of positive power-law 
functions of $\phi$ or $X$ in $G_{2,3,4,5}$ besides 
the GB couplings, we also find new classes of 
BH solutions endowed with additional hair which are 
free from ghost or Laplacian instabilities. 
Since all these hairy BH solutions disappear without 
the coupling~$\alpha_5(\phi) \ln |X|\subset G_5$, 
the presence of such a logarithmic quintic interaction is 
crucial for realizing asymptotically Minkowski BHs 
with scalar hair.
We will also show that $F(R_{\rm GB}^2)$~gravity~\cite{Nojiri:2005jg,DeFelice:2007zq,Li:2007jm,DeFelice:2008wz,DeFelice:2009aj} rewritten in terms 
of Horndeski theories gives rise to hairy BHs, which, however, 
are subject to a ghost instability of even-parity perturbations.
The extension of this $F(R_{\rm GB}^2)$-equivalent Horndeski
theory to that containing  
a canonical kinetic term of the scalar field leads to 
no-hair asymptotically Minkowski BH solutions.

The rest of this paper is organized as follows.
In Sec.~\ref{sec:background}, we present the background 
equations of motion and conditions for the absence 
of ghost/Laplacian instabilities
for static and spherically symmetric BHs
in full Horndeski theories.
In Sec.~\ref{sec:expansion}, we prove the generic 
instability of BHs for theories in which 
$X$ is analytic and has
a nonvanishing value on the horizon.
In Sec.~\ref{nohairsec}, we show the absence of 
hairy asymptotically Minkowski BHs for theories 
containing coupling functions~$G_I\supset \alpha_I (\phi) F_I(X)$, where $\alpha_I (\phi)$ and $F_I(X)$
are analytic functions of $\phi$ and $X$, respectively.
Then, we search for the possibility for realizing 
hairy BH solutions in full Horndeski theories.
In Sec.~\ref{quinsec}, we investigate the existence 
of hairy BHs and the issues of ghost/Laplacian instabilities
for the GB coupling~$\alpha \xi(\phi)  R_{\rm GB}^2$. 
We then extend the analysis to theories in which 
positive power-law functions of $\phi$ or $X$ 
in $G_{2,3,4,5}$ are present besides the GB couplings.
In Sec.~\ref{sec:fG}, we study BH solutions in 
$F(R_{\rm GB}^2)$~gravity and its extensions.
Section~\ref{sec:conc} is devoted to conclusions.

\section{Background equations and linear stability conditions}
\label{sec:background}

We consider full 
Horndeski theories~\cite{Horndeski,Def11,KYY,Charmousis:2011bf} 
given by the action 
\be
{\cal S}=\int {\rm d}^4 x \sqrt{-g}\,{\cal L}_H\,,
\label{action}
\ee
where $g$ is the determinant of the metric tensor~$g_{\mu \nu}$ and 
\ba
{\cal L}_H
&=&
G_2(\phi,X)-G_{3}(\phi,X)\square\phi 
+G_{4}(\phi,X)R +G_{4,X}(\phi,X)\left[ (\square \phi)^{2}
-(\nabla_{\mu}\nabla_{\nu} \phi)
(\nabla^{\mu}\nabla^{\nu} \phi) \right]
+G_{5}(\phi,X)G_{\mu \nu} \nabla^{\mu}\nabla^{\nu} \phi
\notag\\
&&
-\frac{1}{6}G_{5,X}(\phi,X)
\left[ (\square \phi )^{3}-3(\square \phi)\,
(\nabla_{\mu}\nabla_{\nu} \phi)
(\nabla^{\mu}\nabla^{\nu} \phi)
+2(\nabla^{\mu}\nabla_{\alpha} \phi)
(\nabla^{\alpha}\nabla_{\beta} \phi)
(\nabla^{\beta}\nabla_{\mu} \phi) \right]\,,
\label{LH}
\ea
with $R$ and $G_{\mu \nu}$ being the Ricci scalar and 
Einstein tensor associated with the metric 
$g_{\mu\nu}$, respectively.
The four functions~$G_{I}$'s ($I=2,3,4,5$) depend on  
the scalar field~$\phi$ and 
its canonical kinetic term~$X=-g^{\mu\nu}\nabla_{\mu}\phi\nabla_{\nu}\phi/2$.
We also use the notations~$\square \phi \equiv \nabla^{\mu}\nabla_{\mu} \phi$ 
and $G_{I,\phi} \equiv \partial G_I/\partial \phi$, 
$G_{I,X} \equiv \partial G_I/\partial X$, 
$G_{I,\phi X} \equiv \partial^2 G_I/(\partial X \partial \phi)$, 
etc. 

We assume a static and spherically symmetric 
background metric and 
scalar field
\begin{eqnarray}
\rd s^2&=&-f(r) \rd t^{2} +h^{-1}(r) \rd r^{2}
+ r^{2} \left(\rd \theta^{2}+\sin^{2}\theta\,\rd\varphi^{2} 
\right)\,,
\label{BGmetric}
\\
\phi&=&\phi(r)\,,
\end{eqnarray}
where $f(r)$, $h(r)$, and $\phi(r)$ are functions of 
the radial coordinate~$r$.
On this background, the 
scalar-field kinetic term is given by
$X=-h \phi'^2/2$, where the prime represents 
the derivative with respect to $r$.
The $tt$-, $rr$-, $\theta \theta$-components 
of gravitational field equations are given, respectively, by~\cite{Kobayashi:2012kh,Kobayashi:2014wsa,Kase:2021mix}
\ba
\mE_{tt} 
&\equiv&
\left(A_1+\frac{A_2}{r}+\frac{A_3}{r^2}\right)\phi''
+\left(\frac{\phi'}{2h}A_1+\frac{A_4}{r}+\frac{A_5}{r^2}\right)h'
+A_6+\frac{A_7}{r}+\frac{A_8}{r^2}
=0\,,\label{back1}\\
\mE_{rr} 
&\equiv&
-\left(\frac{\phi'}{2h}A_1+\frac{A_4}{r}+\frac{A_5}{r^2}\right) \frac{hf'}{f}
+A_9-\frac{2\phi'}{r}A_1-\frac{1}{r^2}
\left[\frac{\phi'}{2h}A_2+(h-1)A_4\right]=0
\,,\label{back2}\\
\mE_{\theta \theta} 
&\equiv&
\left\{\left[A_2+\frac{(2h-1)\phi'A_3+2hA_5}{h\phi' r}\right]\frac{f'}{4f}+A_1+\frac{A_2}{2r}\right\}\phi''
+\frac{1}{4f}\left(2hA_4-\phi'A_2+\frac{2hA_5-\phi'A_3}{r}\right)\left(f''-\frac{f'^2}{2f}\right)\notag\\
&&
+\left[A_4+\frac{2h(2h+1)A_5-\phi'A_3}{2h^2r}\right]\frac{f'h'}{4f}
+\left(\frac{A_7}{4}+\frac{A_{10}}{r}\right)\frac{f'}{f}
+\left(\frac{\phi'}{h}A_1+\frac{A_4}{r}\right)\frac{h'}{2}+A_6+\frac{A_7}{2r}
=0\,,\label{back3}
\ea
where 
\begin{equation}
\begin{split}
A_1&=-h^2 (G_{3,X}-2 G_{4,\phi X} ) \phi'^2-2 G_{4,\phi} h\,,\\
A_2&=2 h^3 ( 2 G_{4,XX}-G_{5,\phi X} ) \phi'^3-4 h^2 ( G_{4,X}-G_{5,\phi} ) \phi'\,,\\
A_3&=-h^4G_{5,XX} \phi'^4+h^2G_{5,X}  ( 3 h-1 ) \phi'^2\,,\\
A_4&=h^2 ( 2 G_{4,XX}-G_{5,\phi X} ) \phi'^4+h ( 3 G_{5,\phi}-4 G_{4,X} ) \phi'^2-2 G_4\,,\\
A_5&=-\frac12 \left[G_{5,XX} h^3{\phi'}^{5}- hG_{5,X}  ( 5 h-1 ) \phi'^3\right]\,,\\
A_6&=h ( G_{3,\phi}-2 G_{4,\phi\phi} ) \phi'^2+G_2\,,\\
A_7&=-2 h^2 ( 2 G_{4,\phi X}-G_{5,\phi\phi} ) \phi'^3-4 G_{4,\phi} h\phi'\,,\\
A_8&=G_{5,\phi X} h^3\phi'^4-h ( 2 G_{4,X} h-G_{5,\phi} h-G_{5,\phi} ) \phi'^2-2 G_4  ( h-1 )\,,\\
A_9&=-h ( G_{2,X}-G_{3,\phi} ) \phi'^2-G_2\,,\\
A_{10}&=\frac12 G_{5,\phi X} h^3\phi'^4-\frac12 h^2 ( 2 G_{4,X}-G_{5,\phi} ) \phi'^2-G_4 h\,.
\end{split}
\end{equation}
Varying the action~\eqref{action} with respect 
to $\phi$, it follows that
\be
\frac{1}{r^2} \sqrt{\frac{h}{f}} 
\left( r^2 \sqrt{\frac{f}{h}} J^r 
\right)'+P_{\phi}=0\,,
\label{Ephi2}
\ee
where
\ba
J^r &=&
h \phi' \biggl[
G_{2,X}-\left( \frac{2}{r}+\frac{f'}{2f} 
\right) h \phi' G_{3,X}+2 \left( \frac{1-h}{r^2}
-\frac{h f'}{rf} \right) G_{4,X}+2h \phi'^2 
\left( \frac{h}{r^2}+\frac{hf'}{rf} \right) G_{4,XX} 
\nonumber \\
&&-\frac{f'}{2r^2f} (1-3h)h \phi' G_{5,X}
-\frac{f' h^3 \phi'^3}{2r^2 f}G_{5,XX} \biggr]\,,
\label{Jrc}\\
P_{\phi} &=& G_{2,\phi}+\lambda_1 G_{3,\phi}+\lambda_2 G_{3,\phi \phi}
+\lambda_3 G_{3,\phi X}+\lambda_4 G_{4,\phi}+\lambda_5 G_{4,\phi X}
+\lambda_6 G_{4,\phi \phi X}+\lambda_7 G_{4,\phi XX} \nonumber \\
&&
+\lambda_8 G_{5,\phi}
+\lambda_9 G_{5,\phi \phi}+\lambda_{10} G_{5,\phi X}+
\lambda_{11} G_{5,\phi \phi X}+\lambda_{12}G_{5,\phi X X}\,.
\label{Pphide}
\ea
The coefficients~$\lambda_1$--$\lambda_{12}$ 
in $P_{\phi}$ are given in Appendix~\ref{AppA}. 
We note that Eq.~(\ref{Ephi2}) is 
equivalent to the following equation:
\be
{\cal E}_{\phi} \equiv -\frac{2}{\phi'}
\left[\frac{f'}{2f}{\cal E}_{tt}+{\cal E}_{rr}'
+\left(\frac{f'}{2f}+\frac{2}{r}\right){\cal E}_{rr}
+\frac{2}{r}{\cal E}_{\theta \theta} \right]=0\,.
\label{back5}
\ee
This shows that the scalar-field equation is not 
independent of other Eqs.~(\ref{back1})--(\ref{back3}), which is always the case for theories with general covariance~\cite{Motohashi:2016prk}.

In Refs.~\cite{Kobayashi:2012kh,Kobayashi:2014wsa,Kase:2021mix}, 
the odd- and even-parity perturbation theories about the static and spherically-symmetric solutions have been formulated 
in full Horndeski theories.
In the following, we will briefly summarize conditions for 
the absence of ghost/Laplacian instabilities derived in 
these papers (see also Ref.~\cite{Minamitsuji:2022mlv} for a brief summary of linear stability conditions in
shift-symmetric Horndeski theories).
Readers who are interested in the derivation of them should 
refer to Refs.~\cite{Kobayashi:2012kh,Kobayashi:2014wsa,Kase:2021mix}.

The stability against odd-parity perturbations 
is ensured under the following three 
conditions~\cite{Kobayashi:2012kh}:
\ba
{\cal F} &\equiv& 2G_4+h\phi'^2G_{5,\phi}-h\phi'^2  \left( \frac12 h' \phi'+h \phi'' \right) 
G_{5,X}>0\,,
\label{cFdef}\\
{\cal G} &\equiv& 2G_4+2 h\phi'^2G_{4,X}
-h\phi'^2 \left( G_{5,\phi}+{\frac {f' h\phi' G_{5,X}}{2f}} \right)>0 \,,
\label{cGdef}\\
{\cal H} &\equiv& 2G_4+2 h\phi'^2G_{4,X}-h\phi'^2G_{5,\phi}
-\frac{h^2 \phi'^3 G_{5,X}}{r}
>0\,.
\label{cHdef}
\ea
The ghost is absent under the inequality~(\ref{cGdef}). The squared propagation speeds of 
odd-parity perturbations along the radial and angular
directions are given, respectively, by 
\be
c_{r,{\rm odd}}^2=\frac{{\cal G}}{{\cal F}}\,,
\qquad 
c_{\Omega,{\rm odd}}^2=\frac{{\cal G}}{{\cal H}}\,,
\ee
which are both positive under the 
conditions~(\ref{cFdef})--(\ref{cHdef}).

In the even-parity sector, the no-ghost condition 
is quantified as~\cite{Kobayashi:2014wsa}
\be
{\cal K} \equiv 2{\cal P}_1-{\cal F}>0\,,
\label{Kcon}
\ee
with 
\be
{\cal P}_1 \equiv \frac{h \mu}{2fr^2 {\cal H}^2} 
\left( 
\frac{fr^4 {\cal H}^4}{\mu^2 h} \right)'\,,\qquad
\mu \equiv \frac{2(\phi' a_1+r\sqrt{fh}{\cal H})}{\sqrt{fh}}\,,
\label{defP1}
\ee
where $a_1$ is given in Appendix~\ref{AppB}.
For the multipoles~$\ell \geq 2$, the even-parity sector 
consists of two dynamical degrees of freedom.
One is the perturbation of the scalar field~$\delta\phi$, while the other, which we denote by $\psi$, can be regarded as the gravitational perturbation
(see 
Refs.~\cite{Kobayashi:2014wsa,Kase:2021mix,Kase:2020qvz} 
for the definition of $\psi$).
In the limit of high frequencies, 
the conditions for the absence of Laplacian 
instabilities of 
$\psi$ and $\delta \phi$ along the 
radial direction are given, 
respectively, by 
\ba
c_{r1,{\rm even}}^2 
&=& \frac{\mG}{\mF}>0\,,
\label{cr1even}\\
c_{r2,{\rm even}}^2
&=&
\frac{2\phi'[ 4r^2 (fh)^{3/2} {\cal H} c_4 
(2\phi' a_1+r\sqrt{fh}\,{\cal H})
-2a_1^2 f^{3/2} \sqrt{h} 
\phi' {\cal G} 
+( a_1 f'+2 c_2 f ) r^2 fh 
{\cal H}^2]}{f^{5/2} h^{3/2} 
(2{\cal P}_1-{\cal F}) \mu^2}>0
\,,\label{cr2}
\ea
where $c_2$ and $c_4$ are presented 
in Appendix~\ref{AppB}. 
Since $c_{r1,{\rm even}}^2$ is the same as 
$c_{r,{\rm odd}}^2$, only the second propagation 
speed squared~$c_{r2,{\rm even}}^2$  
provides an additional stability condition.

For the monopole mode ($\ell=0$), there is no propagation 
for the gravitational perturbation~$\psi$, while the scalar-field perturbation~$\delta \phi$ 
propagates with the same radial 
velocity as Eq.~(\ref{cr2}). 
For the dipole mode ($\ell=1$), there is 
a gauge degree 
of freedom for fixing $\delta \phi=0$, 
under which the perturbation~$\psi$ propagates 
with the same radial speed squared as Eq.~(\ref{cr2}). 

In the limit of large multipoles~$\ell$, 
the conditions associated with 
the squared angular propagation speeds of even-parity 
perturbations are~\cite{Kase:2021mix}
\be
c_{\Omega \pm}^2=-B_1\pm\sqrt{B_1^2-B_2}>0\,,
\label{cosq}
\ee
where we present the explicit form of 
$B_1$ and $B_2$ in Appendix~\ref{AppB}. 
These conditions are satisfied if
\be
B_1^2 \geq B_2>0 \quad {\rm and} 
\quad B_1<0\,.
\label{B12con}
\ee
The conditions (\ref{cFdef}), (\ref{cGdef}), (\ref{cHdef}), 
(\ref{Kcon}), (\ref{cr2}), and (\ref{B12con}) 
ensure the existence of consistent 
hyperbolic evolution of perturbations, which are essential for 
formulating the well-posed initial value problems.
For the complete proof of the stability of BHs,
as in the case of GR and conventional scalar-tensor theories,
on top of these conditions, we further need to clarify
the absence of mode instabilities and the stability 
at the nonlinear level,
which will not be addressed in this paper.

\section{Generic instabilities of black holes 
with nonvanishing scalar kinetic term on the horizon}
\label{sec:expansion}

In this section, we will show the presence of 
ghost or Laplacian instabilities of BHs 
for a nonvanishing kinetic term on the BH horizon,
\be
X_s\equiv X(r_s) \neq 0\,,
\ee
where $r_s$ denotes the radius of the BH horizon. 
We assume that $X$ is an analytic function of $r$ 
around $r=r_s$. 
Here, we will focus on the asymptotically Minkowski BHs 
and assume that the metric solution contains only 
a single horizon, namely the BH event horizon.
Since the metric component~$h$ vanishes at $r=r_s$,
$X_s$ can be finite only when
the derivative of the scalar field~$\phi'$ diverges on the horizon. 
In shift-symmetric Horndeski theories 
like $G_4 \supset X$ 
or $G_4 \supset (-X)^{1/2}$ 
with $G_2=\eta X-\Lambda$, where $\eta$ and 
$\Lambda$ are constants, 
there are some exact non-asymptotically Minkowski BHs 
with $X_s \neq 0$~\cite{Rinaldi:2012vy,Anabalon:2013oea,Minamitsuji:2013ura,Cisterna:2014nua,Babichev:2017guv}. 

Since the sign of 
the scalar-field kinetic term is $X<0$ for $r>r_s$ 
and $X>0$ for $r<r_s$, the scalar field is spacelike 
outside the horizon and timelike inside the horizon, 
respectively. This means that $X$ undergoes a sudden 
change of the sign across the horizon, so the BH horizon 
corresponds to a singular hypersurface. 
Then, we can define a BH solution only outside the horizon 
in which the scalar field is spacelike.
In Ref.~\cite{Minamitsuji:2022mlv}, it was shown that 
BH solutions with $X_s\neq 0$ generically suffer from 
either ghost or Laplacian instabilities in the domain
where the solution can exist, and hence 
such solutions could not be realistic. 
We will show that the same instability problem 
persists in full Horndeski theories with analytic 
coupling functions when $X_s\ne 0$.

\subsection{
Generic instabilities}

We expand the background metric components 
around $r=r_s$ as
\ba
f&=& f_1(r-r_s)+f_2(r-r_s)^2+\cdots\,, 
\label{fexpan}\\
h&=& h_1(r-r_s)+h_2(r-r_s)^2+\cdots\,,
\label{hexpan}
\ea
where $f_j$ and $h_j$ ($j=1,2,3, \cdots$) are constants. 
Here and in the following, we focus on the standard case in which $h$ and $f$ 
simultaneously approach 0 as $r \to r_s$. 
Note that 
there are some spherically symmetric solutions where $f$ 
does not vanish as $h \to 0$ in
specific Lorentz violating scalar-tensor theories~\cite{Chagoya:2018yna}, 
but we will not consider such cases in the context of Horndeski theories.
Since both $f$ and $h$ are positive outside the horizon, 
we have $f_1>0$ and $h_1>0$. 
We are assuming that $X$ is an analytic 
function of $r$ around the horizon, 
so we can expand $X$ and $\phi$ in the forms
\ba
X &=& X_s+X_1 (r-r_s)+X_2 (r-r_s)^2+\cdots\,,
\label{Xexpan}\\
\phi &=& \phi_s+\phi_1(r-r_s)^{1/2}
+\phi_2(r-r_s)^{3/2}+\cdots\,,
\label{phiexpan}
\ea
where $X_j$, $\hat{\phi}_j$, and $\phi_s$ 
are constants, with 
\be
X_s=-\frac{1}{8} h_1 \phi_1^2\,.
\label{Xsv}
\ee
The expansion of the scalar field~(\ref{phiexpan}) is valid 
only outside the horizon ($r>r_s$), in which regime 
$X_s<0$ for $\phi_1 \neq 0$. 
As we already mentioned, the BH solution can be defined 
only in the domain outside the horizon. 
Hence, it is enough to show instabilities only outside 
the horizon in order to exclude these BH solutions.
Since the expansion of the scalar field~\eqref{phiexpan} 
is valid for $X_s\neq 0$, in this section we focus on 
the solution with $X_s<0$.

In the vicinity of $r=r_s$, the left-hand sides of the background equations~(\ref{back1})--(\ref{back3}) 
reduce, respectively, to 
\begin{align}
    \mE_{tt}&=G_2-2X_sG_{3,\phi}+\fr{2(1-h_1r_s)}{r_s^2}G_4+\fr{4h_1X_s}{r_s}G_{4,X}+4X_sG_{4,\phi\phi}-\fr{2(1+h_1r_s)X_s}{r_s^2}G_{5,\phi}+\mO((r-r_s)^{1/2})\,, \\
    \mE_{rr}&=\sqrt{-2h_1 X_s}\bra{-X_sG_{3,X}+G_{4,\phi}+2X_sG_{4,\phi X}-\fr{X_s}{r_s^2}G_{5,X}}(r-r_s)^{-1/2}+\mO((r-r_s)^0)\,, \\
    \mE_{\theta\theta}&=-\sqrt{\frac{-h_1 X_s}{2}}
    \bra{2G_{4,\phi}-4X_sG_{4,\phi X}+\fr{h_1X_s}{r_s}G_{5,X}+2X_sG_{5,\phi\phi}}(r-r_s)^{-1/2}
    +\mO((r-r_s)^{0})\,,
\end{align}
where the coupling functions~$G_{2,3,4,5}$ and their derivatives 
should be evaluated on the horizon. 
Then, the leading-order terms obey 
the following relations:
\ba
& &
G_2-2X_sG_{3,\phi}+\fr{2(1-h_1r_s)}{r_s^2}G_4+\fr{4h_1X_s}{r_s}G_{4,X}+4X_sG_{4,\phi\phi}-\fr{2(1+h_1r_s)X_s}{r_s^2}G_{5,\phi}
=0\,,\label{backle1}\\
& &
\sqrt{-h_1 X_s} \left( 
-X_sG_{3,X}+G_{4,\phi}+2X_sG_{4,\phi X}
-\fr{X_s}{r_s^2}G_{5,X} \right)=0\,,
\label{backle2}\\
& &
\sqrt{-h_1 X_s} \bra{2G_{4,\phi}-4X_sG_{4,\phi X}+\fr{h_1X_s}{r_s}G_{5,X}+2X_sG_{5,\phi\phi}}=0\,.
\label{backle3}
\ea
For given functions~$G_{2,3,4,5}$, the values 
of $h_1$, $\phi_s$, and $X_s$ are 
fixed by 
solving Eqs.~(\ref{backle1})--(\ref{backle3}) in general. 
Since we are now interested in 
solutions with $X_s \neq 0$, we will focus on the 
case in which the terms inside the parentheses of 
Eqs.~(\ref{backle2}) and (\ref{backle3}) vanish. 
Depending on the coupling functions $G_{3,4,5}$, there are specific 
cases in which the left-hand sides of Eqs.~(\ref{backle2}) or 
(\ref{backle3}) vanish identically.
In such cases, we need to compute their next-to-leading-order 
terms. For example, the next-order contributions to 
$\mE_{rr}$ are given by 
\ba
\mE_{rr}^{(2)}
&\equiv &
-\frac{1}{r_s^2} \{G_2 r_s^2+2G_4 (1-h_1 r_s)
+2X_s [ G_{5,\phi}-3G_{5,\phi} h_1 r_s 
-(G_{2,X} -G_{3,\phi} - 2G_{4,\phi \phi}) r_s^2 
- 2G_{4,X}(1-2h_1r_s) ]
\nonumber \\
& &
-4X_s^2 [ G_{5,\phi X} - (2G_{4,XX}-G_{5,\phi X})
h_1r_s 
+ (G_{3,\phi X} - 2G_{4,\phi \phi X})r_s^2]\}\,.
\label{nextto}
\ea
Later, we will consider specific theories in which 
the equation $\mE_{rr}^{(2)}=0$ needs to be used.

Around the BH horizon, the quantities in Eqs.~(\ref{cFdef})--(\ref{cHdef}) are expanded as 
\begin{align}
    \mF&=2(G_4-X_sG_{5,\phi})+\mO((r-r_s)^{1/2})\,, \\
    \mG&=-\sqrt{2h_1}(-X_s)^{3/2} G_{5,X}(r-r_s)^{-1/2}
    +2(G_4-2X_sG_{4,X}+X_sG_{5,\phi}-2X_s^2G_{5,\phi X})+\mO((r-r_s)^{1/2})\,, \\
    \mH&=2(G_4-2X_sG_{4,X}+X_sG_{5,\phi})+\mO((r-r_s)^{1/2})\,.
\end{align}
Provided that $G_{5,X}(\phi_s,X_s)\ne 0$, we have $c_{r,{\rm odd}}^2=c_{r1,{\rm even}}^2 
={\cal G}/{\cal F} \to \infty$
in the limit of $r\to r_s$. 
Since this signals the strong coupling, 
we require the 
condition
\be
G_{5,X}(\phi_s,X_s)=0\,,
\label{G5Xcon}
\ee
to realize finite values of
$c_{r,{\rm odd}}^2$ and $c_{r1,{\rm even}}^2$. 
We exploit the condition~(\ref{G5Xcon}) 
in the following discussion.
We note that the unusual divergence of $\mG$ for 
$G_{5,X}(\phi_s,X_s) \neq 0$ arises from 
the assumption of 
$X_s \neq 0$ 
with the scalar field expansion~(\ref{phiexpan}).

For the computation of $c_{r2,{\rm even}}^2$, 
we resort to the expansions~(\ref{fexpan})--(\ref{phiexpan}) around $r=r_s$ 
as well as the leading-order background 
Eqs.~(\ref{backle2}) and (\ref{backle3}) with $X_s \neq 0$ 
to eliminate the terms~$G_{4,\phi X}(\phi_s, X_s)$ 
and $G_{5,\phi \phi}(\phi_s, X_s)$.
Then, the radial propagation speed squared 
of the scalar field perturbation~$\delta \phi$ reads
\be
c_{r2,{\rm even}}^2=
\frac{2h_1 X_s \kappa_r}{\zeta_r (r-r_s)}
+{\cal O} ((r-r_s)^{0})\,,
\label{cr2evend}
\ee
where we have defined
\ba
\kappa_r &\equiv & X_s r_s^2 (2X_s G_{3,XX}-G_{3,X})+r_s^2 
(3G_{4,\phi}-4 X_s^2 G_{4,\phi XX})
+2X_s^2 G_{5,XX}\,, \label{kappa_r} \\
\zeta_r &\equiv &  X_s^3 [ 8r_s^2(2G_{4,\phi \phi \phi X}
-G_{3,\phi \phi X})+
16 h_1 r_s (2 G_{4,\phi XX}-G_{5,\phi \phi X})
-4h_1^2 G_{5,XX}-8 G_{5,\phi \phi X}] 
\nonumber \\
& & +X_s^2 \{[ 2h_1 X_1 (G_{3,XX}-2G_{4,\phi XX})+8 G_{4,\phi \phi \phi}]r_s^2
-8 h_1 r_s G_{3,X}+2X_1 h_1 G_{5,XX}\}\nonumber \\
& &
-X_s h_1r_s (G_{3,X}X_1 r_s-24 G_{4,\phi})
+3 h_1 r_s^2 X_1 G_{4,\phi}\,.
\ea
The product~${\cal K}c_{r2,{\rm even}}^2$ is 
expanded as 
\be
{\cal K}c_{r2,{\rm even}}^2=
\frac{\sqrt{-2X_s}h_1^{3/2}r_s^2 
(G_4-2X_s G_{4,X}+G_{5,\phi}X_s)^2 \kappa_r}
{2\zeta^2 (r-r_s)^{3/2}}
+{\cal O} ((r-r_s)^{-1/2})\,,
\label{Kcr2}
\ee
where
\ba
\zeta &\equiv & 2 G_{3,\phi X}
X_s^2 r_s^2+r_s (G_4 h_1 
-2G_{4,\phi \phi}X_s r_s
-4G_{4,X} X_s h_1-4 G_{4,XX} 
X_s^2 h_1-4G_{4,\phi \phi X}X_s^2r_s) \nonumber \\
& &+X_s [3G_{5,\phi}h_1 r_s
+2G_{5,\phi X} X_s (1+h_1 r_s)]\,.
\ea
The necessary condition for avoiding ghost 
or Laplacian instabilities of 
even-parity perturbations is that  
the leading-order contribution to 
${\cal K}c_{r2,{\rm even}}^2$ is positive. 
Since the term~$G_4-2X_s G_{4,X}+G_{5,\phi}X_s$ 
in Eq.~(\ref{Kcr2}) corresponds to the leading-order term 
of ${\cal H}/2$ on the horizon, we require the condition~$G_4-2X_s G_{4,X}+G_{5,\phi}X_s>0$.
Then, the positivity of ${\cal K}c_{r2,{\rm even}}^2$
amounts to the inequality~$\kappa_r>0$, 
under which there is the divergence of 
$c_{r2,{\rm even}}^2$ on the horizon, 
which signals the strong coupling problem.
For $\kappa_r<0$, there is either ghost or 
Laplacian instability along the radial direction. 
Then, so long as
$\kappa_r \neq 0$
on the horizon, we encounter either strong coupling or
ghost/Laplacian instability of even-parity perturbations.
If $\kappa_r=0$, then the leading-order contribution to 
$c_{r2,{\rm even}}^2$ in Eq.~(\ref{cr2evend}) vanishes, 
in which case it may be possible to avoid the 
strong coupling problem.
Even in this case, however, we further require that the 
next-to-leading-order terms of 
Eqs.~(\ref{cr2evend}) and (\ref{Kcr2})
are both positive for the absence of ghost
and Laplacian instabilities.

Around $r=r_s$, the product~${\cal F}{\cal K}B_2$ 
can be expanded as  
\be
{\cal F}{\cal K}B_2=-\frac{4h_1^2 X_s^4 
r_s^4 \kappa^2}{\zeta^2 (r-r_s)^2}
+{\cal O} ((r-r_s)^{-1})\,,
\label{prod}
\ee
where
\be
\kappa \equiv G_4 G_{4,XX}+G_{4,X}^2
-G_{5,\phi X} (G_4-X_s G_{4,X})
-G_{5,\phi} (2G_{4,X}+X_s G_{4,XX}-G_{5,\phi})\,.
\label{xi}
\ee
As long as
$\kappa \neq 0$
on the horizon, the leading-order term of 
Eq.~(\ref{prod}) is negative, i.e., 
\be
{\cal F}{\cal K}B_2<0
\quad {\rm as} \quad 
r \to r_s\,.
\ee
Since one of the quantities~${\cal F}$, ${\cal K}$, 
and $B_2$ must be negative, we cannot avoid either 
ghost or Laplacian instabilities. 
The angular propagation of even-parity perturbations 
plays a crucial role for reaching the conclusion of 
generic instabilities of BH solutions with 
$X_s \neq 0$ and $\kappa \neq 0$. 
For $\kappa=0$, there is a possibility for avoiding the 
above instability, 
in which case the next-to-leading-order 
term of the product~${\cal F}{\cal K}B_2$ must be positive.

In summary, as long as $\kappa_r \neq 0$ or $\kappa \neq 0$, 
the BH solutions with $X_s\ne 0$ in Horndeski theories 
having analytic coupling functions
are subject to either 
intrinsic ghost/Laplacian instabilities 
or strong coupling problems. 

\subsection{\texorpdfstring{Concrete models having BH solutions with $X_s \neq 0$}{Concrete models having BH solutions with nonvanishing Xs}}
\label{BHinsec}

Let us proceed to the discussion of concrete models 
giving rise to BH solutions with $X_s \neq 0$.
In such cases, the BH solutions 
with $X_s \neq 0$ are excluded by the 
instability around the horizon. 

We first discuss the following example in the 
framework of shift-symmetric Horndeski theories:
\be
G_2=\eta X\,,\qquad G_3=\alpha_3 X\,,\qquad 
G_4=\frac{\Mpl^2}{2}+\alpha_4 X\,,\qquad
G_5=\alpha_5 X\,,
\label{linearHo}
\ee
where $\eta$ and $\alpha_{3,4,5}$ are nonvanishing 
constants and $\Mpl$ is the reduced Planck mass.
If all $\alpha_3$, $\alpha_4$, and $\alpha_5$ are nonzero, 
there is no solution with $X_s\neq 0$ for Eqs.~(\ref{backle1})--(\ref{backle3}).
The same conclusion persists if either 
$\alpha_3$ or $\alpha_5$ is vanishing.
On the other hand, if both $\alpha_3$ and $\alpha_5$ 
are zero in Eq.~(\ref{linearHo}), the left-hand sides of 
Eqs.~(\ref{backle2}) and (\ref{backle3}) vanish 
identically, so we exploit the next-to-leading-order 
equation~$\mE_{rr}^{(2)}=0$ together with 
Eq.~(\ref{backle1}). 
Then, there is the following solution:
\be
X_s=\frac{\eta \Mpl^2 r_s^2}
{4 \alpha_4 (\eta r_s^2+2\alpha_4)}\,,\qquad
h_1=\frac{\eta r_s^2+2\alpha_4}{2\alpha_4 r_s}\,,
\ee
which corresponds to non-asymptotically Minkowski hairy
BHs studied in Refs.~\cite{Rinaldi:2012vy,Anabalon:2013oea,Minamitsuji:2013ura,Cisterna:2014nua}. 
Since the coupling functions~$G_2$ and $G_4$
are both analytic functions of $X$, we can resort to the 
analytic expansion of $X$ used in Eq.~(\ref{Xexpan}) 
around the horizon. 
The quantities~$\kappa_r$ and $\kappa$ are given, 
respectively, by 
\be
\kappa_r=0\,,\qquad \kappa=\alpha_4^2\,.
\ee
As long as $\eta \neq 0$ and $\alpha_4 \neq 0$, 
we have $X_s \neq 0$ and $\kappa \neq 0$. 
Then, this solution inevitably suffers from the
ghost or Laplacian instability around the horizon. 
This conclusion agrees with what was found for 
shift- and reflection-symmetric Horndeski theories containing 
the functional dependence~$G_2(X)$ 
and $G_4(X)$ with $G_3=G_5=0$~\cite{Minamitsuji:2022mlv}.

As our second example, let us consider 
the following model:
\be
G_2=\eta X\,,\qquad G_3=\alpha_3 X\,,\qquad
G_4=\frac{\Mpl^2}{2}+\alpha_4 \phi^2+\beta_4 
\phi^2 X\,,\qquad G_5=\alpha_5 \phi^2\,,
\label{secondmo}
\ee
where $\eta$, $\alpha_{3,4,5}$, and $\beta_4$ 
are nonzero constants. We note that all the coupling functions 
in Eq.~(\ref{secondmo}) are analytic functions of 
$\phi$ and $X$.
Solving the leading-order background equations for 
$X_s$, $\phi_s$, and $h_1$,
we find that there is the solution
\be
X_s=\frac{\alpha_4 (\alpha_3+2\alpha_5)}
{\beta_4 (\alpha_3-6\alpha_5)}\,,\qquad 
\phi_s=\frac{\alpha_3+2\alpha_5}{8\beta_4}\,,
\ee
which exists for $\beta_4\neq 0$ and 
$\alpha_3-6\alpha_5\ne 0$.
Here, the explicit form of $h_1$ is 
not shown due to its complexity. 
For this solution, the quantities~$\kappa_r$ 
and $\kappa$ reduce, respectively, to 
\be
\kappa_r=\frac{\alpha_4 (\alpha_3+2\alpha_5)r_s^2}
{2\beta_4}\,,\qquad 
\kappa=\frac{(\alpha_3+2\alpha_5)^2 
(\alpha_3-14\alpha_5)^2}{4096 \beta_4^2}\,.
\ee
The existence of the 
solution with $X_s \neq 0$ requires
that $\alpha_4 (\alpha_3+2\alpha_5) \neq 0$, under 
which $\kappa_r \neq 0$. 
Also, we have $\kappa\ne 0$ unless $\alpha_3=14\alpha_5$.
We note that, even when $\alpha_5=0$, 
there are solutions with $X_s \neq 0$ plagued by 
the instability problem. 

\section{Theories with no-hair Schwarzschild black holes}
\label{nohairsec}

Given the existence of instabilities for the BH solutions 
with $X_s \neq 0$, we are now interested in 
solutions with $X_s=0$. 
For example, the models~(\ref{linearHo}) and (\ref{secondmo}) 
give rise to the branch~$X_s=0$ besides the branch~$X_s \neq 0$ 
discussed in Sec.~\ref{sec:expansion}. 
For the solution with $X_s=0$, there are in general 
two possibilities. 
One is a trivial field profile with 
$\phi(r)=\phi_s={\rm constant}$ at any radius.  
The other is a hairy solution where 
the scalar field varies 
as a function of $r$. For this hairy solution 
with $X_s=0$, the scalar field regular around the horizon 
can be expanded as 
\be
\phi(r)=\phi_s+\phi_1 (r-r_s)
+\phi_2 (r-r_s)^2+\cdots\,.
\label{phise}
\ee
Since this is different from Eq.~(\ref{phiexpan}), 
we need to handle this case separately for the discussion of 
ghost/Laplacian instabilities.
Furthermore, the expansion (\ref{phise}) is insufficient 
to ensure the BH stability 
throughout the horizon exterior.
Thus, in this and subsequent sections, 
we will derive perturbative 
BH solutions with respect to a small parameter
arising from the coupling functions~$G_{2,3,4,5}$ 
and explore the issue of ghost/Laplacian instabilities 
of BHs in the region outside the horizon.

In the horizon limit~$r\to r_s$, the hairy BH solutions 
with $X_s=0$ should satisfy the following boundary conditions:
\be
\label{bc1}
f(r)\to 0\,,
\quad 
h(r)\to 0
\quad 
\text{with}
\quad 
\frac{h(r)}{f(r)}
\to
{\rm finite}\,,
\quad 
\text{and}
\quad
|\phi'(r)| \to 
|\phi_1|<\infty\,.
\ee
On the other hand, at spatial infinity~$r\to \infty$, 
we impose that the metric is asymptotically Minkowski and the scalar-field derivative vanishes, i.e.,
\ba
\label{bc2}
\label{bc_infinity}
f(r)\to {\rm constant}\,,
\quad 
h(r)\to 1\,,
\quad
\phi'(r)\to 0\,.
\ea
This means that the scalar field~$\phi(r)$ approaches a constant value. Here, we do not impose that the asymptotic constant value of the scalar field is a specific value, e.g., zero,
but we allow an arbitrary constant value
as long as it does not conflict with the asymptotically Minkowski metric.
We will construct hairy BH solutions under the boundary 
conditions~\eqref{bc1} and \eqref{bc2}.

Now, we search for the possibility of  
BH solutions with $X_s=0$ which are not 
prone to the generic problem of ghost/Laplacian instabilities found in the previous section.
We recall that the scalar-field equation of motion is given by Eq.~(\ref{Ephi2}).
Outside the horizon, the solution to Eq.~(\ref{Ephi2}) can be expressed in an integrated form
\be
J^r=\frac{Q}{r^2} \sqrt{\frac{h}{f}}
-\frac{1}{r^2} \sqrt{\frac{h}{f}} \int_{r_s}^r
r^2 \sqrt{\frac{f}{h}} P_\phi\,{\rm d}r\,,
\label{Jrso}
\ee
where $Q$ is an integration constant and $J^r$ is the radial current component defined in Eq.~\eqref{Jrc}, which we repeat here for the reader's convenience:
\ba
J^r &=&
h \phi' \biggl[
G_{2,X}-\left( \frac{2}{r}+\frac{f'}{2f} 
\right) h \phi' G_{3,X}+2 \left( \frac{1-h}{r^2}
-\frac{h f'}{rf} \right) G_{4,X}+2h \phi'^2 
\left( \frac{h}{r^2}+\frac{hf'}{rf} \right) G_{4,XX} 
\nonumber \\
&&-\frac{f'}{2r^2f} (1-3h)h \phi' G_{5,X}
-\frac{f' h^3 \phi'^3}{2r^2 f}G_{5,XX} \biggr]\,.
\label{Jrc_re}
\ea
In shift-symmetric Horndeski theories, we have $P_{\phi}=0$, 
and hence the second term on the right-hand side of Eq.~(\ref{Jrso}) vanishes. In non-shift-symmetric Horndeski theories containing the $\phi$-dependence in $G_{2,3,4,5}$, the integral containing $P_{\phi}$ contributes to $J^r$.
Throughout the discussion below, we include the kinetic term~$\eta X$ in $G_2$ and the Einstein-Hilbert term $\Mpl^2/2$ in $G_4$, i.e., 
\be
G_2 \supset \eta X\,,\qquad 
G_4 \supset \frac{\Mpl^2}{2}\,,
\label{G24}
\ee
with $\eta$ being a constant.
\subsection{Shift-symmetric theories}

In shift-symmetric theories, the radial current component 
is given by 
\be
J^r=\frac{Q}{r^2} \sqrt{\frac{h}{f}}\,. \label{Jr_shift-sym}
\ee
Let us first consider 
the simplest case of GR with a linear kinetic term of the scalar field, i.e.,
\be
G_2=\eta X\,, \qquad
G_3=0\,, \qquad
G_4=\frac{\Mpl^2}{2}\,, \qquad
G_5=0\,.
\ee
Since $J^r=\eta h \phi'$ in this case, 
the scalar-field derivative is given by 
\be
\phi'(r)=\frac{Q}{\eta r^2} \frac{1}{\sqrt{fh}}\,.
\ee
As $r$ approaches the horizon radius~$r_s$, 
there is the divergence of $\phi'$. 
To avoid this behavior, we require that $Q=0$ and hence 
\be
\phi'(r)=0\,,
\ee
which corresponds to a no-hair solution. 

We can generalize the above argument to more general theories 
where the coupling functions~$G_{2,3,4,5}$ are analytic functions of $X$, i.e., 
\be
G_I(X)=\sum_{p\geq 0} (\alpha_{I})_p (-X)^p
\qquad (I=2,3,4,5)\,,
\ee
where $(\alpha_{I})_p$ are constants and 
the sum is taken over all integers $p\geq 0$.
Taking into account the terms in Eq.~(\ref{G24}), 
we consider the following coupling functions:
\be
\begin{split}
&
G_2=\eta X+\sum_{p_2\geq 2} (\alpha_2)_{p_2} (-X)^{p_2}\,,\qquad
G_3=\sum_{p_3\geq 1} (\alpha_3)_{p_3} (-X)^{p_3}\,,\\
&
G_4=\frac{\Mpl^2}{2}+\sum_{p_4\geq 1} (\alpha_4)_{p_4} (-X)^{p_4}\,,\qquad
G_5=\sum_{p_5\geq 1} (\alpha_5)_{p_5} (-X)^{p_5}\,.
\label{Gpower}
\end{split}
\ee
We drop constant terms in each $G_{I}$ apart from $\Mpl^2/2$ 
in $G_4$, as they are irrelevant to the existence of asymptotically Minkowski hairy BH solutions.
Using the expansions~(\ref{fexpan}) and (\ref{hexpan})
around the horizon, it follows that the metric functions~$f$, $h$, and their derivatives do not cause 
divergences for the terms appearing in the  
square brackets in Eq.~(\ref{Jrc_re}).
This is also the case at spatial infinity 
where $f$ and $h$ approach 
constants with $\phi'(r) \to 0$ for asymptotically Minkowski solutions.
The radial current component~\eqref{Jrc_re} can be written in the form $J^r=h\phi'[\eta+F(\phi')]$, where $F(\phi')$ is a regular power-law function of $\phi'$ containing only positive powers. 
Note that the constant terms arising from 
the other parts of the coupling 
functions [i.e., $\sum (\alpha_I)_{p_I} (-X)^{p_I}$] 
have been absorbed into $\eta$, and hence $F(0)=0$.
We can rewrite $J^r=(Q/r^2)\sqrt{h/f}$ in the form
\be
h=\frac{1}{\phi'[\eta+F(\phi')]}\frac{Q}{r^2} 
\sqrt{\frac{h}{f}}\,.
\ee
To realize $h=0$ on the horizon for a finite or vanishing 
value of $\phi'$, we require that $Q=0$. Then, we obtain
\be
\label{jr0}
J^r=h\phi'[\eta+F(\phi')]=0\,.
\ee
Under the boundary condition~$\phi'(\infty)=0$, 
the function~$F(\phi')$ approaches zero 
at spatial infinity.
Then, so long as $\eta\ne 0$, we have to choose 
the branch~$\phi'(r)=0$.
This means that, for theories with the coupling functions of the form~(\ref{Gpower}), we end up with no-hair BH solutions.
This fact was already recognized in Ref.~\cite{Hui:2012qt}.

In the above discussion, the main reason for reaching 
the no-hair conclusion is that the dominant contribution to 
$J^r$ in the limit $\phi' \to 0$ is the linear term in $\phi'$. 
This behavior can be avoided by considering the following
nonanalytic coupling functions~\cite{Sotiriou:2013qea,Babichev:2017guv}:
\be
G_2=\eta X+\alpha_2 \sqrt{-X}\,,\qquad 
G_3=\alpha_3 \ln |X|\,,\qquad 
G_4=\frac{\Mpl^2}{2}+\alpha_4 \sqrt{-X}\,,\qquad 
G_5=\alpha_5 \ln |X|\,,
\label{Gchoice}
\ee
with $\alpha_{2,3,4,5}$ being constants.
For $\phi'>0$, we have 
\be
J^r=\eta h \phi'-\sqrt{\frac{h}{2}}\alpha_2+\left( \frac{f'}{f}
+\frac{4}{r} \right)h \alpha_3-\frac{\sqrt{2h}}{r^2} \alpha_4
-\frac{f' h(h-1)}{fr^2} \alpha_5\,.
\label{Jrcon}
\ee
Apart from the first term, there are 
no $\phi'$-dependent terms. 
For any coupling functions with stronger divergence as $X\to 0$ than the choice~\eqref{Gchoice}, $J^r$ diverges as $\phi'\to 0$. 
In such theories, we cannot take the proper Minkowski limit at large 
distances~\cite{Creminelli:2020lxn}.
In this sense, the choice of the coupling functions~(\ref{Gchoice})  
is unique for the realization of hairy BH solutions 
in shift-symmetric Horndeski theories.
From Eq.~\eqref{Jr_shift-sym}, the scalar-field derivative is now expressed as 
\be
\phi'=\frac{1}{\eta h}
\left[ \sqrt{\frac{h}{2}}\alpha_2-\left( \frac{f'}{f}
+\frac{4}{r} \right)h \alpha_3+\frac{\sqrt{2h}}{r^2} \alpha_4
+\frac{f' h(h-1)}{fr^2} \alpha_5+\frac{Q}{r^2} \sqrt{\frac{h}{f}} \right]\,.
\label{phihairy}
\ee
The terms associated with $\alpha_2$ and $\alpha_4$ are 
dominated over the term~$(Q/r^2)\sqrt{h/f}$ around the horizon. 

In the case of $\alpha_3=\alpha_5=0$,
to avoid the divergence of $\phi'$ 
induced by the term~$(Q/r^2)\sqrt{h/f}$ 
for $\alpha_2 \neq 0$ or $\alpha_4 \neq 0$,  
we need to set $Q=0$.
However, even under $Q=0$, $\phi'(r)$ still diverges on the horizon. 
Indeed, the scalar-field kinetic term 
has the following dependence:
\ba
X &=&
-\frac{\alpha_2^2}{4\eta^2} \qquad (\alpha_2 \neq 0,~\alpha_4=0)\,,\\
X &=& -\frac{\alpha_4^2}{\eta^2 r^4}
\qquad (\alpha_4 \neq 0,~\alpha_2=0)\,.
\ea
In both cases, 
$X$ is an analytic function of $r$  
which is nonvanishing at $r=r_s$. Hence, these BH solutions 
are excluded by the intrinsic instability problem discussed in Sec.~\ref{sec:expansion}.

In the case of $\alpha_2=\alpha_4= 0$,
the terms associated with 
$\alpha_3$ and $\alpha_5$ as well as the term~$(Q/r^2)\sqrt{h/f}$ in the square brackets in Eq.~(\ref{phihairy})
approach constants as $r \to r_s$. 
In such cases, we can choose 
$Q$ such that $\phi'$ becomes 
regular on the horizon. 
Using the expansions~(\ref{fexpan}) and (\ref{hexpan}), 
the values of $Q$ and $\phi'$ at $r=r_s$ 
are given by  
\ba
Q &=& \alpha_3 \sqrt{f_1 h_1}r_s^2\,,\qquad 
\phi'(r_s)=-\frac{\alpha_3 (f_1 h_2 r_s+3f_2 h_1 r_s+12f_1 h_1)}{2\eta f_1 h_1 r_s}
\qquad (\alpha_3 \neq 0,~\alpha_5=0)\,,
\label{cubicQ}
\\
Q &=& \alpha_5 \sqrt{f_1 h_1}
\,,\qquad 
\phi'(r_s)=\frac{\alpha_5 (2f_1 h_1^2-f_1 h_2-3f_2 h_1)}{2\eta f_1 h_1 r_s^2}
\qquad
(\alpha_5 \neq 0,~\alpha_3=0)\,.
\label{quinticQ}
\ea
In both cases, the scalar-field derivative is finite on the horizon 
and hence $X_s=0$. 
As we will show in Sec.~\ref{cusec}, 
the BHs present for the cubic coupling 
case~(\ref{cubicQ}) 
are not asymptotically Minkowski.
On the other hand, the quintic coupling 
case~(\ref{quinticQ}) realizes 
asymptotically Minkowski hairy BH solutions. 
Indeed, this is equivalent to the 
scalar field linearly coupled to the Gauss-Bonnet term studied in 
Refs.~\cite{Sotiriou:2013qea,Sotiriou:2014pfa}.

\subsection{Non-shift-symmetric theories}
\label{nonshiftsec}

Let us investigate the possibility for realizing hairy BH 
solutions in non-shift-symmetric theories containing the 
dependence of $\phi$ as well as $X$ in $G_{2,3,4,5}$. 
For the $X$-dependent part of the couplings, we take one 
of the powers~$(-X)^{p_I}$ in Eq.~(\ref{Gpower}) 
for each $I=2,3,4,5$ for simplicity.
Multiplying such terms with $\phi$-dependent analytic 
functions~$\alpha_I(\phi)$, 
we can consider the following couplings:
\be
G_2=\eta X+\alpha_2(\phi) (-X)^{p_2}\,,\qquad
G_3=\alpha_3(\phi) (-X)^{p_3} \,,\qquad
G_4=\frac{\Mpl^2}{2}+\alpha_4(\phi) (-X)^{p_4}\,,\qquad
G_5=\alpha_5(\phi) (-X)^{p_5}\,,
\label{Gpower2}
\ee
where $p_{2,3,4,5} \geq 0$ are integers. 
Since $\alpha_{2,3,4,5}(\phi)$ are analytic functions 
of $\phi$, they can be expanded around 
some $\phi_0$ as
\be
\alpha_I(\phi)=\sum_{q_I\geq 0} (\alpha_I)_{q_I} 
\left( \phi-\phi_0 \right)^{q_I}
\qquad (I=2,3,4,5)\,,
\label{alphaI}
\ee
where $(\alpha_I)_{q_I}$ are constants.
For $p_{2,3,4,5}=0$, the purely $\phi$-dependent couplings in 
$G_{2,3,4,5}$ can be accommodated.

Analogous to the discussion in shift-symmetric theories~\cite{Minamitsuji:2022mlv}, 
the radial current component for the couplings~(\ref{Gpower2}) can be expressed in the form~$J^r=h\phi'[\eta+\tilde{F}(\phi',\phi)]$, 
where $\tilde{F}(\phi',\phi)$ is
an analytic function containing the positive power-law 
dependence of $\phi'$ and $\phi$. 
Assuming that $P_{\phi}$ is finite in the limit of the horizon, which will be confirmed in the perturbative approach later,
the last integral in Eq.~(\ref{Jrso}) vanishes as $r \to r_s$. 
Then, on the horizon, there is the relation
\be
h=\frac{1}{\phi'_s[\eta+\tilde{F}(\phi'_s,\phi_s)]}
\frac{Q}{r_s^2} \sqrt{\frac{h}{f}}\,,
\ee
with $\phi'_s\equiv \phi'(r_s)$.
Since $h$ is vanishing at $r=r_s$ with finite 
values of $\phi'_s$ and $\phi_s$, we require that $Q=0$. 
Then, from Eq.~(\ref{Jrso}), we obtain 
\be
h\phi'[\eta+\tilde{F}(\phi',\phi)]
=-\frac{1}{r^2} \sqrt{\frac{h}{f}} \int_{r_s}^r
r^2 \sqrt{\frac{f}{h}} P_\phi\,{\rm d}r\,.
\label{Jreq}
\ee
Regarding the finiteness of $P_\phi$ in the horizon limit~$r\to r_s$, we note that 
$P_{\phi}$ contains the second scalar-field derivative~$\phi''$ as well as the contributions from $f$, $h$, and their derivatives [see Eq.~(\ref{Pphide}) with Eq.~(\ref{lambda})].
This means that we need to integrate Eq.~(\ref{Ephi2}) together with the other background equations~(\ref{back1}) and (\ref{back2}) to determine the value of $P_{\phi}$.
We will explicitly see this by using 
perturbative solutions in the small 
coupling regime.

For the purpose of deriving perturbative BH solutions, we write 
the coupling functions~$\alpha_{I}$ ($I=2,3,4,5$) 
in Eq.~(\ref{Gpower2}) as 
\be
\alpha_I(\phi)=\alpha \tilde{\alpha}_I(\phi)\,,
\ee
where $\alpha$ is a dimensionless coupling constant 
and $\tilde{\alpha}_{2,3,4,5}(\phi)$ are analytic 
functions of $\phi$. 
This ansatz allows us to control the deviation from GR with 
a canonical scalar field by a single parameter~$\alpha$.
Since we are considering BH solutions with the vanishing kinetic term on the event horizon, 
$X_s=0$,
the scalar-field derivative~$\phi'(r)$ (and the scalar field itself) is finite and regular on the horizon. 
We perform the perturbative expansions of $f$, $h$, and $\phi$ with respect to 
the small coupling constant $\alpha$ around the Schwarzschild 
background given by the metric components 
$f=h=1-2m/r$, where $m$ is a constant.
Namely, we consider the metric and scalar field given by
\ba
& &
f(r)=
\left(1-\frac{2m}{r}\right)
\left[
1+\sum_{j\ge 1}
\hat{f}_j(r) \alpha^j
\right]^2\,,
\qquad
h(r)
=
\left(1-\frac{2m}{r}\right)
\left[
1+\sum_{j\ge 1}
\hat{h}_j(r)\alpha^j
\right]^{-2}\,,
\label{sch_pert}\\
& &
\phi(r)= \phi_0(r)+
\sum_{j\ge 1}
\hat{\phi}_j(r) \alpha^j\,,
\label{sch_pert2}
\ea
where the coefficients~$\hat{f}_j$, $\hat{h}_j$, $\phi_0$, 
and $\hat{\phi}_j$ are functions of $r$. 
We note that, for the perturbative 
ansatze~\eqref{sch_pert} and \eqref{sch_pert2}, 
the horizon distance~$r_s$ is given by $r_s=2m$.
We require the validity of the perturbative expansion~\eqref{sch_pert} and \eqref{sch_pert2} with respect to a 
small coupling constant $|\alpha|\ll 1$
and impose that the coefficients~$\hat{f}_j(r)$, $\hat{h}_j(r)$, and $\hat{\phi}_j(r)$ are regular $r=2m$.
Moreover, since $\phi$ is a scalar quantity,
its value does not depend on the choice of the coordinates.
If $\phi$ is divergent at $r=2m$ in the original coordinates,
it would also diverge at the corresponding position in the new coordinates.
On the other hand, 
a divergence of the metric at $r=2m$ might be 
removed by an appropriate coordinate transformation.
However, since the metric and scalar field are coupled in our system,
a choice of the integration constants different from the below would result in the 
divergence of the scalar field.
Thus, the constructed perturbative solution~\eqref{sch_pert} and \eqref{sch_pert2}
indeed exists, irrespective of the choice of the coordinates.
The coupling functions~$\tilde{\alpha}_I(\phi)$ ($I=2,3,4,5$) 
can be expanded as 
\be
\tilde{\alpha}_I(\phi)=\tilde{\alpha}_I(\phi_0)
+\sum_{n\geq 1} \tilde{\alpha}^{(n)}_I(\phi_0)
\frac{(\phi-\phi_0)^n}{n!}\,,
\label{tial}
\ee
where $\tilde{\alpha}^{(n)}_I(\phi_0)
\equiv {\rm d}^n \tilde{\alpha}_I/{\rm d}\phi^n |_{\phi=\phi_0}$.
We derive the BH solutions perturbatively by using the dimensionless 
parameter~$\alpha$ arising from each coupling~$\alpha_{2,3,4,5}$.
By construction, the no-hair argument given below is valid 
in the regime of the small coupling constant, $|\alpha|\ll 1$.

In Eqs.~(\ref{Jrc_re}) and (\ref{Pphide}), the zeroth-order terms 
in $\alpha$ are $J^r=\eta h \phi'=(1-2m/r) \eta \phi'$ and $P_{\phi}=0$. 
From Eq.~(\ref{Ephi2}), the zeroth-order scalar field 
obeys the differential equation 
\be
\left[ r^2 \left( 1 -\frac{2m}{r} \right) \phi_0'(r) \right]'=0\,,
\label{phidif}
\ee
whose solution is given by 
\be
\phi_0(r)=\tilde{\phi}_0+\frac{D_0}{2m} \ln \left( 1-\frac{2m}{r} 
\right)\,,
\ee
where $\tilde{\phi}_0$ and $D_0$ are integration constants. 
To avoid the divergent behavior of $\phi_0(r)$ on the horizon at $r=2m$, we 
require that $D_0=0$ and hence 
$\phi_0(r)=\tilde{\phi}_0={\rm constant}$. 
In the following, we discuss two different cases: 
1) $p_2\neq 0$ and 
2) $p_2=0$, in turn, 
where $p_2$ is the power 
appearing in the coupling function~$G_2$ in Eq.~(\ref{Gpower2}).

\subsubsection{\texorpdfstring{$p_2 \neq 0$}{Nonvanishing p2}}

For $p_2 \neq 0$, the first-order expanded solutions in $\alpha$ are given by 
\be
\hat{\phi}_1(r)=\tilde{\phi}_1+\frac{D_1}{2m} \ln 
\left( 1-\frac{2m}{r} \right)\,,\qquad 
\hat{h}_1(r)=\frac{C_1}{r-2m}\,,\qquad 
\hat{f}_1(r)=-\frac{C_1}{r-2m}+C_2\,,\qquad 
\label{h1ex}
\ee
where $D_1, C_1, C_2$ are integration constants. 
For the regularity at $r=2m$, we require that 
$D_1=0$ and $C_1=0$, and we also set $C_2=0$ 
by a suitable time reparametrization. 
Then, the first-order solution is given by $\hat{\phi}_1(r)=\tilde{\phi}_1$, 
$\hat{h}_1(r)=0$, and $\hat{f}_1(r)=0$.
One can show that the $j$th-order perturbative 
solutions ($j \geq 2$) are of the same form 
as Eq.~(\ref{h1ex}).
Then, the solutions regular on the horizon are
\be
\hat{\phi}_j(r)=\tilde{\phi}_j\,,\qquad 
\hat{h}_j(r)=0\,,\qquad 
\hat{f}_j(r)=0\,,
\label{hathf}
\ee
where $\tilde{\phi}_j$ are constants.
Since these relations hold for all integers~$j$, this 
corresponds to the no-hair BH solution [i.e., $\phi(r)={\rm constant}$] with the Schwarzschild metric components~$f(r)=h(r)=1-2m/r$.

\subsubsection{\texorpdfstring{$p_2=0$}{Vanishing p2}}

Let us consider the theories with 
$p_2=0$, i.e., the coupling~$\alpha_2(\phi)=\alpha \tilde{\alpha}_2(\phi)$ in $G_2$.
The first-order solutions of $h$ and $f$ regular at $r=2m$ 
are given by
\be
\hat{h}_1(r)=-\frac{4m^2+2mr+r^2}{6\Mpl^2} 
\tilde{\alpha}_2(\phi_0)\,,\qquad 
\hat{f}_1(r)=\frac{(2m+r)r}{6\Mpl^2}
\tilde{\alpha}_2(\phi_0)\,,
\ee
where we have set the integration constant 
in $\hat{f}_1(r)$ to be zero.
To realize the asymptotically Minkowski metric, we require that 
$\tilde{\alpha}_2(\phi_0)=0$.
The first-order solution to the scalar field is given by 
\be
\hat{\phi}_1(r)=\tilde{\phi}_1+\frac{D_1}{2m} \ln 
\left( 1-\frac{2m}{r} \right)
-\frac{[r^2+4mr+8m^2 
\ln (r/2m)]}{6 \eta}
\tilde{\alpha}_{2}^{(1)}(\phi_0)\,,
\ee
with $D_1$ being an integration constant.
We impose $\tilde{\alpha}_{2}^{(1)}(\phi_0) = 0$, as otherwise 
the last term leads to the divergence of $\hat{\phi}_1(r)$ 
at spatial infinity and the 
condition~\eqref{bc2}
[i.e., $\hat{\phi}_1'(\infty)=0$]
is not respected.
Then, the regularity of $\hat{\phi}_1'(r)$ at $r=2m$ requires that 
$D_1=0$, and hence $\hat{\phi}_1(r)=\tilde{\phi}_1={\rm constant}$.
To avoid the divergences of $\hat{\phi}_j'(r)$, $\hat{h}_j(r)$, and $\hat{f}_j(r)$ at spatial infinity for higher-order 
solutions ($j \geq 2$), we require 
that $\tilde{\alpha}_2^{(n)}(\phi_0)=0$ for all $n~(\geq 1)$. 
Hence, we end up with the no hair solution characterized by $\hat{\phi}_j(r)=\tilde{\phi}_j={\rm constant}$ 
and $\tilde{\alpha}_2(\phi)=\tilde{\alpha}_2(\phi_0)=0$.

The above discussion shows that, in theories with 
the coupling functions~(\ref{Gpower2}), the asymptotically Minkowski BH solutions 
respecting the regularity on the horizon 
are restricted to be no-hair solutions
with $\phi(r)={\rm constant}$. 
Substituting $\phi'(r)=0$ and $\phi''(r)=0$ into 
the expression of $P_\phi$ given in Eq.~(\ref{Pphide}) 
and using the property that the $\phi$- and $X$-derivatives 
of the couplings~$G_{2,3,4,5}$ 
do not contain negative powers of $\phi'$, 
it follows that $P_{\phi}=G_{2,\phi}
+\lambda_4 G_{4,\phi}$, where $\lambda_4$ 
is defined in Eq.~(\ref{lambda}).
Since the background geometry is the Schwarzschild metric, 
the quantity~$\lambda_4$ vanishes. 
The coupling~$\alpha_2(\phi)(-X)^{p_2}$ with 
$p_2=0$ gives rise to $\phi$-dependent contributions 
in $G_{2,\phi}$, but they vanish due to the property~$\tilde{\alpha}_2^{(n)}(\phi_0)=0$ 
derived above.
Then, we have $P_{\phi}=0$, so the right-hand side of 
Eq.~(\ref{Jreq}) vanishes. Since $\tilde{F}(\phi'_s,\phi_s)$ 
does not contain negative powers of $\phi'$, 
the no-hair solution with $\phi'(r)=0$ is consistent 
with Eq.~(\ref{Jreq}) everywhere outside the horizon. 

The above results show that, for the theories 
characterized by the coupling functions~(\ref{Gpower2}) 
with (\ref{alphaI}), there are no asymptotically Minkowski 
BHs with scalar hair. 
Such theories include couplings of the forms~$G_I  \supset \phi^{q_I}(-X)^{p_I}$
($I=2,3,4,5$)\,,
with integers~$q_I \geq 0 $ and $p_I \geq 0$.
The no-hair property persists for the product of 
two analytic functions~$\alpha_I(\phi)$ and 
$F_I(X)$, i.e., $G_I \supset \alpha_I(\phi)F_I(X)$.

As in the case of shift-symmetric theories, the possibility 
for evading the no-hair property of BHs is 
to choose couplings with specific 
non-analytic functions of $X$. 
In shift-symmetric theories, 
for the coupling functions~\eqref{Jrcon},
the $X$-dependences in $G_{2,3,4,5}$ are uniquely fixed in 
such a way that they give rise to terms without containing 
the $\phi'$-dependence in $J^r$, except for $\eta h \phi'$~\cite{Sotiriou:2013qea,Babichev:2017guv}. 
This can be straightforwardly extended to non-shift-symmetric 
theories by multiplying analytic functions 
of $\phi$ to each non-analytic functions of $X$ as
\be
G_2=\eta X+\alpha_2(\phi) \sqrt{-X}\,,\qquad 
G_3=\alpha_3(\phi) \ln |X|\,,\qquad 
G_4=\frac{\Mpl^2}{2}+\alpha_4(\phi) \sqrt{-X}\,,\qquad 
G_5=\alpha_5(\phi) \ln |X|\,,
\label{Gchoicef}
\ee
where $\alpha_I (\phi)$'s ($I=2,3,4,5$) are 
analytic functions of $\phi$.
From Eq.~(\ref{Jrso}), we obtain
\be
\phi'=\frac{1}{\eta h}
\left[ \sqrt{\frac{h}{2}}\alpha_2(\phi)-\left( \frac{f'}{f}
+\frac{4}{r} \right)h \alpha_3(\phi)+\frac{\sqrt{2h}}{r^2} \alpha_4(\phi)
+\frac{f' h(h-1)}{fr^2} \alpha_5(\phi)+\frac{Q+Q_{\phi}}{r^2} \sqrt{\frac{h}{f}} \right]\,,
\label{phihairy2}
\ee
where 
\be
Q_{\phi}=-\int_{r_s}^r
r^2 \sqrt{\frac{f}{h}} P_{\phi}\,{\rm d}r\,.
\label{Qphi}
\ee
Around the horizon, we can use the expansions~(\ref{fexpan}) 
and \eqref{hexpan} of the metric 
components and the expansion~(\ref{phise})
of the scalar field which is valid for $X_s=0$. 
Then, the leading-order terms of $P_{\phi}$ 
for the couplings~$G_2$, $G_3$, $G_4$, $G_5$ 
in Eq.~(\ref{Gchoicef}) are 
proportional to $\sqrt{r-r_s}$, $\ln (r-r_s)$, 
$1/\sqrt{r-r_s}$, and $\ln (r-r_s)$ in the vicinity of $r=r_s$,
respectively. 
Even in those cases, however, the integral~(\ref{Qphi}) 
vanishes for $r \to r_s$, i.e., $Q_{\phi}=0$.
This means that the discussion 
performed in shift-symmetric theories 
can be applied to the present $\phi$-dependent couplings 
as well. 

For the quadratic and quartic couplings, we need to choose $Q=0$ 
as in shift-symmetric theories, but the scalar-field kinetic term on the horizon reduces to 
$X_s=-\alpha_2^2(\phi_s)/(4\eta^2)$ and $X_s=-\alpha_4^2(\phi_s)/(\eta^2 r_s^4)$, 
respectively. The results of Sec.~\ref{sec:expansion} show that these solutions 
suffer from ghost or Laplacian instabilities around the horizon.
We note that, for $X_s\neq 0$ 
where we employ the expansion~(\ref{phiexpan}), 
the leading-order terms of $P_{\phi}$
for the couplings~$G_2$ and $G_4$ in Eq.~(\ref{Gchoicef}) are 
proportional to $(r-r_s)^0$ and $1/\sqrt{r-r_s}$ in the vicinity of $r=r_s$, respectively. 
As in the case of $X_s=0$, all these contributions lead to 
$Q_{\phi}=0$ in the limit of $r\to r_s$.

For the cubic and quintic couplings, the charge~$Q$ 
should be chosen to 
realize the regular behavior of $\phi'$ on the horizon. 
Indeed, we obtain the same expressions of $Q$ and $\phi'(r_s)$
as those given by Eqs.~(\ref{cubicQ}) and (\ref{quinticQ}) 
for the cubic and quintic couplings, respectively, 
with the replacements~$\alpha_3 \to \alpha_3 (\phi_s)$ and 
$\alpha_5 \to \alpha_5 (\phi_s)$.
At least in the vicinity of the horizon, these solutions have scalar hair 
characterized by finite values of $\phi'(r_s)$ and $X_s~(=0)$, 
so they are not subject to the instability problem discussed in Sec.~\ref{sec:expansion}.
However, this is not enough to ensure the existence of asymptotically Minkowski
hairy BH solutions throughout the horizon exterior. 

In Sec.~\ref{cusec}, we will study hairy BH solutions for the cubic 
coupling $G_3=\alpha_3(\phi) \ln |X|$ and show that they 
are not asymptotically Minkowski in general.
For the quintic coupling~$G_5=\alpha_5(\phi) \ln |X|$, 
there exist asymptotically Minkowski hairy BH solutions
for constant $\alpha_5$. In non-shift-symmetric theories, 
we need other couplings besides $G_5=\alpha_5(\phi) \ln |X|$
for the realization of regular BH solutions with scalar hair.
We will address these issues in Secs.~\ref{quinlogsec} and \ref{quinsec}.

\subsection{Cubic logarithmic couplings}
\label{cusec}

We study the possibility for realizing asymptotically Minkowski BH 
solutions in theories containing cubic logarithmic couplings 
given by 
\be
G_2=\eta X\,,\qquad 
G_3=
\alpha_3(\phi) \ln |X|\,,\qquad 
G_4=\frac{\Mpl^2}{2}\,,\qquad 
G_5=0\,,
\label{logG3}
\ee
where
$\alpha_3(\phi)$ is an analytic function of $\phi$.
In what follows, we write $\alpha_3(\phi)=\alpha \gamma(\phi)$, where $\alpha$ is a constant which we assume to be small.
Performing the expansions~(\ref{sch_pert}) and (\ref{sch_pert2}) 
with respect to the small parameter~$\alpha$, the leading-order 
solution to the scalar field is $\phi_0(r)=\phi_0={\rm constant}$. 
The first-order solutions in $\alpha$ are given by 
\be
\hat{h}_1(r)=\frac{C_1}{r-2m}\,,\qquad 
\hat{f}_1(r)=-\frac{C_1}{r-2m}+C_2\,,\qquad 
\hat{\phi}_1(r)=\tilde{\phi}_1+\frac{C_3}{2m} \ln 
\left( 1-\frac{2m}{r} \right)
-\frac{4\gamma (\phi_0)}{\eta} 
\ln\bra{\fr{r}{2m}}\,,
\label{h1ex2}
\ee
where $C_{1,2,3}$ and $\tilde{\phi}_1$ are integration constants.
The regularity of $\hat{h}_1(r)$, $\hat{f}_1(r)$, 
and $\hat{\phi}_1(r)$ at $r=2m$ imposes that 
$C_1=0$ and $C_3=0$,
and a suitable time reparametrization allows 
us to choose $C_2=0$.
We then obtain 
\be
\hat{h}_1(r)=0\,,\qquad 
\hat{f}_1(r)=0\,,\qquad 
\hat{\phi}_1(r)=\tilde{\phi}_1-\frac{4\gamma (\phi_0)}{\eta} 
\ln\bra{\fr{r}{2m}}\,.
\label{h1ex3}
\ee
At second order in $\alpha$, the integrated solutions 
of the metric components are given by 
\be
\hat{h}_2(r)=\frac{4m \gamma(\phi_0)^2}
{\eta \Mpl^2 (r-2m)}\ln \left( \frac{r}{2m} \right)\,,
\qquad
\hat{f}_2(r)=
-\frac{4 \gamma(\phi_0)^2(r-m)}
{\eta \Mpl^2 (r-2m)} \ln\bra{\frac{r}{2m}}
\,.
\label{f2log}
\ee
In the following, we will discuss shift-symmetric and 
non-shift-symmetric theories separately.

\subsubsection{Shift-symmetric theories}

We first consider the case 
\be
\gamma(\phi)=\gamma_0={\rm constant}\,.
\ee
Then, the metric components~(\ref{f2log}) 
have the asymptotic behavior~$\hat{h}_2(r) \to 0$ and  
$\hat{f}_2(r) \to -[4\gamma_0^2/(\eta \Mpl^2)] \ln [r/(2m)]$ 
at spatial infinity.
The logarithmic divergence of $\hat{f}_2(r)$ 
can be eliminated by imposing $\gamma_0=0$, 
but in this case we end up with 
the no-hair Schwarzschild solution.

\subsubsection{Non-shift-symmetric theories}

Let us next proceed to the case in which $\gamma(\phi)$ is a nontrivial analytic function of $\phi$. 
Analogous to Eq.~(\ref{tial}), we can 
expand $\gamma(\phi)$ around $\phi=\phi_0$ as
\be
\gamma(\phi)=\gamma(\phi_0)
+\sum_{n\geq 1}\gamma^{(n)}(\phi_0)
\frac{(\phi-\phi_0)^n}{n!}\,,
\label{tial2}
\ee
where $\gamma^{(n)}(\phi_0)
\equiv {\rm d}^n \gamma/{\rm d}\phi^n |_{\phi=\phi_0}$. 
From Eq.~(\ref{f2log}), the metric can be asymptotically Minkowski only if
\be
\gamma(\phi_0)=0\,.
\label{tial2a}
\ee
Deriving the higher-order solutions in $\alpha$, 
we find that the regularity of $\hat{\phi}_j$ ($j \geq 2$) 
on the horizon requires the conditions 
\be
\gamma^{(n)}(\phi_0)=0 \qquad (n \geq 1)\,.
\label{tial2b}
\ee
Namely, the $G_3$~term in the action must be absent.
In this case, we obtain the following regular solutions 
\be
\hat{h}_j(r)=0\,,\qquad 
\hat{f}_j(r)=0\,,\qquad 
\hat{\phi}_j(r)=\tilde{\phi}_j={\rm constant}\,,
\ee
for all $j$. 
Then, we obtain  
the no-hair Schwarzschild solution with $\phi={\rm constant}$ 
due to the absence of the $G_3$~term in the action.

\vspace{0.3cm}

We thus showed that, for both the shift-symmetric and non-shift-symmetric 
forms of the cubic logarithmic couplings~(\ref{logG3}), 
asymptotically Minkowski hairy BH solutions cannot be realized.

\subsection{Quintic logarithmic couplings}
\label{quinlogsec}

Let us finally discuss the model with the quintic 
logarithmic coupling only,
\be
G_2=\eta X\,,\qquad 
G_3=0\,,\qquad 
G_4=\frac{\Mpl^2}{2}\,,\qquad 
G_5=\alpha_5(\phi) \ln |X|\,,
\label{logG5}
\ee
where $\alpha_5(\phi)$ is a regular function of $\phi$. 
Except for shift-symmetric theories with $\alpha_5={\rm constant}$, 
this model generally gives rise to terms of the  
form~$(-X)^p \ln |X|$ ($p \geq 1$) 
in the background equations of motion. Although they themselves 
vanish in the limit~$X \to 0$, higher-order $X$-derivatives 
of them diverge for $X \to 0$. 
In this case, we may have the problem of instability 
at the level of higher-order perturbations. 
In the process of deriving perturbative solutions with 
the expansions~(\ref{sch_pert}) and (\ref{sch_pert2}), 
we also encounter terms like $\alpha^p \ln |\alpha|$, 
so the power-law expansions with respect to $\alpha$ 
lose their validity.
The quintic couplings like 
$G_5=[1+\sum_{p=1}C_p (-X)^p]\alpha_5(\phi) 
\ln |X|$, which reduce to $G_5 \to \alpha_5(\phi) \ln |X|$ 
in the limit~$X \to 0$,
again generate terms 
of the form~$(-X)^p \ln |X|$ in the background equations. 

As we will see in Sec.~\ref{quinsec},
the scalar-GB coupling~$\xi(\phi)R_{\rm GB}^2$, which amounts to the coupling functions of the form~\eqref{Ggauss},
corresponds to the only exceptional case in which 
$\ln |X|$-dependent terms completely 
disappear from the background equations.
In this case, since we do not face the aforementioned
problem, we can resort to the expansions~(\ref{sch_pert}) and (\ref{sch_pert2}) to derive BH solutions perturbatively. 
As long as all the coupling functions in Eq.~(\ref{Ggauss}) are present, 
we can add other regular functions like $(-X)^p$ and $\phi^q$ 
in the coupling functions
to see how the structure of hairy BH solutions 
is modified.
We will also address this issue in Sec.~\ref{GBother}.

\section{BHs in the presence of 
the Gauss-Bonnet coupling}
\label{quinsec}

In this section, we study the existence of asymptotically Minkowski
BH solutions related to the GB coupling~$\xi(\phi) R_{\rm GB}^2$, 
where $\xi(\phi)$ is a function of $\phi$ and $R_{\rm GB}^2$ 
is the GB term defined by 
\be
\label{def_gb}
R_{\rm GB}^2
\equiv R^2-4R_{\alpha\beta}R^{\alpha\beta}
+R_{\alpha\beta\mu\nu}R^{\alpha\beta\mu\nu}\,,
\ee
with $R_{\alpha\beta}$ and $R_{\alpha\beta\mu\nu}$ being the 
Ricci and Riemann tensors associated with the metric~$g_{\mu\nu}$, respectively.
In the language of Horndeski theories, the Lagrangian~$\xi (\phi)R_{\rm GB}^2$ is equivalent to the combination 
of the following couplings~\cite{KYY,Langlois:2022eta}:
\ba
& &
G_2=8 \xi^{(4)}(\phi) X^2 (3-\ln |X|)\,,\qquad 
G_3=4\xi^{(3)}(\phi) X (7-3\ln |X|)\,,\nonumber \\
& &
G_4=4\xi^{(2)}(\phi) X (2-\ln |X|)\,,\qquad
G_5=-4\xi^{(1)}(\phi) \ln |X|\,,
\label{Ggauss}
\ea
where $\xi^{(n)}(\phi) \equiv {\rm d}^n \xi(\phi)/{\rm d} \phi^n$.\footnote{Note that the Horndeski Lagrangian with $G_2=-8 \xi^{(4)}(\phi) X^2$, $G_3=-12\xi^{(3)}(\phi) X$, $G_4=-4\xi^{(2)}(\phi) X$, and $G_5=-4\xi^{(1)}(\phi)$ is a total derivative for any smooth function~$\xi(\phi)$, and hence a simultaneous multiplication of a constant in each logarithmic function in Eq.~\eqref{Ggauss} does not matter.
Therefore, it is legitimate to replace $\ln|X|\to \ln|X/X_0|$ with $X_0$ being a constant of mass dimension four to make the argument of the logarithmic function dimensionless.}
We note that the form of $G_5$ is identical to that 
given in Eq.~\eqref{Gchoicef}.
The linear GB coupling~$\xi (\phi)=\alpha\phi$, where 
$\alpha$ is constant, corresponds 
to the quintic interaction~$G_5=-4 \alpha \ln |X|$
with $G_{2,3,4}=0$. In this class of shift-symmetric theories, 
it is known that there exist asymptotically Minkowski
BHs endowed with scalar hair~\cite{Sotiriou:2013qea,Sotiriou:2014pfa}.
The recent analysis of Ref.~\cite{Minamitsuji:2022mlv} showed that 
these BH solutions satisfy the conditions for the absence of 
ghost and Laplacian instabilities of odd- and 
even-parity perturbations.
Moreover, the propagation speeds of all the perturbation modes 
approach unity in the asymptotic infinity ($r\to \infty$).

For power-law GB couplings given by $\xi(\phi)=\alpha\phi^n$ 
with $n\geq 2$, which no longer respect the shift symmetry,
asymptotically Minkowski hairy BHs have been obtained 
numerically~\cite{Antoniou:2017acq}. 
For more general GB couplings where $\xi(\phi)$ is a generic 
analytic function of $\phi$, 
we will construct solutions by using the method of perturbative expansions~(\ref{sch_pert}) and (\ref{sch_pert2})
valid for the small dimensionless coupling constant~$|\alpha|\ll 1$.
Although general couplings accommodate models of BH spontaneous scalarization~\cite{Doneva:2017bvd,Silva:2017uqg,Antoniou:2017acq,Blazquez-Salcedo:2018jnn,Minamitsuji:2018xde,Silva:2018qhn,Macedo:2019sem}
and nonlinear scalarization~\cite{Doneva:2021tvn} as well,
our construction with the ansatze~(\ref{sch_pert}) 
and (\ref{sch_pert2}) does not incorporate such 
scalarized BHs which can be realized  
only in a nonperturbative regime 
with $|\alpha|={\cal O}(1)$.
Our purpose is rather to address the issue of ghost/Laplacian 
instabilities for hairy BHs realized as the consequence 
of perturbative deviation from the Schwarzschild solution
in non-shift-symmetric Horndeski theories.

\subsection{General GB couplings and BH stabilities}
\label{geneGB}

We first consider scalar-GB theories in the presence of 
a kinetic term~$\eta X$ and the Einstein-Hilbert 
term in the action, i.e., 
\be
{\cal S}=\int {\rm d}^4 x \sqrt{-g} \left[ 
\frac{\Mpl^2}{2}R+\eta X
+\alpha \xi(\phi) R_{\rm GB}^2 \right]\,,
\label{GBaction}
\ee
where $\xi(\phi)$ is an analytic function of $\phi$ and $\alpha$ is a dimensionless coupling constant which we assume to be small.
We perform the expansions~(\ref{sch_pert}) and 
(\ref{sch_pert2}) with respect to $\alpha$. 
Analogous to Eq.~(\ref{tial}), the GB coupling 
function~$\xi(\phi)$ is Taylor-expanded around a constant scalar 
field value~$\phi_0$.
The first-order solutions in $\alpha$, which are 
regular on the BH horizon at $r=2m$, are given by 
\be
\hat{h}_1(r)=0\,,\qquad 
\hat{f}_1(r)=0\,,\qquad
\hat{\phi}_1(r)=\tilde{\phi}_1
+\frac{2(3\hr^2+3\hr+4)\xi^{(1)}(\phi_0)}
{3 \eta m^2 \hr^3} \,,
\label{phi1le}
\ee
where $\tilde{\phi}_1$ is a constant
and we have introduced the dimensionless 
coordinate~$\hat{r}\equiv r/m$.
At spatial infinity, $\hat{\phi}_1(r)$ approaches a constant~$\tilde{\phi}_1$
with the derivative~$\hat{\phi}_1'(r)$ proportional to $\hat{r}^{-2}$.

Similarly, the regular second-order solutions are
\ba
\hspace{-0.7cm}
\hat{h}_2(r)
=\hat{h}_{2{\rm GB}}(r) &\equiv&
\frac{\left( 147\hr^5 + 174 \hr^4 +228 \hr^3 -1624 \hr^2 
- 3488\hr -7360 \right)\xi^{(1)}(\phi_0)^2}
{120 \eta m^4 \Mpl^2 \hr^6} \,,
\label{pert2a}\\
\hspace{-0.7cm}
\hat{f}_2(r)
=\hat{f}_{2{\rm GB}}(r) &\equiv&
-\frac{\left( 147\hr^5 + 294 \hr^4 +548 \hr^3+56 \hr^2 
-416\hr -1600 \right)\xi^{(1)}(\phi_0)^2}{120 \eta m^4 \Mpl^2 \hr^6}\,,
\label{pert2b}
\\
\hspace{-0.7cm}
\hat{\phi}_2(r)
=\hat{\phi}_{2{\rm GB}}(r) &\equiv&
\tilde{\phi}_2+\frac{2(3\hr^2+3\hr+4) \xi^{(2)}(\phi_0)
\tilde{\phi}_1}{3 \eta m^2 \hr^3}\nonumber \\
& &
+\frac{\xi^{(1)}(\phi_0) \xi^{(2)}(\phi_0)
[1095 (\hr^5+\hr^4)+1460\hr^3+2190\hr^2+1344\hr+800]}
{450 \eta^2 m^4 \hr^6}\,,
\label{pert2c}
\ea
where $\tilde{\phi}_2$ is a constant.
Since $\hat{h}_2(r) \propto \hat{r}^{-1}$, 
$\hat{f}_2(r) \propto \hat{r}^{-1}$, 
and 
$\hat{\phi}'_2(r) \propto \hat{r}^{-2}$ as $\hat{r}\to \infty$, 
the first-order solutions are consistent with the asymptotically Minkowski metric.
For the linear coupling~$\xi(\phi) \propto \phi$, we have 
$\xi^{(2)}(\phi_0)=0$, and hence 
$\hat{\phi}_2(r)=\tilde{\phi}_2$.

The regular third-order solutions are
\ba
\hat{h}_3(r)
&=& \hat{h}_{3{\rm GB}}(r) \equiv 
(66319 \hr^8+86648 \hr^7+127306 \hr^6-174628 \hr^5
-1046036 \hr^4-2874280 \hr^3-680960 \hr^2
-5948320 \hr \nonumber \\
& & 
-4659200)
\frac{\xi^{(1)}(\phi_0)^2\,\xi^{(2)}(\phi_0)}{18900 \eta^2 m^6 \Mpl^2 \hr^9}
+(147\hr^5 + 174\hr^4 + 228\hr^3 - 1624 \hr^2 - 3488\hr - 7360)
\frac{\xi^{(1)}(\phi_0)\,\xi^{(2)}(\phi_0)}{60\eta m^4 \Mpl^2 \hr^6}\,,
\label{hath3} \\
\hat{f}_3(r)
&=& \hat{f}_{3{\rm GB}}(r) \equiv
-(66319 \hr^8+132638 \hr^7+249946 \hr^6+285272 \hr^5
+199852 \hr^4-98920 \hr^3-981200 \hr^2
-847840 \hr \nonumber \\
& & 
-716800)
\frac{\xi^{(1)}(\phi_0)^2\,\xi^{(2)}(\phi_0)}{18900 \eta^2 m^6 \Mpl^2 \hr^9}
+(147\hr^5 + 294\hr^4 + 548\hr^3 +56 \hr^2 -416\hr - 1600)
\frac{\xi^{(1)}(\phi_0)\,\xi^{(2)}(\phi_0)}{60\eta m^4 \Mpl^2 \hr^6}\,,
\label{hatf3} \\
\hat{\phi}_3(r) &=&\hat{\phi}_{3{\rm GB}}(r) \equiv \tilde{\phi}_3
+\frac{\varphi_{3{\rm GB}}(\hr)}{2381400m^6 \Mpl^2 \eta^3 \hr^9}\,,
\label{hat3}
\ea
where $\tilde{\phi}_3$ is a constant and 
$\varphi_{3{\rm GB}}(\hr)$ is an eighth-degree 
polynomial of ${\hat r}$.
For the linear coupling~$\xi(\phi) \propto \phi$,
$\hat{h}_3(r)$ and $\hat{f}_3(r)$ vanish identically, which is consistent with the result of \cite{Sotiriou:2013qea,Sotiriou:2014pfa}.
Deriving higher-order solutions, 
we find that, in the limit~$r\to\infty$, 
the metric components and scalar field behave as
\be
\hat{h}_j(r) \propto \hat{r}^{-1}\,\qquad 
\hat{f}_j(r) \propto \hat{r}^{-1}\,,\qquad
\hat{\phi}'_j(r) \propto \hat{r}^{-2} \qquad (j \geq 2)\,.
\ee
Hence, the metric is asymptotically Minkowski at all orders.
We note that the constant parts of the scalar field $\phi_0$, 
$\tilde{\phi}_1$, $\cdots$ are determined by the boundary conditions 
at spatial infinity.

For the linear GB coupling which respects the shift symmetry, it follows that 
$\xi^{(1)}(\phi_0)={\rm constant}$ and $\xi^{(n)}(\phi_0)=0$  
for $n \geq 2$. Then, the $\phi_0$-dependence disappears from all 
the expressions of $\hat{h}_j(r)$, $\hat{f}_j(r)$, and 
$\hat{\phi}'_j(r)$ with $j \geq 1$.
This reflects the property of shift-symmetric theories 
in which the field value itself does not matter, so that 
we can set $\tilde{\phi}_j=0$ ($j \geq 1$). 
In this case, we realize asymptotically Minkowski hairy BH solutions 
where only the even-order ($j=2,4,\cdots$) terms of metric components 
and the odd-order ($j=1,3,\cdots$) terms of scalar field 
are nonvanishing \cite{Sotiriou:2014pfa,Minamitsuji:2022mlv}.

For general non-shift-symmetric GB couplings, 
the $\phi_0$-dependence remains 
in the metric components and scalar field. 
For positive power-law couplings~$\xi(\phi) \propto \phi^n$ with integer~$n~(\geq 2)$, in the limit~$\phi_0 \to 0$, both $\hat{h}_j$ and $\hat{f}_j$ vanish for all $j \geq 1$
and the Schwarzschild solution with a constant 
scalar field is recovered.
Thus, in contrast to the case of $n=1$,
a nonvanishing value of $\phi_0$ is necessary 
to realize hairy BH solutions. 
Provided that $\phi_0 \neq 0$, the metric components 
$\hat{h}_j$ and $\hat{f}_j$ ($j \geq 2$) are 
subject to deviations from those in the 
Schwarzschild metric with the nonvanishing field derivative $\hat{\phi}_j'(r)$. 

The quantities associated with the conditions for the absence of ghost or Laplacian instabilities of odd-parity perturbations are estimated as
\ba
{\cal F}
&=&
\Mpl^2-\frac{16( 2\hr^3+\hr^2+2\hr-36)\xi^{(1)}(\phi_0)^2 }
{\eta m^4 \hr^6}\alpha^2+{\cal O}
\left(\alpha^3 \right)\,,\label{Fex}\\
{\cal G}
&=&
\Mpl^2
+\frac{16(\hr^2+2\hr+4) \xi^{(1)}(\phi_0)^2}{\eta m^4 \hr^6}
\alpha^2
+{\cal O}
\left(\alpha^3 \right)\,,\label{Gex}\\
{\cal H}
&=&
\Mpl^2
+
\frac{16(\hr^3-8)\xi^{(1)}(\phi_0)^2 }{\eta m^4\hr^6}
\alpha^2
+{\cal O}
\left( \alpha^3 \right)\,.
\label{Hex}
\ea
The next-to-leading-order terms of ${\cal F}$, ${\cal G}$, and 
${\cal H}$ are at most of order 
$\xi^{(1)}(\phi_0)^2\alpha^2/(\eta m^4)$.
Provided that 
\be
\frac{\xi^{(1)}(\phi_0)^2}
{\eta m^4 \Mpl^2}\alpha^2 \ll 1\,,
\label{alup}
\ee
there are neither ghost nor 
Laplacian instabilities in the odd-parity sector 
due to the dominance of the term $\Mpl^2$ in ${\cal F}$, ${\cal G}$, and ${\cal H}$. 

In the even-parity sector, the quantity associated with the no-ghost condition is estimated as
\be
{\cal K}=
\frac{2(\hr^2+2\hr+4)^2 \xi^{(1)}(\phi_0)^2}{\eta m^4 \hr^6}
\alpha^2
+{\cal O}
\left( \alpha^3 \right)\,,
\label{calKe}
\ee
and hence the ghost is absent for $\eta>0$. 
The radial propagation speed squared~$c_{r1,{\rm even}}^2$
of the gravitational perturbation, which is equivalent to 
$c_{r,{\rm odd}}^2={\cal G}/{\cal F}$ in the odd-parity 
sector, is given by 
\be
c_{r1,{\rm even}}^2=c_{r,{\rm odd}}^2=1+
\frac{32(\hr-2)(\hr^2+3\hr+8)\xi^{(1)}(\phi_0)^2}{\eta 
m^4 \Mpl^2 \hr^6}
\alpha^2
+{\cal O}
\left( \alpha^3 \right)\,.
\label{crGB}
\ee
On the horizon ($\hr=2$), the next-to-leading-order term of
Eq.~(\ref{crGB}) vanishes, so $c_{r1,{\rm even}}^2$ is 
close to the (squared) speed of light. 
In the vicinity of the horizon, 
$c_{r1,{\rm even}}^2$ deviates from 
unity, but it quickly decreases as 
$|c_{r1,{\rm even}}^2-1| \propto \hr^{-3}$.
Under the condition~(\ref{alup}), 
it is possible to satisfy the  
bound of speed of GWs given in Refs.~\cite{LIGOScientific:2017vwq,LIGOScientific:2017ync,LIGOScientific:2017zic}.
For the GB couplings satisfying $\xi^{(2)}(\phi_0) \neq 0$, 
the squared propagation speed of the scalar field perturbation~$\delta \phi$ 
in the even-parity sector is generally of order unity,
\be
c_{r2,{\rm even}}^2=1+{\cal O}(\alpha^2)\,.
\label{cr2even}
\ee
The linear coupling~$\xi(\phi) \propto \phi$ gives rise to 
further suppression for the deviation of 
$c_{r2,{\rm even}}^2$ from unity, 
such that $c_{r2,{\rm even}}^2=1+{\cal O}(\alpha^4)$~\cite{Minamitsuji:2022mlv}. 
The squared angular propagation speeds in the even-parity sector 
can be estimated as 
\be
c_{\Omega,\pm}^2
=1 \pm \frac{24\xi^{(1)}(\phi_0)} 
{m^2 \Mpl \hr^3} 
\sqrt{\frac{2}{\eta}} |\alpha|+{\cal O}
\left( \alpha^2 \right)\,,
\label{cOmeex}
\ee
where the double signs are in the same order and we have
used the no-ghost condition~$\eta>0$.
The above results show that, in the limit $|\alpha| \ll 1$ 
with $\eta>0$, all the conditions for the absence of 
ghost/Laplacian instabilities against odd- and 
even-parity sectors are consistently satisfied for 
asymptotically Minkowski hairy BH solutions 
present for the models of the form~(\ref{GBaction}).

As mentioned previously, the above solutions~\eqref{phi1le}--\eqref{hat3} accommodate 
asymptotically Minkowski BH solutions for the linear coupling~$\xi(\phi) \propto \phi$ in the small coupling limit~$|\alpha|\ll 1$~\cite{Sotiriou:2013qea,Sotiriou:2014pfa}.
On the other hand, for the couplings of the form~$\xi(\phi)=\sum_{j\geq 1} c_j \phi^{2j}$ with 
$c_j$ being constants, including
$\xi(\phi)\propto c_2\phi^2+c_4 \phi^4$ (with $c_2>0$) \cite{Minamitsuji:2018xde,Silva:2018qhn} 
or 
$\xi (\phi)\propto 1-{\rm e}^{-k \phi^2}$ ($k>0$)
\cite{Doneva:2017bvd},
the solutions~\eqref{phi1le}--\eqref{hat3} 
do not incorporate BHs realized as the consequence of spontaneous 
scalarization, by reflecting the fact that 
scalarized BHs can be obtained only nonperturbatively 
for $|\alpha|={\cal O} (1)$.

\subsection{GB couplings with other interactions}
\label{GBother}

We also study how the hairy BH solutions discussed in 
Sec.~\ref{geneGB} are subject to modifications by taking 
into account power-law coupling functions~$(-X)^p$ or $\phi^q$ to the GB theory 
given by Eq.~(\ref{GBaction}). 
For simplicity, we consider lowest-order power-law functions in most cases, but we will also study theories containing the couplings~$G_3 \supset \gamma_3 \ln (-X)$ and $G_4 \supset \gamma_4 \sqrt{-X}$. 
In scalar-GB theories 
with the quadratic potential~$V(\phi)=\mu_2 \phi^2$ 
in $G_2$, the presence of hairy BH solutions was 
numerically confirmed in Ref.~\cite{Doneva:2019vuh}. 
Hence, we do not analyze the same model here.
In the presence of the term~$\alpha \tilde{\alpha}_2 
X^2$ in $G_2$, 
using the expansions~(\ref{sch_pert}) and (\ref{sch_pert2}) with respect to the small coupling constant~$\alpha$ 
shows that there are no corrections to $\hat{h}_j$, $\hat{f}_j$, and $\hat{\phi}_j$ derived in GB theories~(\ref{GBaction}) up to the order~$j=3$. Similarly, in the presence of 
$G_2\supset \alpha_2 X^n$ ($n\geq 3$),
the nontrivial corrections do not appear up to the order~$j=n+1$.

\subsubsection{Cubic and GB couplings}
\label{cubsub}

The $\phi$-dependent cubic coupling~$G_3(\phi)$ is equivalent 
to the term $-2X G_{3,\phi}$ in $G_2$~\cite{KYY}, so adding the linear coupling~$\alpha \mu_3 \phi$ to $G_3$
does not modify the structure of the theory~(\ref{GBaction}).
We then consider
\be
G_3(\phi) \supset \alpha \mu_3 \phi^2\,,
\label{G3phis}
\ee
with $\mu_3$ being a nonvanishing constant, which is equivalent 
to $-4\alpha \mu_3 \phi X$ in $G_2$.
We perform the expansions (\ref{sch_pert}) and (\ref{sch_pert2}) 
in terms of the small coupling constant $\alpha$.
The first-order solutions regular on the horizon are
equivalent to those in Eq.~(\ref{phi1le}), while 
the second- and third-order solutions are 
\ba
& &
\hat{h}_2(r)=\hat{h}_{2{\rm GB}}(r)\,,\qquad
\hat{f}_2(r)=\hat{f}_{2{\rm GB}}(r)\,,\qquad
\hat{\phi}_2(r)=\hat{\phi}_{2{\rm GB}}(r)+\frac{8
(3\hr^2+3\hr+4)\phi_0 \xi^{(1)}(\phi_0) \mu_3}
{3\eta^2 m^2 \hr^3}
\,,\\
& & 
\hat{h}_3(r)=\hat{h}_{3{\rm GB}}(r)+\frac{ 
(147 \hr^5 + 174 \hr^4 + 228 \hr^3 - 1624 \hr^2 - 3488 \hr - 7360)
\phi_0 \xi^{(1)}(\phi_0)^2 \mu_3}
{30 \eta^2 m^4 \Mpl^2 \hr^6}\,,\\
& & 
\hat{f}_3(r)=\hat{f}_{3{\rm GB}}(r)-\frac{
(147 \hr^5 + 294 \hr^4 + 548 \hr^3 +56 \hr^2 -416 \hr -1600)
\phi_0 \xi^{(1)}(\phi_0)^2 \mu_3}
{30 \eta^2 m^4 \Mpl^2 \hr^6}\,,\\
& &
\hat{\phi}_3(r)=\hat{\phi}_{3{\rm GB}}(r)
+\frac{\mu_3}{\eta^3 m^4 \hr^6}\varphi_3 (\hr)\,,
\ea
where $\varphi_3(\hr)$ is the fifth degree polynomial of $\hr$.
The cubic coupling~(\ref{G3phis}) gives rise to modifications in $\hat{\phi}_2(r)$, $\hat{h}_3(r)$, $\hat{f}_3(r)$, and $\hat{\phi}_3(r)$ 
in comparison to those derived for the GB couplings. 
At large distances, these new terms have 
the same radial dependence as 
their leading-order terms. 

Using the expanded solutions with $|\phi_0|$ at most of order~$M_{\rm pl}$, it follows that the conditions 
for the absence of ghost/Laplacian instabilities
against odd- and even-parity perturbations 
are also satisfied for 
$|M_{\rm pl} \mu_3| \lesssim 1$, $|\alpha| \ll 1$, 
and $\eta>0$.
Then, the cubic coupling~$G_3(\phi)\supset \alpha \mu_3 \phi^2$ 
besides the GB coupling~$\alpha \xi(\phi) G$ 
leads to the existence of hairy BH solutions free from ghost or Laplacian instabilities.
Similarly, for more general $\phi$-dependent 
cubic coupling~$G_3 \supset \alpha \mu_3(\phi)$, 
nontrivial corrections to 
the metric functions and those to the scalar field
show up at the orders~$j=3$ and $j=2$, respectively.

Second, we discuss the case in which 
the cubic Galileon coupling 
\be
G_3(X)\supset \alpha \alpha_3 X\,,
\ee
with $\alpha_3$ being a nonvanishing constant,
is present besides GB coupling~$\alpha \xi(\phi) R_{\rm GB}^2$. 
Then, the corrections to 
the BH solutions in GB theories, 
up to the order~$j=3$, appear only in $\hat{\phi}_3(r)$ as
\be
\hat{\phi}_3(r)=\hat{\phi}_{3{\rm GB}}(r)-\frac{2
(21 \hr^5+42 \hr^4+84 \hr^3-24 \hr^2-84 \hr -224)
\xi^{(1)}(\phi_0)^2 \alpha_3}
{21 \eta^3 m^6 \hr^9}\,. 
\label{phi3Gali}
\ee
At large distances, the correction to $\hat{\phi}_3'(r)$ arising from 
the cubic Galileon is proportional to $\hat{r}^{-5}$, which 
decays faster than $\hat{\phi}_{3{\rm GB}}'(r) \propto \hat{r}^{-2}$. 
Since the metric components are not modified  
up to the order~$j=3$, the cubic Galileon does not 
induce strong modifications to hairy GB BHs in comparison 
to the coupling $G_3(\phi) \supset \alpha \mu_3 \phi^2$. 
The absence of ghost/Laplacian instabilities
of BHs is also ensured for 
$|M_{\rm pl}\alpha_3/m^2| \lesssim 1$, 
$|\alpha| \ll 1$, and $\eta>0$.
Similarly, in the presence of $G_3\supset \alpha 
\alpha_3 X^n$ ($n\geq 2$), nontrivial corrections 
to the scalar field 
show up at the order~$j=n+2$.

The next example is the cubic logarithmic 
interaction given by 
\be
G_3 \supset \alpha \gamma_3 \ln (-X)\,,
\ee
with $\gamma_3$ being a nonvanishing constant,
which belongs to  
the couplings in Eq.~(\ref{Gchoice}).
The first-order solutions in $\alpha$ regular on the horizon ($\hat{r}=2$) are
\be
\hat{h}_1(r)=0\,,\qquad
\hat{f}_1(r)=0\,,\qquad 
\hat{\phi}'_1(r)=-\frac{2}{\eta m^3 \hr^4}
\left[ (\hr^2+2\hr+4) \xi^{(1)}(\phi_0)+2m^2 
\hr^3 \gamma_3 \right]\,.
\label{phi1le2}
\ee
At large distances, the leading-order contributions to 
$\hat{\phi}'_1(r)$ arise from the cubic logarithmic coupling.
The second-order solutions regular 
on the horizon are given by 
\ba
\hat{h}_2(r) &=& 
\hat{h}_{2{\rm GB}}(r)+\frac{\gamma_3}{2\eta \Mpl^2} \left[ 
\frac{(3\hr^2+10\hr-40)\xi^{(1)}(\phi_0) }{m^2 \hr^3}
+\frac{8\gamma_3}{\hr-2} \ln \left( \frac{\hr}{2} \right) \right]
\,,\\
\hat{f}_2(r) &=&
\hat{f}_{2{\rm GB}}(r)+\frac{\gamma_3}{6\eta \Mpl^2} 
-\left[\frac{(9 \hr^2+18 \hr-88)\xi^{(1)}(\phi_0)}{m^2 \hr^3}
+24\gamma_3\fr{\hr-1}{\hr-2}\ln\bra{\fr{\hr}{2}}
 \right]
\,,\\ 
\hat{\phi}_2'(r)&=&\phi'_{2{\rm GB}}(r)-\fr{64\xi^{(2)}(\phi_0)\gamma_3}{\eta^2m^3\hr^4(\hr-2)}\ln\bra{\fr{\hr}{2}}\,. 
\ea
At spatial infinity, the metric component~$\hat{f}_2(r)$ exhibits the logarithmic divergence
\be
\hat{f}_2(r) \to -\frac{4 \gamma_3^2}
{\eta \Mpl^2} \ln \hat{r} \quad {\rm as}\quad 
\hat{r} \to \infty\,.
\ee
This means that, even in the presence of the GB couplings, the cubic  
logarithmic coupling prevents the realization of asymptotically Minkowski hairy BH solutions.

\subsubsection{Quartic and GB couplings}
\label{quasec}

We proceed to the model of a linear nonminimal coupling 
\be
G_4 \supset \alpha \mu_4 \phi\,,
\label{G4qua}
\ee
with $\mu_4$ being a nonvanishing constant,
besides the GB coupling~$\alpha \xi(\phi) R_{\rm GB}^2$.
Then, we find that the first-order solutions in $\alpha$ 
regular on the horizon
are the same as those derived in Eq.~(\ref{phi1le}).
The second-order solutions are given by 
\ba
\hat{h}_2(r) &=& 
\hat{h}_{2{\rm GB}}(r)-\frac{
(\hr+4)(\hr+10)\xi^{(1)}(\phi_0)\mu_4}
{6 \eta m^2 \Mpl^2 \hr^3}
\,,\\
\hat{f}_2(r) &=&
\hat{f}_{2{\rm GB}}(r)-\frac{
(23 \hr^2+22\hr+24)\xi^{(1)}(\phi_0) 
\mu_4}{6 \eta m^2 \Mpl^2 \hr^3}\,,\\ 
\hat{\phi}_2(r) &=& 
\hat{\phi}_{2{\rm GB}}(r)\,.
\ea
The linear nonminimal coupling affects $\hat{h}_2(r)$ 
and $\hat{f}_2(r)$, 
while its effect does not appear in $\hat{\phi}_2(r)$. 
In comparison to the cubic-order interactions discussed in Sec.~\ref{cubsub}, the modifications to the background 
geometry arising from the quartic coupling~$\alpha \mu_4 \phi$ 
in $G_4$ already appear at the order of $j=2$.
Higher-order solutions of $\hat{h}_j (r)$, $\hat{f}_j(r)$, and 
$\hat{\phi}_j(r)$ ($j \geq 3$) also receive corrections from 
the linear nonminimal coupling. 
Since all 
$\hat{f}_j(r)$, $\hat{h}_j(r)$, and $\hat{\phi}_j'(r)$ ($j \geq 1$)  
vanish at spatial infinity, the resulting hairy BH solutions are asymptotically Minkowski. 
Similarly, for the $\phi$-dependent coupling function~$G_4\supset \alpha \mu_4 (\phi)$,
nontrivial corrections to the metric components
show up at the order~$j=2$.

Using the expanded solutions, the quantities associated with
odd-parity perturbations are given by 
\be
{\cal F}=
\Mpl^2+2 \phi_0 \mu_4 \alpha
+{\cal O}
\left(\alpha^2 \right)\,,\quad 
{\cal G}=
\Mpl^2+2 \phi_0 \mu_4 \alpha
+{\cal O}
\left(\alpha^2 \right)\,,\quad 
{\cal H}=
\Mpl^2+2 \phi_0 \mu_4 \alpha
+{\cal O}
\left(\alpha^2 \right)\,,
\ee
so that the linear nonminimal coupling 
gives rise to corrections of order $\alpha$. 
We can avoid the ghost and Laplacian instabilities 
under the condition
\be
\Mpl^2+2 \phi_0 \mu_4 \alpha>0\,.
\ee
As long as $|\alpha| \ll 1$ and $|\mu_4/M_{\rm pl}| \lesssim 1$, 
this condition is satisfied for $|\phi_0| \lesssim \Mpl$. 
We note that the radial and angular propagation speed 
squares for the odd modes, $c_{r,{\rm odd}}^2={\cal G}/{\cal F}$ and 
$c_{\Omega,{\rm odd}}^2={\cal G}/{\cal H}$, are 
both $1+{\cal O}(\alpha^2)$.

Up to the order of $\alpha^2$, the quantity~${\cal K}$ 
is the same as that given in Eq.~(\ref{calKe}).
The propagation speed squared of the scalar field 
perturbation~$\delta \phi$ in the even-parity sector is estimated as 
$c_{r2,{\rm even}}^2=1+{\cal O}(\alpha^2)$ for theories 
with $\xi^{(2)}(\phi) \neq 0$.
For the linear GB coupling~$\xi(\phi) \propto \phi$, we have 
$c_{r2,{\rm even}}^2=1+{\cal O}(\alpha^3)$, where 
the terms of order~$\alpha^3$ arise from 
$\alpha \mu_4 \phi$ in $G_4$.
Up to the linear order in $\alpha$, the squared angular 
propagation speeds~$c_{\Omega,\pm}^2$ in the even-parity sector are 
identical to those in Eq.~(\ref{cOmeex}).
These discussions show that, provided $|\alpha| \ll 1$, 
$|\mu_4/\Mpl| \lesssim 1$, $\eta>0$, and $|\phi_0| \lesssim \Mpl$,
there are neither ghost nor Laplacian instabilities 
for hairy BHs discussed above.

Instead of the linear nonminimal coupling of 
the form~$\beta_4 \phi R$, we can also consider 
nonminimal couplings with higher-order powers, i.e., 
$\alpha \mu_{4}\phi^p R$ with $p \geq 2$, besides the GB 
couplings~\cite{Antoniou:2021zoy}. 
As in the case of $p=1$, the contributions 
to $f$ and $h$ from $\mu_4$ appear at the order 
of $j=2$, while the scalar-field derivative starts to
receive corrections from the third order.
In such models, we can also realize asymptotically Minkowski hairy BHs  
satisfying all the conditions for the absence of ghost/Laplacian instabilities.

Let us next study the model of quartic derivative couplings 
of the form
\be
G_4 \supset \alpha \alpha_4 X\,,
\label{G4qua2}
\ee
with $\alpha_4$ being a nonvanishing constant,
besides the GB coupling.
Performing the expansions~(\ref{sch_pert}) and (\ref{sch_pert2}),
we obtain the same first- and second-order
regular solutions as those given in 
Eqs.~(\ref{phi1le}) and (\ref{pert2a})--(\ref{pert2c}). 
The third-order solutions are 
\be
\hat{h}_3(r)=\hat{h}_{3{\rm GB}}(r)+
\frac{4(\hr^2+2\hr+4)^2(\hr-2)\xi^{(1)}
(\phi_0)^2 \alpha_4}{\eta^2 m^6 \Mpl^2 \hr^9}\,,\qquad 
\hat{f}_3(r)=\hat{f}_{3{\rm GB}}(r)\,,\qquad
\hat{\phi}_3(r)=\hat{\phi}_{3{\rm GB}}(r)\,.
\label{hath3GB}
\ee
At this order, the effect of the quartic 
derivative coupling appears only in the expression 
of $\hat{h}_3(r)$. 
This correction vanishes on the horizon, 
with the asymptotic 
behavior~$\hat{h}_3(r)-\hat{h}_{3{\rm GB}}(r) \propto 
\hat{r}^{-4}$ at spatial infinity. 

We recall that the BH solutions with $X_s \neq 0$ 
realized by quartic derivative interactions 
without GB couplings are prone to 
the instability problem around 
the horizon~\cite{Minamitsuji:2022mlv}. 
On the other hand, the GB coupling besides 
the term~$\alpha \alpha_4 X$ 
in $G_4$ gives rise to asymptotically Minkowski hairy BH solutions. 
The leading-order  
corrections due to $\alpha$ to 
${\cal F}$, ${\cal G}$, ${\cal H}$, ${\cal K}$, 
$c_{r2,{\rm even}}^2$, 
and $c_{\Omega,\pm}^2$ 
are the same as those given in 
Eqs.~(\ref{Fex}), (\ref{Gex}), (\ref{Hex}), 
(\ref{calKe}), (\ref{cr2even}), and (\ref{cOmeex}), respectively, so 
these hairy solutions can satisfy
all the conditions for the absence of 
ghost/Laplacian instabilities.

For quartic derivative interactions with higher-order powers, 
i.e., $G_4 \supset \alpha \alpha_4 X^n$ ($n\geq 2$), 
nontrivial corrections to the metric components 
or the scalar field show up at the order $j=n+2$.
In such models there are BH solutions consistent with 
conditions for the absence of ghost/Laplacian instabilities 
and strong coupling problems, but it is difficult to distinguish 
them from those realized by GB couplings alone.

We also study the model given by   
\be
G_4 \supset \alpha \gamma_4 \sqrt{-X}\,,
\label{G4sq}
\ee
with $\gamma_4$ being a nonvanishing constant,
which belongs to couplings in Eq.~(\ref{Gchoice}).
In the absence of the GB coupling, this model gives rise to 
an exact BH solution~\cite{Babichev:2017guv}, but 
it is unstable due to the property~$X_s \neq 0$ on 
the horizon~\cite{Minamitsuji:2022mlv}.
The first-order solutions with respect to the GB 
coupling constant~$\alpha$ are 
\be 
\hat{h}_1(r)=0\,,\qquad \hat{f}_1(r)=0\,,\qquad 
\hat{\phi}_1'(r)=\frac{C\eta 
m \hr^3+m\hr^{5/2}
\sqrt{2(\hr-2)} \gamma_4+16  \xi^{(1)}(\phi_0)}
{\eta m^3 \hr^4 (\hr-2)}\,,
\ee
where $C$ is an integration constant. 
The numerator of $\hat{\phi}_1'(r)$ needs to vanish 
for its regularity at $r=2m$, which gives 
$C=-2\xi^{(1)}(\phi_0)/(\eta m)$. 
Then, we obtain the following solution 
\be
\hat{\phi}'_1(r)=
-\frac{2(\hr^2+2\hr+4)
\xi^{(1)}(\phi_0)}
{\eta m^3 \hr^4}+\frac{\sqrt{2} \gamma_4}
{\eta m^2 \hr^{3/2} \sqrt{\hr-2}}\,.
\ee
For $\gamma_4 \neq 0$, there is still the divergence 
of $\hat{\phi}_1'(r)$ at $\hr=2$. 
The leading-order term of $X$ on the horizon is 
a nonvanishing constant given by 
$X_s=-\gamma_4^2/(16 \eta^2 m^4)$.
Hence, even in the presence of the GB term, the quartic coupling~$\alpha \gamma_4 \sqrt{-X}$ in $G_4$ violates
the conditions for the absence of ghost/Laplacian 
instabilities of hairy BH solutions.

\subsubsection{Quintic and GB couplings}

The linear coupling~$\mu_5 \phi$ in $G_5$ is equivalent 
to the quartic derivative coupling~$-\mu_5 X$
in $G_4$ \cite{KYY}, 
so we already studied such a case 
in Sec.~\ref{quasec}. 
Let us then consider the coupling 
\be
G_5 \supset \alpha \mu_5 \phi^2\,,
\ee
with $\mu_5$ being a nonvanishing constant,
besides the GB coupling. 
The first- and second-order 
regular solutions are equivalent to those 
derived in Eqs.~(\ref{phi1le}) and (\ref{pert2a})--(\ref{pert2c}), while the third-order solutions are given by 
\be
\hat{h}_3(r)=\hat{h}_{3{\rm GB}}(r)
-\frac{8(\hr^2+2\hr+4)^2 (\hr-2)\phi_0 
\xi^{(1)}(\phi_0)^2 \mu_5}
{\eta^2 m^6 \Mpl^2 \hr^9}\,,\qquad 
\hat{f}_3(r)=\hat{f}_{3{\rm GB}}(r)\,,\qquad
\hat{\phi}_3(r)=\hat{\phi}_{3{\rm GB}}(r)\,.
\ee
They are similar to those derived for the quartic derivative 
coupling~$G_4 \supset \alpha \alpha_4 X$ [see Eq.~(\ref{hath3GB})]. 
Up to the order~$j=3$, the quintic 
coupling~$\alpha \mu_5 \phi^2$ affects only $\hat{h}_3(r)$. 
Similarly, for more general $\phi$-dependent 
quintic coupling $G_5\supset \alpha \mu_5(\phi)$,
nontrivial corrections to the metric component~$h(r)$
show up at the order~$j=3$.
Using these expanded solutions and computing the 
quantities associated with the linear stability of perturbations, it follows 
that the hairy BH solutions can satisfy
all the conditions for the absence of ghost/Laplacian 
instabilities 
for $|\mu_5 \Mpl/m^2| \lesssim 1$, $|\alpha| \ll 1$, 
and $\eta>0$. 

Finally, we consider the quintic derivative 
coupling given by 
\be
G_5(X)\supset \alpha \alpha_5 X\,,
\ee
with $\alpha_5$ being a nonvanishing constant,
besides the GB coupling. 
The corrections to BH solutions in GB theories, 
up to the order~$j=3$, arise only in $\hat{\phi}_3(r)$ as
\be
\hat{\phi}_3(r)=\hat{\phi}_{3{\rm GB}}(r)
+\frac{(88 \hr^5+77 \hr^4
-1232 \hr^2-1792 \hr-2464)\xi^{(1)}(\phi_0)^2 \alpha_5}
{77\eta^3 m^8\hr^{12}}
\,.
\ee
Similarly, in the presence of 
the term~$\alpha \alpha_5 X^n$ ($n\geq 2$) 
in $G_5$, nontrivial corrections to the metric components 
or the scalar field show up at 
higher order.
This property is similar to that for the cubic derivative 
coupling~$G_3 \supset \alpha \alpha_3 X$ [see Eq.~(\ref{phi3Gali})]. 
In comparison to the cubic coupling, the scalar-field derivative~$\hat{\phi}_3'(r)$ is more strongly 
suppressed at large distances 
[$\hat{\phi}_3'(r) \propto \hat{r}^{-8}$].
Although there are hairy BH solutions satisfying
all the conditions for the absence of ghost/Laplacian instabilities 
for $|\alpha_5 \Mpl/m^4| \lesssim 1$, 
$|\alpha| \ll 1$, and $\eta>0$, 
it would be challenging
to distinguish them from those present for the pure GB theories.

\section{\texorpdfstring{Black holes in $F(R_{\rm GB}^2)$~gravity}{Black holes in F(GB) gravity}}
\label{sec:fG}

In this section, we explore the BH solutions in gravitational theories where the Lagrangian contains an arbitrary function~$F(R_{\rm GB}^2)$ 
of the GB curvature invariant~$R_{\rm GB}^2$, besides the Einstein-Hilbert term. 
As was pointed out in \cite{KYY}, $F(R_{\rm GB}^2)$~gravity can be embedded in Horndeski theories.
We then generalize 
this $F(R_{\rm GB}^2)$-equivalent Horndeski theory
by adding a canonical kinetic term of the scalar 
field. 

\subsection{\texorpdfstring{$F(R_{\rm GB}^2)$ gravity}{F(GB) gravity}}

Let us consider theories given by the Lagrangian 
\be
\mL=\frac{\Mpl^2}{2}R+F(R_{\rm GB}^2)\,,
\label{LHG}
\ee
which can be equivalently expressed as
\be
\mL=\frac{\Mpl^2}{2}R
+F_{,\varphi}R_{\rm GB}^2-V(\varphi)\,,
\label{LHG2}
\ee
where $\varphi$ is a new scalar degree of freedom associated 
with the GB term and we have defined
\be
V(\varphi) \equiv F_{,\varphi}\varphi -F(\varphi)\,.
\ee
Indeed, varying the Lagrangian~(\ref{LHG2}) 
with respect to $\varphi$ leads to 
\be
(\varphi-R_{\rm GB}^2) F_{,\varphi \varphi}=0\,.
\ee
Therefore, provided that $F_{,\varphi \varphi} 
\neq 0$, we have $\varphi=R_{\rm GB}^2$, and hence 
the Lagrangian~(\ref{LHG2}) 
reduces to the original one in Eq.~(\ref{LHG}). 
From Eq.~(\ref{LHG2}), we find that 
a scalar field $\varphi$ with the potential $V(\varphi)$
couples to the GB term of the form $F_{,\varphi}R_{\rm GB}^2$.

We introduce the following quantities:
\be
\phi \equiv \Mpl m^4 \varphi\,,
\qquad \xi(\phi) \equiv F_{,\varphi}\,,
\label{phicor}
\ee
where $2m$ corresponds to the 
horizon radius of a BH solution (if it exists). 
The quantity~$\phi$ has mass dimension one, which we identify as the scalar field in Horndeski theories.
Then, the Lagrangian for $F(R_{\rm GB}^2)$ gravity is 
equivalent to
\be
\mL=\frac{\Mpl^2}{2}R
+\xi(\phi) R_{\rm GB}^2-V(\phi)\,,\qquad {\rm where} 
\qquad V(\phi)=\xi \varphi-F\,.
\label{LHG3}
\ee
For a given function~$F(R_{\rm GB}^2)$, the GB 
coupling~$\xi(\phi)$ and the scalar potential~$V(\phi)$ are 
fixed by using the correspondence~(\ref{phicor}) with $\varphi=R_{\rm GB}^2$.
In the language of Horndeski theories, 
the theory~(\ref{LHG3}) corresponds to the 
following choice of the coupling functions~\cite{KYY,Langlois:2022eta}:
\ba
& &
G_2=-V(\phi)+8 \xi^{(4)}(\phi) X^2 (3-\ln |X|)\,,\qquad G_3=4\xi^{(3)}(\phi) X (7-3\ln |X|)\,,\nonumber \\
& &
G_4=\frac{\Mpl^2}{2}+4\xi^{(2)}(\phi) X (2-\ln |X|)\,,\qquad
G_5=-4\xi^{(1)}(\phi) \ln |X|\,.
\ea

In the following, for concreteness, we consider the power-law $F(R_{\rm GB}^2)$~models given by 
\be
F(R_{\rm GB}^2)=\beta (R_{\rm GB}^2)^n\,,
\label{fGpower}
\ee
where $\beta$ and $n$ are constants. 
Introducing the dimensionless coupling~$\alpha=2m^{2-4n} \Mpl^{-2} \beta$ and 
performing the expansions~(\ref{sch_pert}) of 
metric components with respect to the small parameter~$|\alpha|\ll1$, the scalar potential is given by 
\be
V(\phi)=\alpha (n-1) \frac{\Mpl^2}{2m^2} 
\left( \frac{\phi}{\Mpl} \right)^n\,.
\label{Vpower}
\ee
Apart from the specific powers~$n=0$ and $n=1$, 
the scalar field has a nonvanishing effective mass squared~$M_{\phi}^2\equiv V_{,\phi \phi}$. 
For $n=0$, we have 
$F=\beta={\rm constant}$ 
and hence the resulting solution is the 
Schwarzschild--(anti-)de Sitter spacetime. 
When $n=1$, we have $\xi=\beta={\rm constant}$ 
and $V=0$, so we end up with the no-hair 
Schwarzschild solution.
Thus, we will focus on integer powers 
with $n \geq 2$.

The scalar-field equation at first order 
in $\alpha$ gives the relation
\be
n (n-1) \left[ r^6 \phi_0(r)-48 m^6 \Mpl 
\right] \phi_0(r)^{n-2}=0\,.
\label{fiGBeq}
\ee
When $n=2$, we have only the following solution:
\be
\phi_0(r)
=\frac{48 m^6 \Mpl}{r^6}\,,
\label{phi0GB}
\ee
which corresponds to the GB term in 
the Schwarzschild spacetime. 
In this case, we obtain
\ba
{\hat h}_1(r)
&=&\frac{5\hr(\hr+2)(\hr^2+4)(\hr^4+16)-17152}{8 \hr^9}\,,
\nonumber\\
{\hat f}_1(r)
&=&
-\frac{ 
5\hr(\hr+2)(\hr^2+4)(\hr^4+16)-2816}{8\hr^9}\,,
\nonumber
\\
\hat{\phi}_1(r)
&=&\frac{12( 5\hr^9-101376\hr+217088 )\Mpl}
{\hr^{15}}\,,
\ea
where we recall that $\hat{r}=r/m$. Deriving the higher-order solutions as well, 
we find that the metric is asymptotically Minkowski at 
all orders. Using such expanded solutions for $n=2$, 
the dominant terms in ${\cal F}$, ${\cal G}$, and ${\cal H}$ 
are $\Mpl^2$ with the corrections of order~$\alpha$, 
and hence the BH is free from ghost/Laplacian instabilities 
against odd-parity perturbations. 
However, the 
quantity associated with the no-ghost condition in the even-parity sector 
is given by 
\be
{\cal K}=-\frac{15925248 \Mpl^2}{\hr^{18}}\alpha^2
+{\cal O}(\alpha^3)\,.
\ee
Since the leading-order term in ${\cal K}$ is negative, 
there is the ghost instability for even-parity perturbations. 
While $c_{r2,{\rm even}}^2=1+{\cal O}(\alpha)$, the quantities 
associated with the angular propagation speeds in the even-parity sector are $B_1=1/2+{\cal O}(\alpha)$ and $B_2=1+{\cal O}(\alpha)$, 
so the conditions~(\ref{B12con}) are also violated.

For $n \geq 3$, we have the same branch as Eq.~(\ref{phi0GB}) 
besides the branch~$\phi_0(r)=0$. In such cases, the ghost arises 
in the even-parity sector, with 
a similar behavior of $B_1$ and 
$B_2$ as in the case of $n=2$.  When $n=3$, for example, we find 
\be
{\cal K}=-\frac{330225942528 \Mpl^2}{\hr^{30}}\alpha^2
+{\cal O}(\alpha^3)\,,
\ee
whose leading-order term is negative, and hence 
the hairy BH solutions with $\phi_0(r)=48 m^6 \Mpl/r^6$ 
are prone to ghost instabilities.
When $n \geq 3$, Eq.~(\ref{fiGBeq}) admits the other branch 
$\phi_0(r)=0$. In this case, the $j$th-order expanded 
solutions ($j \geq 1$) of Eqs.~(\ref{sch_pert}) and 
(\ref{sch_pert2}) are given by 
\be
\hat{h}_j(r)=0\,,\qquad 
\hat{f}_j(r)=0\,,\qquad 
\hat{\phi}_{j}(r)=0\,,
\ee
which correspond to no-hair BHs.

We have thus shown that the power-law $F(R_{\rm GB}^2)$~models with (\ref{fGpower}) do not give rise to nontrivial 
BH solutions with scalar hair satisfying 
all the conditions for the absence of 
ghost/Laplacian instabilities. 
We also studied the logarithmic model with
$F(R_{\rm GB}^2)=\beta \ln |R_{\rm GB}^2|$ and 
reached the same conclusion.

\subsection{\texorpdfstring{$F(R_{\rm GB}^2)$-equivalent Horndeski theories with a scalar-field kinetic term}{F(GB)-equivalent Horndeski theories with a scalar-field kinetic term}}

We also study BHs in theories where 
the kinetic term~$\eta X$ is added to the $F(R_{\rm GB}^2)$-equivalent 
Horndeski theories~(\ref{LHG3}), i.e.,
\be
\mL=\frac{\Mpl^2}{2}R
+\eta X
+\xi(\phi) R_{\rm GB}^2-V(\phi)\,,
\label{LHG4}
\ee
where $\xi(\phi)$ and $V(\phi)$ are given by 
Eqs.~\eqref{phicor} and \eqref{LHG3}, respectively.

Considering the power-law coupling functions of the form~$F(\varphi)=\beta \varphi^n$ analogous to Eq.~(\ref{fGpower})
and introducing the dimensionless parameter~$\alpha =2m^{2-4n} \Mpl^{-2} \beta$, 
the scalar potential is given by the same form as that in Eq.~\eqref{Vpower}.
We first investigate the case of $n=2$,
which corresponds to the quadratic potential~$V(\phi)= \alpha \phi^2/(2m^2)$ with the linear
GB coupling~$\xi(\phi)=\alpha m^2 \Mpl \phi$. 
Although $V(\phi)$ and $\xi(\phi)$ 
are similar to those in the model studied 
in Ref.~\cite{Doneva:2019vuh},
both $V(\phi)$ and $\xi(\phi)$ are proportional 
to $\alpha$ in our case.
This fact affects the resulting BH solutions 
derived by using the expansions~(\ref{sch_pert}) 
and (\ref{sch_pert2}) with respect to $\alpha$.

As discussed in Sec.~\ref{nonshiftsec}, 
the zeroth-order equation for the scalar field reduces to 
the differential equation~(\ref{phidif}). 
The solution to $\phi_0(r)$ regular on the horizon is 
$\phi_0(r)=\phi_0={\rm constant}$. 
The first-order regular solutions are given by  
\be
\hat{h}_1(r)=\frac{(\hr^2+2\hr+4)\phi_0^2}{12\Mpl^2}\,,
\qquad
\hat{f}_1(r)=-\frac{\hr( \hr+2)\phi_0^2}{12\Mpl^2}\,,
\qquad  
\hat{\phi}_1'(r)=\frac{(\phi_0 \hr^3-6\Mpl)(\hr^2+2\hr+4)}
{3\eta m \hr^4}\,.
\ee
The asymptotically Minkowski metric can be realized 
only if $\phi_0=0$, under which both 
$\hat{h}_1(r)$ and $\hat{f}_1(r)$ vanish with 
the dependence~$\hat{\phi}_1'(r) \propto \hat{r}^{-2}$ 
at large distances.

At second order, the metric components~$\hat{h}_2(r)$ and 
$\hat{f}_2(r)$ are equivalent to those derived by 
taking the limit of $\xi^{(1)}(\phi_0)^2/(m^4 \Mpl^2) \to 1$ in
Eqs.~(\ref{pert2a}) and (\ref{pert2b}), respectively.
The scalar-field derivative consistent 
with the regularity on the horizon 
($\hat{r}=2$) yields
\be
\hat{\phi}_2'(r)=\frac{\Mpl}{3\eta^2 m \hr( \hr-2)}
\left[ 3(\hr+4)(\hr-2)+8 \ln \left( \frac{\hr}{2} \right) \right]\,,
\ee
which has the dependence~$\hat{\phi}_2'(r) \to \Mpl/(\eta^2 m)$ as 
$\hat{r}\to \infty$. 
Then, the scalar field does not satisfy the boundary 
condition~$\phi'(r) \to 0$ at spatial infinity [see Eq.~\eqref{bc2}].

For $n \geq 3$, again the asymptotically Minkowski metric is realized only if $\phi_0=0$, under which $\hat{h}_1(r)=0$ and $\hat{f}_1(r)=0$. 
The first-order solution to the scalar-field derivative 
regular on the horizon is $\hat{\phi}_1'(r)=0$, 
so that $\hat{\phi}_1(r)=\tilde{\phi}_1={\rm constant}$ for arbitrary $r$. 
When $n=3$, the second-order regular solutions are given by 
\be
\hat{h}_2(r)=0\,,\qquad \hat{f}_2(r)=0\,,\qquad 
\hat{\phi}_2 (r)=\tilde{\phi}_2+\frac{2(3\hr^2+3\hr+4)}
{\eta \hr^3}\tilde{\phi}_1\,,
\ee
where $\tilde{\phi}_2$ is a constant. 
Similarly, the metric components of 
third-order solutions are $\hat{h}_3(r)=0$ and $\hat{f}_3(r)=0$, 
while the leading-order contribution to $\hat{\phi}'_3(r)$ at spatial 
infinity is $\tilde{\phi}_1^2 \hat{r}/(\eta m \Mpl)$. To satisfy the 
condition~$\hat{\phi}_3'(r) \to 0$ as $r \to \infty$, we require that 
$\tilde{\phi}_1=0$. 
We also find that the solutions compatible with the boundary conditions~\eqref{bc2}, $\hat{\phi}_j'(r) \to 0$ at spatial 
infinity, are given by 
\be
\hat{h}_j(r)=0\,,\qquad 
\hat{f}_j(r)=0\,,\qquad 
\hat{\phi}_j(r)=0 \qquad (j \geq 1)\,.
\ee
The same conclusion holds also for $n \geq 4$.
Thus, for $n \geq 3$, we only have the  
Schwarzschild BH solutions without scalar hair.
These results show the absence of asymptotically Minkowski hairy BH solutions,
at least as long as the perturbative ansatze~\eqref{sch_pert} 
and \eqref{sch_pert2} are valid in the small coupling limit.
This does not exclude the possibility for the existence of asymptotically Minkowski hairy BH solutions beyond 
the perturbative regime.

For a massive scalar field with the potential~$V(\phi)=M_{\phi}^2 \phi^2/2$, the property of 
BHs was studied in Ref.~\cite{Doneva:2019vuh} 
for the linear GB coupling~$\xi(\phi)=\alpha \phi$.
Since in this case the mass~$M_{\phi}$ is not related
to the GB coupling~$\alpha$, the resulting BH 
solution is different from that discussed above for $n=2$.
Indeed, the second-order differential equation for 
$\hat{\phi}_1(r)$ contains 
a mass term~$-M_{\phi}^2 \hat{\phi}_1(r)$. In such a case, we do not have an analytic solution for $\hat{\phi}_1(r)$, 
so it requires numerical integration as performed in Ref.~\cite{Doneva:2019vuh}.
At spatial infinity, the scalar field solution is approximately given by the form~$\hat{\phi}_1(r) \simeq C_1e^{M_{\phi}r}/r+C_2 e^{-M_{\phi}r}/r$, with $C_1$ and $C_2$ being constants.
In order to satisfy the boundary 
condition~\eqref{bc2}, namely
$\hat{\phi}_1'(\infty)=0$,
we need to choose the coefficient~$C_1$ to be 
zero.
The existence of hairy BH solutions was numerically confirmed
for the case of quadratic potential~$V(\phi)=M_{\phi}^2 \phi^2/2$ with several different choices 
of GB couplings including the linear coupling~$\xi(\phi)=\alpha \phi$.

\section{Conclusions}
\label{sec:conc}

In this paper, we scrutinized the existence and 
linear stability of static and spherically symmetric 
BH solutions with a static scalar field
in full Horndeski theories 
without imposing the shift symmetry from the outset.
For this purpose, we employed a perturbative method of deriving 
BH solutions, which is valid in the regime of small coupling 
constant(s). 
We then exploited the conditions for the absence of ghost/Laplacian 
instabilities against odd- and even-parity perturbations derived in 
Refs.~\cite{Kobayashi:2012kh,Kobayashi:2014wsa,Kase:2021mix}, 
which are summarized in Sec.~\ref{sec:background}.
In particular, the angular propagation speed of 
even-parity perturbations plays an important role for 
ruling out some of the BH solutions by 
the Laplacian instability around the BH horizon. 
In shift-symmetric Horndeski theories, it was shown in 
Ref.~\cite{Minamitsuji:2022mlv} that hairy BH solutions present 
for theories with the k-essence Lagrangian~$G_2(X)$ and 
a nonminimal derivative coupling~$G_4(X)R$~\cite{Rinaldi:2012vy,Anabalon:2013oea,Minamitsuji:2013ura,Cisterna:2014nua} 
are subject to this generic instability around the horizon.

In Sec.~\ref{sec:expansion}, we extended the linear stability analysis for BHs in shift-symmetric theories performed in \cite{Minamitsuji:2022mlv} to full Horndeski theories. 
For hairy BHs where the scalar-field kinetic term~$X$ 
is an analytic function of $r$ with a nonvanishing value 
on the horizon ($X_s \neq 0$), the product~${\cal F}{\cal K}B_2$ 
was shown to be negative
for nonzero values of $\kappa$ defined by Eq.~(\ref{xi}). 
This implies that the linear stability conditions summarized in Sec.~\ref{sec:background} cannot be satisfied simultaneously, and hence
BHs with $X_s \neq 0$ are generally subject to either ghost or Laplacian instability. We also found that, as long as $X_s \neq 0$ with $\kappa_r \neq 0$, there is the divergence of the radial propagation speed of scalar-field perturbations. 
In Sec.~\ref{BHinsec}, we presented examples of theories 
that give rise to the branch of unstable hairy BH 
solutions with $X_s \neq 0$. 
Our results show that, even in full Horndeski theories, 
BH solutions free from instabilities should not 
have a nonvanishing $X_s$ in general.
Under this condition, there is also a jump of $X$ across 
the horizon, so the BH solutions are physically unacceptable.

Given the generic instability for BHs with $X_s\ne 0$, for the search of hairy BHs which are free from ghost or Laplacian instabilities, we focused on theories leading to the solutions with $X_s=0$, i.e., a finite scalar-field derivative~$\phi'(r_s)$ on the horizon.
Using the scalar-field equation of motion for theories containing 
the scalar-field kinetic term~$\eta X$ and the Einstein-Hilbert term~$\Mpl^2/2$ 
in the action, we discussed the possibility 
for realizing such hairy BHs in both shift-symmetric and non-shift-symmetric Horndeski theories in Sec.~\ref{nohairsec}. 
For the couplings of the form~$G_{I} \supset \alpha_I (\phi) F_I(X)$ 
($I=2,3,4,5$) with $\alpha_I(\phi)$ and $F_I(X)$ 
being arbitrary regular functions,
the asymptotically Minkowski solutions respecting the regularity on the horizon are restricted to be no-hair solutions with $\phi'(r)=0$.  
There are possibilities for evading this no-hair feature of BHs 
in theories given by the coupling functions~(\ref{Gchoicef}), 
which are not analytic at $X=0$. 
However, the couplings~$G_2 \supset \alpha_2(\phi) \sqrt{-X}$ and 
$G_4 \supset \alpha_4(\phi) \sqrt{-X}$ result in BH solutions 
with $X_s \neq 0$, so they are excluded in terms of the gradient or Laplacian instabilities around the horizon.
For the cubic logarithmic coupling~$G_3 \supset 
\alpha_3(\phi) \ln |X|$, where $\alpha_3(\phi)$ is an analytic function of $\phi$,  
we showed the absence of asymptotically Minkowski hairy BH solutions. 

The remaining theories allowing for the existence of hairy BH solutions should possess a quintic coupling of the form $G_5=\alpha_5(\phi)\ln |X|$. 
With this coupling only, the background equations contain 
terms like $X^p \ln |X|$, which causes the breakdown of 
our perturbative analysis
by using the expansions (\ref{sch_pert}) and (\ref{sch_pert2}) 
with respect to a small coupling parameter.
The scalar field coupled to the GB curvature invariant, 
$\xi(\phi) R_{\rm GB}^2$, which is equivalent to the Horndeski functions~(\ref{Ggauss}), is the only exceptional case in which $\ln |X|$-dependent terms disappear from 
the background equations due to the presence of the other specific 
couplings~$G_{2,3,4}$ besides $G_5=\alpha_5(\phi)\ln |X|$.
 
In Sec.~\ref{geneGB}, we derived the solutions to 
the metric components and the scalar field for the GB term~$\alpha \xi(\phi) R_{\rm GB}^2$ with an analytic 
function~$\xi(\phi)$ by resorting to the expansions with 
respect to the small GB coupling~$\alpha$. 
We also showed that ghost and Laplacian instabilities
of hairy BHs against odd- and even-parity 
perturbations are absent for the small GB coupling with $\eta>0$.
In Sec.~\ref{GBother}, we implemented positive 
power-law functions of $\phi$ or $X$ 
in $G_{2,3,4,5}$ besides the GB coupling~$\alpha \xi(\phi) R_{\rm GB}^2$ and obtained new classes of hairy BH solutions free from 
ghost or Laplacian instabilities.
Since all such hairy BH solutions disappear in the absence of 
the term~$\alpha_5(\phi)\ln |X|$ in $G_5$,
the existence of this form of quintic couplings is 
crucial for realizing asymptotically Minkowski hairy BH solutions. We also found that, even 
in the presence of the GB term~$\alpha \xi(\phi) R_{\rm GB}^2$, 
the couplings~$G_3 \supset \gamma_3 \ln (-X)$ or 
$G_4 \supset \gamma_4 \sqrt{-X}$
prevent the existence of asymptotically Minkowski solutions with scalar hair which do not suffer from ghost or Laplacian instabilities.

In Sec.~\ref{sec:fG}, we studied whether BH solutions 
free from ghost or Laplacian instabilities are present
in $F(R_{\rm GB}^2)$~gravity where $F$ is a regular function 
of the GB term~$R_{\rm GB}^2$. 
For power-law couplings~$F(R_{\rm GB}^2)=\beta (R_{\rm GB}^2)^n$ 
with $n \geq 2$, we found a new class of asymptotically Minkowski BH solutions where the GB term plays a role of the new scalar degree of freedom. However, they are prone to the ghost instability of even-parity perturbations. 
Although we also studied $F(R_{\rm GB}^2)$-equivalent 
Horndeski theories with a scalar-field kinetic 
term~$\eta X$, there are no hairy BH solutions
with asymptotically Minkowski metric. 
These results show that the presence of 
the GB coupling~$\alpha \xi(\phi) R_{\rm GB}^2$ besides the kinetic 
term~$\eta X$ play a prominent role for realizing 
asymptotically Minkowski hairy BHs free from ghost or Laplacian instabilities in full Horndeski theories.

In \hyperref[table]{Table}, we summarize the existence of 
hairy BH solutions and ghost/Laplacian instabilities 
for the theories studied in this paper.
Theories~(F) and (G) are the examples leading to 
asymptotically Minkowski BHs that can satisfy all the 
conditions for the absence of ghost or Laplacian instabilities
against odd- and even-parity perturbations.
It should be noted that both these theories contain 
the scalar-GB coupling~$\alpha \xi(\phi) 
R_{\rm GB}^2$.\footnote{Our result is reminiscent of that in Ref.~\cite{Minamitsuji:2019iwp}, where the authors showed that scalar-GB gravity is effectively the only theory within the Horndeski class that accommodates spontaneous scalarization of Schwarzschild BHs.} 
It will be of interest to compute the sensitivity parameters  
as well as the quasinormal modes for such 
surviving BH solutions for the purpose of detecting signatures 
of the modification of gravity in future observations of GWs.

\begin{table}[t]
\renewcommand\thetable{\!\!}
\begin{center}
\caption[crit]{Existence and the linear stability 
of asymptotically Minkowski hairy BHs 
for nine subclasses of Horndeski theories. 
Except for the theory~(H), we assumed 
the presence of the canonical kinetic term~$\eta X$ in $G_2$ and 
the Einstein-Hilbert term~$\Mpl^2/2$ in $G_4$. 
In the third column, ``AM'' means ``asymptotically Minkowski.''
}
\begin{tabular}{|c|c|c|c|} 
\hline
~Theory~ & Coupling functions & Hairy BHs &  Stability of hairy BHs  \\ 
\hline
\hline
(A) & $G_{2,3,4,5}(X) \supset F_I(X)$ with regular $F_I(X)$
& --
& --
\\
 \hline 
(B) & ~$G_{2,3,4,5}(\phi,X) \supset \alpha_I (\phi) F_I(X)$ 
with regular $\alpha_I(\phi)$ and $F_I (X)$~
&  -- 
 & -- \\
\hline 
(C) & $G_{2}(\phi,X) \supset \alpha_2 (\phi)\sqrt{-X}$ 
& $X_s \neq 0$ 
& 
Unstable 
around the horizon \\
\hline
(D) & $G_3(\phi,X) \supset
\alpha_3(\phi) \ln |X|$  & 
~$X_s=0$, Non-AM~
& -- \\
\hline
(E) & $G_{4}(\phi,X) \supset \alpha_4 (\phi)\sqrt{-X}$
& $X_s \neq 0$ 
& Unstable around the horizon \\
\hline
(F) & GB coupling~$\alpha \xi(\phi) R_{\rm GB}^2$ with regular $\xi(\phi)$
& $X_s=0$, AM
& ~No ghost/Laplacian instability~
\\
\hline 
(G) & ~$\alpha \xi(\phi) R_{\rm GB}^2$ 
plus regular functions of $\phi$ and/or $X$
in $G_{2,3,4,5}$~
& $X_s=0$, AM
& No ghost/Laplacian instability
\\
\hline 
(H) &
$F(R_{\rm GB}^2)\propto (R^2_{\rm GB})^n$ gravity with $n\geq 2$
& AM
& Ghost instability \\
\hline 
(I) &
~$F(R_{\rm GB}^2)$-equivalent
Horndeski theories with $G_2\supset \eta X$~
& -- 
& -- \\
\hline
\end{tabular}
\end{center}
\label{table} 
\end{table}


\acknowledgments{
MM was supported by the Portuguese national fund 
through the Funda\c{c}\~{a}o para a Ci\^encia e a Tecnologia (FCT) in the scope of the framework of the Decree-Law 57/2016 
of August 29, changed by Law 57/2017 of July 19,
and the Centro de Astrof\'{\i}sica e Gravita\c c\~ao (CENTRA) through the Project~No.~UIDB/00099/2020.
K.T.\ was supported by JSPS (Japan Society for the Promotion of Science) 
KAKENHI Grant No.\ JP21J00695.
S.T.\ was supported by the Grant-in-Aid for Scientific Research 
Fund of the JSPS Nos.~19K03854 and 22K03642.
}


\appendix

\section{Coefficients appearing in the background 
scalar-field equation}
\label{AppA}

The coefficients~$\lambda_1$--$\lambda_{12}$ in 
Eq.~(\ref{Ephi2}) are given by 
\ba
& &
\lambda_1= -\left( h'+\frac{4h}{r}+\frac{f' h}{f} \right)\phi'-2h \phi''\,,\qquad 
\lambda_2=- h \phi'^2\,,\qquad 
\lambda_3=\frac12 h \phi'^2 \left( h' \phi'+2h \phi'' \right)\,,\nonumber \\
& &
\lambda_4=\frac{2}{r^2} (1-h -r h')+\frac{hf'^2}{2f^2}
-\frac{r(2f''h+f'h')+4f' h}{2fr}\,,\nonumber \\
& &
\lambda_5=h \phi' \left[ \left( \frac{8h'}{r}+\frac{6h}{r^2}
-\frac{f'^2 h}{2f^2}+\frac{(f''r+6f')h+2rf'h'}{fr} \right)\phi'
+3h \left( \frac{f'}{f}+\frac{4}{r} \right)\phi'' \right]\,,\nonumber \\
& &
\lambda_6=h^2 \phi'^3 \left( \frac{f'}{f}+\frac{4}{r} \right)\,,\qquad 
\lambda_7=-\frac12 h^2 \phi'^3
 \left( \frac{f'}{f}+\frac{4}{r} \right) \left( h' \phi'+2h \phi'' \right)\,,
 \nonumber \\
& &
\lambda_8=\frac{1}{r^2} \left[  h'(3h-1)\phi'+2h (h - 1) \phi''  \right]
-\frac{f'^2 h^2 \phi'}{f^2 r}
+\frac{1}{fr^2} \left[ (2f'' r+3f')h^2 \phi'
+f' h(3rh'-1) \phi'+2f' h^2 r \phi'' \right]\,, \nonumber \\
& &
\lambda_9=\frac{h \phi'^2}{fr^2} \left[ f(h-1)+f' hr \right]\,,
\nonumber \\
& &
\lambda_{10}=-\frac{h \phi'^2}{2r^2} \left[ 10 h^2 \phi''
+h(7 h' \phi'-2\phi'')-h' \phi' \right]
+\frac{f'^2 h^3 \phi'^3}{2f^2 r}
-\frac{h^2 \phi'^2}{2fr^2} \left[ 
( 2f'' r+4f' )h\phi'+10f' hr \phi''+7f'h'r \phi'\right],\nonumber \\
& &
\lambda_{11}=-\frac{h^3 \phi'^4}{fr^2} (rf'+f)\,,\qquad 
\lambda_{12}=\frac{h^3 \phi'^4}{2fr^2}(rf'+f)
(h' \phi'+2h \phi'')\,.
\label{lambda}
\ea

\section{Coefficients associated with perturbations}\label{AppB}

The quantities~$a_1$, $c_2$, and $c_4$ in Eqs.~(\ref{defP1}) 
and (\ref{cr2}) are given by 
\ba
a_1&=&\sqrt{fh} \left\{  \left[ G_{4,\phi}+\frac12 h ( G_{3,X}-2 G_{4,\phi X} ) \phi'^2 \right] r^2
+2 h \phi' \left[ G_{4,X}-G_{5,\phi}-\frac12h ( 2 G_{4,XX}-G_{5,\phi X} ) \phi'^2 \right] r
\right.
\notag\\
&&
\left.
+\frac12 G_{5,XX} h^3\phi'^4
-\frac12 G_{5,X} h ( 3 h-1 ) \phi'^2 
\right\}\,,\\
c_2&=&\sqrt{fh} \left\{  \left[  
\frac{1}{2f}\left( -\frac12 h ( 3 G_{3,X}-8 G_{4,\phi X} ) \phi'^2
+\frac12 h^2 ( G_{3,XX}-2 G_{4,\phi XX} ) \phi'^4
-G_{4,\phi} \right) r^2
\right.\right.
\notag\\
&&
\left.\left.
-{\frac {h\phi'}{f}} \left( 
\frac12 {h^2 ( 2 G_{4,XXX}-G_{5,\phi XX} ) \phi'^4}
-\frac12 {h ( 12 G_{4,XX}-7 G_{5,\phi X} ) \phi'^2}
+3 ( G_{4,X}-G_{5,\phi} ) \right) r
\right.\right.
\notag\\
&&
\left.\left.
+\frac{h\phi'^2}{4f}\left(
G_{5,XXX} h^3\phi'^4
- G_{5,XX} h ( 10 h-1 ) \phi'^2
+3 G_{5,X}  ( 5 h-1 ) 
\right) \right] f'
\right.
\notag\\
&&
\left.
+\phi' \left[ \frac12G_{2,X}-G_{3,\phi}
-\frac12 h ( G_{2,XX}-G_{3,\phi X} ) \phi'^2 \right] r^2
\right.
\notag\\
&&
\left.
+ 2\left[ -\frac12h ( 3 G_{3,X}-8 G_{4,\phi X} ) \phi'^2
+\frac12h^2 ( G_{3,XX}-2 G_{4,\phi XX} ) \phi'^4
-G_{4,\phi} \right] r
-\frac12 h^3 ( 2 G_{4,XXX}-G_{5,\phi XX} ) \phi'^5
\right.
\notag\\
&&
\left.
+\frac12 h \left[ 2\left(6 h-1\right) G_{4,XX}+\left(1-7 h\right)G_{5,\phi X} \right] \phi'^3
- ( 3 h-1 )  ( G_{4,X}-G_{5,\phi} ) \phi' \right\} \,,\\
c_4&=&\frac14 \frac {\sqrt {f}}{\sqrt {h}} 
\left\{ \frac {h\phi'}{f} \left[ 
2 G_{4,X}-2 G_{5,\phi}
-h ( 2 G_{4,XX}-G_{5,\phi X} ) \phi'^2
-{\frac {h\phi'  ( 3 G_{5,X}-G_{5,XX} \phi'^2h ) }{r}} \right]f'
\right.
\notag\\
&&
\left.
+4 G_{4,\phi}
+2 h ( G_{3,X}-2 G_{4,\phi X} ) \phi'^2
+{\frac {4 h ( G_{4,X}-G_{5,\phi} ) \phi'-2 h^2 ( 2 G_{4,XX}-G_{5,\phi X} ) \phi'^3}{r}} \right\} \,.
\ea
The quantities~$B_1$ and $B_2$ in 
Eq.~(\ref{cosq}) are 
\ba
&&
B_1=
\frac {r^3\sqrt {f h} {\cal H} [ 4 h ( \phi' a_1+r\sqrt{fh} {\cal H}) 
\beta_1+\beta_2-4 \phi' a_1 \beta_3] 
-2 fh {\cal G}  [ r \sqrt{fh}( 2 {\cal P}_1-{\cal F}){\cal H}  
( 2\phi' a_1+r\sqrt{fh} {\cal H} )+2\phi'^2a_1^{2}{\cal P}_1 ] }
{4f h ( 2 {\cal P}_1-{\cal F} ){\cal H}
(\phi' a_1+r\sqrt{fh} {\cal H})^2}\,,
\label{B1def}
\notag\\\\
&&
B_2=
-r^2{\frac {r^2h \beta_1 [ 2 fh {\cal F} {\cal G} ( \phi' a_1+r \sqrt{fh}{\cal H} ) 
+r^2\beta_2 ] -{r}^{4}\beta_2 \beta_3
-fh{\cal F} {\cal G}  ( \phi' fh {\cal F} {\cal G}a_1 +2 r^3 \sqrt{fh} {\cal H} \beta_3 ) }
{ fh\phi' a_1 ( 2 {\cal P}_1-{\cal F} ){\cal F}  ( \phi' a_1+r \sqrt{fh}{\cal H} ) ^{2}}}\,,
\label{B2def}
\ea
where
\ba
\beta_1&=&\frac12 \phi'^2 \sqrt{fh} {\cal H}e_4 
-\phi' \left(\sqrt{fh}{\cal H} \right)' c_4 
+ \frac{\sqrt{fh}}{2}\left[ \left( {\frac {f'}{f}}+{\frac {h'}{h}}-\frac{2}{r} \right) {\cal H}
+{\frac {2{\cal F}}{r}} \right] \phi' c_4+{\frac {f{\cal F} {\cal G}}{2r^2}}\,,\\
\beta_2&=& \left[ \frac{\sqrt{fh}{\cal F}}{r^2} \left( 2 hr\phi'^2c_4
+\frac{r \phi' f' \sqrt{h}}{2\sqrt{f}}{\cal H}-\phi' \sqrt{fh}{\cal G} \right)
-\frac{\phi' fh {\cal G}{\cal H}}{r} \left( \frac{{\cal G}'}{{\cal G}}
-\frac{{\cal H}'}{{\cal H}}+\frac{f'}{2f}-\frac{1}{r} \right) \right]a_1 
-\frac{2}{r} (fh)^{3/2}{\cal F}{\cal G}{\cal H}\,, \qquad\,\,\\
\beta_3&=& \frac{\sqrt{fh}{\cal H}}{2}\phi'  
\left( hc_4'+\frac12 h' c_4-\frac{d_3}{2} \right) 
-\frac{\sqrt{fh}}{2} \left( \frac{\cal H}{r}+{\cal H}' \right) 
\left( 2 h \phi'c_4+\frac{\sqrt{fh}{\cal G}}{2r}
+\frac{f'\sqrt{h}{\cal H}}{4\sqrt{f}} \right)\notag\\
&&
+{\frac {\sqrt {fh}{\cal F}}{4r} \left(  2 h \phi'c_4
+\frac{3\sqrt{fh}{\cal G}}{r}
+\frac{f'\sqrt{h}{\cal H}}{2\sqrt{f}}
 \right) }\,,
\ea
with
\ba
e_4 &=&{\frac {1}{\phi'}}c_4'-{\frac {f'}{4f h \phi'^2}} 
\left( \sqrt{fh} {\cal H} \right)'
-{\frac {\sqrt {f}}{2\phi'^2\sqrt {h}r}}{\cal G}'
+{\frac {1}{h\phi' r^2} \left( {\frac {\phi''}{\phi'}}+\frac12 {\frac {h'}{h}} \right) }a_1
\notag\\
&&
+{\frac {\sqrt{f}}{8\sqrt{h}\phi'^2} 
\left[ {\frac { ( f' r-6 f ) f'}{f^2r}}
+\frac {h' ( f' r+4 f ) }{fhr}
-{\frac {4f ( 2 \phi'' h+h' \phi')}
{\phi' h^2r ( f' r-2 f ) }} \right] }{\cal H}
+{\frac {h'}{2h\phi'}}c_4
-\frac{f'r-2f}{4\sqrt{fh}r\phi'}
\frac{\partial {\cal H}}
{\partial \phi}
\notag\\
&&
+{\frac {f' hr-f}{2r^2\sqrt {f}{h}^{3/2}\phi'^2}}{\cal F}
+{\frac {\sqrt {f}}{2r\phi'^2{h}^{3/2}} 
\left[ {\frac {f ( 2 \phi'' h+h' \phi' ) }{h\phi'  ( f' r-2 f ) }}
+{\frac {2 f-f' hr}{2fr}} \right] }{\cal G}
\,, \label{e4}\\
d_3&=&
-{\frac {1}{r^2} \left( {\frac {2\phi''}{\phi'}}+{\frac {h'}{h}} \right) }a_1
+{\frac {f^{3/2}h^{1/2}}{ ( f' r-2 f ) \phi'} \left( 
{\frac {2\phi''}{h\phi' r}}
+ {\frac {{f'}^{2}}{f^2}}
- {\frac {f' h'}{fh}}
-{\frac {2f'}{fr}}
+{\frac {2h'}{hr}}
+ {\frac {h'}{h^2r}} \right) }{\cal H}
\notag\\
&&
+\frac{f'r-2f}{2r} \sqrt{\frac{h}{f}} 
\frac{\partial{\cal H}}{\partial \phi}
+{\frac {\sqrt {f}}{\phi' \sqrt {h}r^2}}{\cal F}
-{\frac {{f}^{3/2}}{\sqrt {h} ( f' r-2 f ) \phi'} 
\left( {\frac {f'}{fr}}+{\frac {2\phi''}{\phi' r}}+{\frac {h'}{hr}}-\frac{2}{r^2} \right) }{\cal G}
\,.
\ea
%


\bibliographystyle{mybibstyle}
\bibliography{bib}

\begin{thebibliography}{113}%
\makeatletter
\providecommand \@ifxundefined [1]{%
 \@ifx{#1\undefined}
}%
\providecommand \@ifnum [1]{%
 \ifnum #1\expandafter \@firstoftwo
 \else \expandafter \@secondoftwo
 \fi
}%
\providecommand \@ifx [1]{%
 \ifx #1\expandafter \@firstoftwo
 \else \expandafter \@secondoftwo
 \fi
}%
\providecommand \natexlab [1]{#1}%
\providecommand \enquote  [1]{``#1''}%
\providecommand \bibnamefont  [1]{#1}%
\providecommand \bibfnamefont [1]{#1}%
\providecommand \citenamefont [1]{#1}%
\providecommand \href@noop [0]{\@secondoftwo}%
\providecommand \href [0]{\begingroup \@sanitize@url \@href}%
\providecommand \@href[1]{\@@startlink{#1}\@@href}%
\providecommand \@@href[1]{\endgroup#1\@@endlink}%
\providecommand \@sanitize@url [0]{\catcode `\\12\catcode `\$12\catcode
  `\&12\catcode `\#12\catcode `\^12\catcode `\_12\catcode `\%12\relax}%
\providecommand \@@startlink[1]{}%
\providecommand \@@endlink[0]{}%
\providecommand \url  [0]{\begingroup\@sanitize@url \@url }%
\providecommand \@url [1]{\endgroup\@href {#1}{\urlprefix }}%
\providecommand \urlprefix  [0]{URL }%
\providecommand \Eprint [0]{\href }%
\providecommand \doibase [0]{http://dx.doi.org/}%
\providecommand \selectlanguage [0]{\@gobble}%
\providecommand \bibinfo  [0]{\@secondoftwo}%
\providecommand \bibfield  [0]{\@secondoftwo}%
\providecommand \translation [1]{[#1]}%
\providecommand \BibitemOpen [0]{}%
\providecommand \bibitemStop [0]{}%
\providecommand \bibitemNoStop [0]{.\EOS\space}%
\providecommand \EOS [0]{\spacefactor3000\relax}%
\providecommand \BibitemShut  [1]{\csname bibitem#1\endcsname}%
\let\auto@bib@innerbib\@empty
\bibitem [{\citenamefont {Will}(2014)}]{Will:2014kxa}%
  \BibitemOpen
  \bibfield  {author} {\bibinfo {author} {\bibfnamefont {C.~M.}\ \bibnamefont
  {Will}},\ }\href {\doibase 10.12942/lrr-2014-4} {\bibfield  {journal}
  {\bibinfo  {journal} {\emph {Living Rev. Rel.}}\ }\textbf {\bibinfo {volume}
  {17}},\ \bibinfo {pages} {4} (\bibinfo {year} {2014})},\ \Eprint
  {http://arxiv.org/abs/1403.7377} {arXiv:1403.7377 [gr-qc]} \BibitemShut
  {NoStop}%
\bibitem [{\citenamefont {Abbott}\ \emph {et~al.}(2016)\citenamefont {Abbott}
  \emph {et~al.}}]{LIGOScientific:2016aoc}%
  \BibitemOpen
  \bibfield  {author} {\bibinfo {author} {\bibfnamefont {B.~P.}\ \bibnamefont
  {Abbott}} \emph {et~al.} (\bibinfo {collaboration} {LIGO Scientific,
  Virgo}),\ }\href {\doibase 10.1103/PhysRevLett.116.061102} {\bibfield
  {journal} {\bibinfo  {journal} {\emph {Phys. Rev. Lett.}}\ }\textbf {\bibinfo
  {volume} {116}},\ \bibinfo {pages} {061102} (\bibinfo {year} {2016})},\
  \Eprint {http://arxiv.org/abs/1602.03837} {arXiv:1602.03837 [gr-qc]}
  \BibitemShut {NoStop}%
\bibitem [{\citenamefont {Akiyama}\ \emph {et~al.}(2019)\citenamefont {Akiyama}
  \emph {et~al.}}]{EventHorizonTelescope:2019dse}%
  \BibitemOpen
  \bibfield  {author} {\bibinfo {author} {\bibfnamefont {K.}~\bibnamefont
  {Akiyama}} \emph {et~al.} (\bibinfo {collaboration} {Event Horizon
  Telescope}),\ }\href {\doibase 10.3847/2041-8213/ab0ec7} {\bibfield
  {journal} {\bibinfo  {journal} {\emph {Astrophys. J. Lett.}}\ }\textbf
  {\bibinfo {volume} {875}},\ \bibinfo {pages} {L1} (\bibinfo {year} {2019})},\
  \Eprint {http://arxiv.org/abs/1906.11238} {arXiv:1906.11238 [astro-ph.GA]}
  \BibitemShut {NoStop}%
\bibitem [{\citenamefont {Berti}\ \emph {et~al.}(2015)\citenamefont {Berti}
  \emph {et~al.}}]{Berti:2015itd}%
  \BibitemOpen
  \bibfield  {author} {\bibinfo {author} {\bibfnamefont {E.}~\bibnamefont
  {Berti}} \emph {et~al.},\ }\href {\doibase 10.1088/0264-9381/32/24/243001}
  {\bibfield  {journal} {\bibinfo  {journal} {\emph {Class. Quant. Grav.}}\
  }\textbf {\bibinfo {volume} {32}},\ \bibinfo {pages} {243001} (\bibinfo
  {year} {2015})},\ \Eprint {http://arxiv.org/abs/1501.07274} {arXiv:1501.07274
  [gr-qc]} \BibitemShut {NoStop}%
\bibitem [{\citenamefont {Barack}\ \emph {et~al.}(2019)\citenamefont {Barack}
  \emph {et~al.}}]{Barack:2018yly}%
  \BibitemOpen
  \bibfield  {author} {\bibinfo {author} {\bibfnamefont {L.}~\bibnamefont
  {Barack}} \emph {et~al.},\ }\href {\doibase 10.1088/1361-6382/ab0587}
  {\bibfield  {journal} {\bibinfo  {journal} {\emph {Class. Quant. Grav.}}\
  }\textbf {\bibinfo {volume} {36}},\ \bibinfo {pages} {143001} (\bibinfo
  {year} {2019})},\ \Eprint {http://arxiv.org/abs/1806.05195} {arXiv:1806.05195
  [gr-qc]} \BibitemShut {NoStop}%
\bibitem [{\citenamefont {Berti}\ \emph
  {et~al.}(2018{\natexlab{a}})\citenamefont {Berti}, \citenamefont {Yagi},\
  and\ \citenamefont {Yunes}}]{Berti:2018cxi}%
  \BibitemOpen
  \bibfield  {author} {\bibinfo {author} {\bibfnamefont {E.}~\bibnamefont
  {Berti}}, \bibinfo {author} {\bibfnamefont {K.}~\bibnamefont {Yagi}},  and
  \bibinfo {author} {\bibfnamefont {N.}~\bibnamefont {Yunes}},\ }\href
  {\doibase 10.1007/s10714-018-2362-8} {\bibfield  {journal} {\bibinfo
  {journal} {\emph {Gen. Rel. Grav.}}\ }\textbf {\bibinfo {volume} {50}},\
  \bibinfo {pages} {46} (\bibinfo {year} {2018}{\natexlab{a}})},\ \Eprint
  {http://arxiv.org/abs/1801.03208} {arXiv:1801.03208 [gr-qc]} \BibitemShut
  {NoStop}%
\bibitem [{\citenamefont {Berti}\ \emph
  {et~al.}(2018{\natexlab{b}})\citenamefont {Berti}, \citenamefont {Yagi},
  \citenamefont {Yang},\ and\ \citenamefont {Yunes}}]{Berti:2018vdi}%
  \BibitemOpen
  \bibfield  {author} {\bibinfo {author} {\bibfnamefont {E.}~\bibnamefont
  {Berti}}, \bibinfo {author} {\bibfnamefont {K.}~\bibnamefont {Yagi}},
  \bibinfo {author} {\bibfnamefont {H.}~\bibnamefont {Yang}},  and \bibinfo
  {author} {\bibfnamefont {N.}~\bibnamefont {Yunes}},\ }\href {\doibase
  10.1007/s10714-018-2372-6} {\bibfield  {journal} {\bibinfo  {journal} {\emph
  {Gen. Rel. Grav.}}\ }\textbf {\bibinfo {volume} {50}},\ \bibinfo {pages} {49}
  (\bibinfo {year} {2018}{\natexlab{b}})},\ \Eprint
  {http://arxiv.org/abs/1801.03587} {arXiv:1801.03587 [gr-qc]} \BibitemShut
  {NoStop}%
\bibitem [{\citenamefont {Riess}\ \emph {et~al.}(1998)\citenamefont {Riess}
  \emph {et~al.}}]{SupernovaSearchTeam:1998fmf}%
  \BibitemOpen
  \bibfield  {author} {\bibinfo {author} {\bibfnamefont {A.~G.}\ \bibnamefont
  {Riess}} \emph {et~al.} (\bibinfo {collaboration} {Supernova Search Team}),\
  }\href {\doibase 10.1086/300499} {\bibfield  {journal} {\bibinfo  {journal}
  {\emph {Astron. J.}}\ }\textbf {\bibinfo {volume} {116}},\ \bibinfo {pages}
  {1009} (\bibinfo {year} {1998})},\ \Eprint
  {http://arxiv.org/abs/astro-ph/9805201} {arXiv:astro-ph/9805201} \BibitemShut
  {NoStop}%
\bibitem [{\citenamefont {Perlmutter}\ \emph {et~al.}(1999)\citenamefont
  {Perlmutter} \emph {et~al.}}]{SupernovaCosmologyProject:1998vns}%
  \BibitemOpen
  \bibfield  {author} {\bibinfo {author} {\bibfnamefont {S.}~\bibnamefont
  {Perlmutter}} \emph {et~al.} (\bibinfo {collaboration} {Supernova Cosmology
  Project}),\ }\href {\doibase 10.1086/307221} {\bibfield  {journal} {\bibinfo
  {journal} {\emph {Astrophys. J.}}\ }\textbf {\bibinfo {volume} {517}},\
  \bibinfo {pages} {565} (\bibinfo {year} {1999})},\ \Eprint
  {http://arxiv.org/abs/astro-ph/9812133} {arXiv:astro-ph/9812133} \BibitemShut
  {NoStop}%
\bibitem [{\citenamefont {Weinberg}(1989)}]{Weinberg:1988cp}%
  \BibitemOpen
  \bibfield  {author} {\bibinfo {author} {\bibfnamefont {S.}~\bibnamefont
  {Weinberg}},\ }\href {\doibase 10.1103/RevModPhys.61.1} {\bibfield  {journal}
  {\bibinfo  {journal} {\emph {Rev. Mod. Phys.}}\ }\textbf {\bibinfo {volume}
  {61}},\ \bibinfo {pages} {1} (\bibinfo {year} {1989})}\BibitemShut {NoStop}%
\bibitem [{\citenamefont {Riess}\ \emph {et~al.}(2019)\citenamefont {Riess},
  \citenamefont {Casertano}, \citenamefont {Yuan}, \citenamefont {Macri},\ and\
  \citenamefont {Scolnic}}]{Riess:2019cxk}%
  \BibitemOpen
  \bibfield  {author} {\bibinfo {author} {\bibfnamefont {A.~G.}\ \bibnamefont
  {Riess}}, \bibinfo {author} {\bibfnamefont {S.}~\bibnamefont {Casertano}},
  \bibinfo {author} {\bibfnamefont {W.}~\bibnamefont {Yuan}}, \bibinfo {author}
  {\bibfnamefont {L.~M.}\ \bibnamefont {Macri}},  and \bibinfo {author}
  {\bibfnamefont {D.}~\bibnamefont {Scolnic}},\ }\href {\doibase
  10.3847/1538-4357/ab1422} {\bibfield  {journal} {\bibinfo  {journal} {\emph
  {Astrophys. J.}}\ }\textbf {\bibinfo {volume} {876}},\ \bibinfo {pages} {85}
  (\bibinfo {year} {2019})},\ \Eprint {http://arxiv.org/abs/1903.07603}
  {arXiv:1903.07603 [astro-ph.CO]} \BibitemShut {NoStop}%
\bibitem [{\citenamefont {Di~Valentino}\ \emph {et~al.}(2021)\citenamefont
  {Di~Valentino}, \citenamefont {Mena}, \citenamefont {Pan}, \citenamefont
  {Visinelli}, \citenamefont {Yang}, \citenamefont {Melchiorri}, \citenamefont
  {Mota}, \citenamefont {Riess},\ and\ \citenamefont
  {Silk}}]{DiValentino:2021izs}%
  \BibitemOpen
  \bibfield  {author} {\bibinfo {author} {\bibfnamefont {E.}~\bibnamefont
  {Di~Valentino}}, \bibinfo {author} {\bibfnamefont {O.}~\bibnamefont {Mena}},
  \bibinfo {author} {\bibfnamefont {S.}~\bibnamefont {Pan}}, \bibinfo {author}
  {\bibfnamefont {L.}~\bibnamefont {Visinelli}}, \bibinfo {author}
  {\bibfnamefont {W.}~\bibnamefont {Yang}}, \bibinfo {author} {\bibfnamefont
  {A.}~\bibnamefont {Melchiorri}}, \bibinfo {author} {\bibfnamefont {D.~F.}\
  \bibnamefont {Mota}}, \bibinfo {author} {\bibfnamefont {A.~G.}\ \bibnamefont
  {Riess}},  and \bibinfo {author} {\bibfnamefont {J.}~\bibnamefont {Silk}},\
  }\href {\doibase 10.1088/1361-6382/ac086d} {\bibfield  {journal} {\bibinfo
  {journal} {\emph {Class. Quant. Grav.}}\ }\textbf {\bibinfo {volume} {38}},\
  \bibinfo {pages} {153001} (\bibinfo {year} {2021})},\ \Eprint
  {http://arxiv.org/abs/2103.01183} {arXiv:2103.01183 [astro-ph.CO]}
  \BibitemShut {NoStop}%
\bibitem [{\citenamefont {Fujii}\ and\ \citenamefont
  {Maeda}(2007)}]{Fujii:2003pa}%
  \BibitemOpen
  \bibfield  {author} {\bibinfo {author} {\bibfnamefont {Y.}~\bibnamefont
  {Fujii}} and \bibinfo {author} {\bibfnamefont {K.}~\bibnamefont {Maeda}},\
  }\href {\doibase 10.1017/CBO9780511535093} {\emph {\bibinfo {title} {{The
  scalar-tensor theory of gravitation}}}},\ Cambridge Monographs on
  Mathematical Physics\ (\bibinfo  {publisher} {Cambridge University Press},\
  \bibinfo {year} {2007})\BibitemShut {NoStop}%
\bibitem [{\citenamefont {Horndeski}(1974)}]{Horndeski}%
  \BibitemOpen
  \bibfield  {author} {\bibinfo {author} {\bibfnamefont {G.~W.}\ \bibnamefont
  {Horndeski}},\ }\href {\doibase 10.1007/BF01807638} {\bibfield  {journal}
  {\bibinfo  {journal} {\emph {Int. J. Theor. Phys.}}\ }\textbf {\bibinfo
  {volume} {10}},\ \bibinfo {pages} {363} (\bibinfo {year} {1974})}\BibitemShut
  {NoStop}%
\bibitem [{\citenamefont {Deffayet}\ \emph {et~al.}(2011)\citenamefont
  {Deffayet}, \citenamefont {Gao}, \citenamefont {Steer},\ and\ \citenamefont
  {Zahariade}}]{Def11}%
  \BibitemOpen
  \bibfield  {author} {\bibinfo {author} {\bibfnamefont {C.}~\bibnamefont
  {Deffayet}}, \bibinfo {author} {\bibfnamefont {X.}~\bibnamefont {Gao}},
  \bibinfo {author} {\bibfnamefont {D.~A.}\ \bibnamefont {Steer}},  and
  \bibinfo {author} {\bibfnamefont {G.}~\bibnamefont {Zahariade}},\ }\href
  {\doibase 10.1103/PhysRevD.84.064039} {\bibfield  {journal} {\bibinfo
  {journal} {\emph {Phys. Rev. D}}\ }\textbf {\bibinfo {volume} {84}},\
  \bibinfo {pages} {064039} (\bibinfo {year} {2011})},\ \Eprint
  {http://arxiv.org/abs/1103.3260} {arXiv:1103.3260 [hep-th]} \BibitemShut
  {NoStop}%
\bibitem [{\citenamefont {Kobayashi}\ \emph {et~al.}(2011)\citenamefont
  {Kobayashi}, \citenamefont {Yamaguchi},\ and\ \citenamefont
  {Yokoyama}}]{KYY}%
  \BibitemOpen
  \bibfield  {author} {\bibinfo {author} {\bibfnamefont {T.}~\bibnamefont
  {Kobayashi}}, \bibinfo {author} {\bibfnamefont {M.}~\bibnamefont
  {Yamaguchi}},  and \bibinfo {author} {\bibfnamefont {J.}~\bibnamefont
  {Yokoyama}},\ }\href {\doibase 10.1143/PTP.126.511} {\bibfield  {journal}
  {\bibinfo  {journal} {\emph {Prog. Theor. Phys.}}\ }\textbf {\bibinfo
  {volume} {126}},\ \bibinfo {pages} {511} (\bibinfo {year} {2011})},\ \Eprint
  {http://arxiv.org/abs/1105.5723} {arXiv:1105.5723 [hep-th]} \BibitemShut
  {NoStop}%
\bibitem [{\citenamefont {Charmousis}\ \emph {et~al.}(2012)\citenamefont
  {Charmousis}, \citenamefont {Copeland}, \citenamefont {Padilla},\ and\
  \citenamefont {Saffin}}]{Charmousis:2011bf}%
  \BibitemOpen
  \bibfield  {author} {\bibinfo {author} {\bibfnamefont {C.}~\bibnamefont
  {Charmousis}}, \bibinfo {author} {\bibfnamefont {E.~J.}\ \bibnamefont
  {Copeland}}, \bibinfo {author} {\bibfnamefont {A.}~\bibnamefont {Padilla}},
  and \bibinfo {author} {\bibfnamefont {P.~M.}\ \bibnamefont {Saffin}},\ }\href
  {\doibase 10.1103/PhysRevLett.108.051101} {\bibfield  {journal} {\bibinfo
  {journal} {\emph {Phys. Rev. Lett.}}\ }\textbf {\bibinfo {volume} {108}},\
  \bibinfo {pages} {051101} (\bibinfo {year} {2012})},\ \Eprint
  {http://arxiv.org/abs/1106.2000} {arXiv:1106.2000 [hep-th]} \BibitemShut
  {NoStop}%
\bibitem [{\citenamefont {Woodard}(2015)}]{Woodard:2015zca}%
  \BibitemOpen
  \bibfield  {author} {\bibinfo {author} {\bibfnamefont {R.~P.}\ \bibnamefont
  {Woodard}},\ }\href {\doibase 10.4249/scholarpedia.32243} {\bibfield
  {journal} {\bibinfo  {journal} {\emph {Scholarpedia}}\ }\textbf {\bibinfo
  {volume} {10}},\ \bibinfo {pages} {32243} (\bibinfo {year} {2015})},\ \Eprint
  {http://arxiv.org/abs/1506.02210} {arXiv:1506.02210 [hep-th]} \BibitemShut
  {NoStop}%
\bibitem [{\citenamefont {Langlois}\ and\ \citenamefont
  {Noui}(2016)}]{Langlois:2015cwa}%
  \BibitemOpen
  \bibfield  {author} {\bibinfo {author} {\bibfnamefont {D.}~\bibnamefont
  {Langlois}} and \bibinfo {author} {\bibfnamefont {K.}~\bibnamefont {Noui}},\
  }\href {\doibase 10.1088/1475-7516/2016/02/034} {\bibfield  {journal}
  {\bibinfo  {journal} {\emph {JCAP}}\ }\textbf {\bibinfo {volume} {02}},\
  \bibinfo {pages} {034} (\bibinfo {year} {2016})},\ \Eprint
  {http://arxiv.org/abs/1510.06930} {arXiv:1510.06930 [gr-qc]} \BibitemShut
  {NoStop}%
\bibitem [{\citenamefont {Crisostomi}\ \emph {et~al.}(2016)\citenamefont
  {Crisostomi}, \citenamefont {Koyama},\ and\ \citenamefont
  {Tasinato}}]{Crisostomi:2016czh}%
  \BibitemOpen
  \bibfield  {author} {\bibinfo {author} {\bibfnamefont {M.}~\bibnamefont
  {Crisostomi}}, \bibinfo {author} {\bibfnamefont {K.}~\bibnamefont {Koyama}},
  and \bibinfo {author} {\bibfnamefont {G.}~\bibnamefont {Tasinato}},\ }\href
  {\doibase 10.1088/1475-7516/2016/04/044} {\bibfield  {journal} {\bibinfo
  {journal} {\emph {JCAP}}\ }\textbf {\bibinfo {volume} {04}},\ \bibinfo
  {pages} {044} (\bibinfo {year} {2016})},\ \Eprint
  {http://arxiv.org/abs/1602.03119} {arXiv:1602.03119 [hep-th]} \BibitemShut
  {NoStop}%
\bibitem [{\citenamefont {Ben~Achour}\ \emph {et~al.}(2016)\citenamefont
  {Ben~Achour}, \citenamefont {Crisostomi}, \citenamefont {Koyama},
  \citenamefont {Langlois}, \citenamefont {Noui},\ and\ \citenamefont
  {Tasinato}}]{BenAchour:2016fzp}%
  \BibitemOpen
  \bibfield  {author} {\bibinfo {author} {\bibfnamefont {J.}~\bibnamefont
  {Ben~Achour}}, \bibinfo {author} {\bibfnamefont {M.}~\bibnamefont
  {Crisostomi}}, \bibinfo {author} {\bibfnamefont {K.}~\bibnamefont {Koyama}},
  \bibinfo {author} {\bibfnamefont {D.}~\bibnamefont {Langlois}}, \bibinfo
  {author} {\bibfnamefont {K.}~\bibnamefont {Noui}},  and \bibinfo {author}
  {\bibfnamefont {G.}~\bibnamefont {Tasinato}},\ }\href {\doibase
  10.1007/JHEP12(2016)100} {\bibfield  {journal} {\bibinfo  {journal} {\emph
  {JHEP}}\ }\textbf {\bibinfo {volume} {12}},\ \bibinfo {pages} {100} (\bibinfo
  {year} {2016})},\ \Eprint {http://arxiv.org/abs/1608.08135} {arXiv:1608.08135
  [hep-th]} \BibitemShut {NoStop}%
\bibitem [{\citenamefont {Takahashi}\ and\ \citenamefont
  {Kobayashi}(2017)}]{Takahashi:2017pje}%
  \BibitemOpen
  \bibfield  {author} {\bibinfo {author} {\bibfnamefont {K.}~\bibnamefont
  {Takahashi}} and \bibinfo {author} {\bibfnamefont {T.}~\bibnamefont
  {Kobayashi}},\ }\href {\doibase 10.1088/1475-7516/2017/11/038} {\bibfield
  {journal} {\bibinfo  {journal} {\emph {JCAP}}\ }\textbf {\bibinfo {volume}
  {11}},\ \bibinfo {pages} {038} (\bibinfo {year} {2017})},\ \Eprint
  {http://arxiv.org/abs/1708.02951} {arXiv:1708.02951 [gr-qc]} \BibitemShut
  {NoStop}%
\bibitem [{\citenamefont {Langlois}\ \emph {et~al.}(2019)\citenamefont
  {Langlois}, \citenamefont {Mancarella}, \citenamefont {Noui},\ and\
  \citenamefont {Vernizzi}}]{Langlois:2018jdg}%
  \BibitemOpen
  \bibfield  {author} {\bibinfo {author} {\bibfnamefont {D.}~\bibnamefont
  {Langlois}}, \bibinfo {author} {\bibfnamefont {M.}~\bibnamefont
  {Mancarella}}, \bibinfo {author} {\bibfnamefont {K.}~\bibnamefont {Noui}},
  and \bibinfo {author} {\bibfnamefont {F.}~\bibnamefont {Vernizzi}},\ }\href
  {\doibase 10.1088/1475-7516/2019/02/036} {\bibfield  {journal} {\bibinfo
  {journal} {\emph {JCAP}}\ }\textbf {\bibinfo {volume} {02}},\ \bibinfo
  {pages} {036} (\bibinfo {year} {2019})},\ \Eprint
  {http://arxiv.org/abs/1802.03394} {arXiv:1802.03394 [gr-qc]} \BibitemShut
  {NoStop}%
\bibitem [{\citenamefont {Takahashi}\ \emph {et~al.}(2022)\citenamefont
  {Takahashi}, \citenamefont {Motohashi},\ and\ \citenamefont
  {Minamitsuji}}]{Takahashi:2021ttd}%
  \BibitemOpen
  \bibfield  {author} {\bibinfo {author} {\bibfnamefont {K.}~\bibnamefont
  {Takahashi}}, \bibinfo {author} {\bibfnamefont {H.}~\bibnamefont
  {Motohashi}},  and \bibinfo {author} {\bibfnamefont {M.}~\bibnamefont
  {Minamitsuji}},\ }\href {\doibase 10.1103/PhysRevD.105.024015} {\bibfield
  {journal} {\bibinfo  {journal} {\emph {Phys. Rev. D}}\ }\textbf {\bibinfo
  {volume} {105}},\ \bibinfo {pages} {024015} (\bibinfo {year} {2022})},\
  \Eprint {http://arxiv.org/abs/2111.11634} {arXiv:2111.11634 [gr-qc]}
  \BibitemShut {NoStop}%
\bibitem [{\citenamefont {Copeland}\ \emph {et~al.}(2006)\citenamefont
  {Copeland}, \citenamefont {Sami},\ and\ \citenamefont
  {Tsujikawa}}]{Copeland:2006wr}%
  \BibitemOpen
  \bibfield  {author} {\bibinfo {author} {\bibfnamefont {E.~J.}\ \bibnamefont
  {Copeland}}, \bibinfo {author} {\bibfnamefont {M.}~\bibnamefont {Sami}},  and
  \bibinfo {author} {\bibfnamefont {S.}~\bibnamefont {Tsujikawa}},\ }\href
  {\doibase 10.1142/S021827180600942X} {\bibfield  {journal} {\bibinfo
  {journal} {\emph {Int. J. Mod. Phys. D}}\ }\textbf {\bibinfo {volume} {15}},\
  \bibinfo {pages} {1753} (\bibinfo {year} {2006})},\ \Eprint
  {http://arxiv.org/abs/hep-th/0603057} {arXiv:hep-th/0603057} \BibitemShut
  {NoStop}%
\bibitem [{\citenamefont {De~Felice}\ and\ \citenamefont
  {Tsujikawa}(2010)}]{DeFelice:2010aj}%
  \BibitemOpen
  \bibfield  {author} {\bibinfo {author} {\bibfnamefont {A.}~\bibnamefont
  {De~Felice}} and \bibinfo {author} {\bibfnamefont {S.}~\bibnamefont
  {Tsujikawa}},\ }\href {\doibase 10.12942/lrr-2010-3} {\bibfield  {journal}
  {\bibinfo  {journal} {\emph {Living Rev. Rel.}}\ }\textbf {\bibinfo {volume}
  {13}},\ \bibinfo {pages} {3} (\bibinfo {year} {2010})},\ \Eprint
  {http://arxiv.org/abs/1002.4928} {arXiv:1002.4928 [gr-qc]} \BibitemShut
  {NoStop}%
\bibitem [{\citenamefont {Clifton}\ \emph {et~al.}(2012)\citenamefont
  {Clifton}, \citenamefont {Ferreira}, \citenamefont {Padilla},\ and\
  \citenamefont {Skordis}}]{Clifton:2011jh}%
  \BibitemOpen
  \bibfield  {author} {\bibinfo {author} {\bibfnamefont {T.}~\bibnamefont
  {Clifton}}, \bibinfo {author} {\bibfnamefont {P.~G.}\ \bibnamefont
  {Ferreira}}, \bibinfo {author} {\bibfnamefont {A.}~\bibnamefont {Padilla}},
  and \bibinfo {author} {\bibfnamefont {C.}~\bibnamefont {Skordis}},\ }\href
  {\doibase 10.1016/j.physrep.2012.01.001} {\bibfield  {journal} {\bibinfo
  {journal} {\emph {Phys. Rept.}}\ }\textbf {\bibinfo {volume} {513}},\
  \bibinfo {pages} {1} (\bibinfo {year} {2012})},\ \Eprint
  {http://arxiv.org/abs/1106.2476} {arXiv:1106.2476 [astro-ph.CO]} \BibitemShut
  {NoStop}%
\bibitem [{\citenamefont {Joyce}\ \emph {et~al.}(2015)\citenamefont {Joyce},
  \citenamefont {Jain}, \citenamefont {Khoury},\ and\ \citenamefont
  {Trodden}}]{Joyce:2014kja}%
  \BibitemOpen
  \bibfield  {author} {\bibinfo {author} {\bibfnamefont {A.}~\bibnamefont
  {Joyce}}, \bibinfo {author} {\bibfnamefont {B.}~\bibnamefont {Jain}},
  \bibinfo {author} {\bibfnamefont {J.}~\bibnamefont {Khoury}},  and \bibinfo
  {author} {\bibfnamefont {M.}~\bibnamefont {Trodden}},\ }\href {\doibase
  10.1016/j.physrep.2014.12.002} {\bibfield  {journal} {\bibinfo  {journal}
  {\emph {Phys. Rept.}}\ }\textbf {\bibinfo {volume} {568}},\ \bibinfo {pages}
  {1} (\bibinfo {year} {2015})},\ \Eprint {http://arxiv.org/abs/1407.0059}
  {arXiv:1407.0059 [astro-ph.CO]} \BibitemShut {NoStop}%
\bibitem [{\citenamefont {Heisenberg}(2019)}]{Heisenberg:2018vsk}%
  \BibitemOpen
  \bibfield  {author} {\bibinfo {author} {\bibfnamefont {L.}~\bibnamefont
  {Heisenberg}},\ }\href {\doibase 10.1016/j.physrep.2018.11.006} {\bibfield
  {journal} {\bibinfo  {journal} {\emph {Phys. Rept.}}\ }\textbf {\bibinfo
  {volume} {796}},\ \bibinfo {pages} {1} (\bibinfo {year} {2019})},\ \Eprint
  {http://arxiv.org/abs/1807.01725} {arXiv:1807.01725 [gr-qc]} \BibitemShut
  {NoStop}%
\bibitem [{\citenamefont {Kase}\ and\ \citenamefont
  {Tsujikawa}(2019)}]{Kase:2018aps}%
  \BibitemOpen
  \bibfield  {author} {\bibinfo {author} {\bibfnamefont {R.}~\bibnamefont
  {Kase}} and \bibinfo {author} {\bibfnamefont {S.}~\bibnamefont {Tsujikawa}},\
  }\href {\doibase 10.1142/S0218271819420057} {\bibfield  {journal} {\bibinfo
  {journal} {\emph {Int. J. Mod. Phys. D}}\ }\textbf {\bibinfo {volume} {28}},\
  \bibinfo {pages} {1942005} (\bibinfo {year} {2019})},\ \Eprint
  {http://arxiv.org/abs/1809.08735} {arXiv:1809.08735 [gr-qc]} \BibitemShut
  {NoStop}%
\bibitem [{\citenamefont {Burrage}\ and\ \citenamefont
  {Seery}(2010)}]{Burrage:2010rs}%
  \BibitemOpen
  \bibfield  {author} {\bibinfo {author} {\bibfnamefont {C.}~\bibnamefont
  {Burrage}} and \bibinfo {author} {\bibfnamefont {D.}~\bibnamefont {Seery}},\
  }\href {\doibase 10.1088/1475-7516/2010/08/011} {\bibfield  {journal}
  {\bibinfo  {journal} {\emph {JCAP}}\ }\textbf {\bibinfo {volume} {08}},\
  \bibinfo {pages} {011} (\bibinfo {year} {2010})},\ \Eprint
  {http://arxiv.org/abs/1005.1927} {arXiv:1005.1927 [astro-ph.CO]} \BibitemShut
  {NoStop}%
\bibitem [{\citenamefont {Kimura}\ \emph {et~al.}(2012)\citenamefont {Kimura},
  \citenamefont {Kobayashi},\ and\ \citenamefont {Yamamoto}}]{Kimura:2011dc}%
  \BibitemOpen
  \bibfield  {author} {\bibinfo {author} {\bibfnamefont {R.}~\bibnamefont
  {Kimura}}, \bibinfo {author} {\bibfnamefont {T.}~\bibnamefont {Kobayashi}},
  and \bibinfo {author} {\bibfnamefont {K.}~\bibnamefont {Yamamoto}},\ }\href
  {\doibase 10.1103/PhysRevD.85.024023} {\bibfield  {journal} {\bibinfo
  {journal} {\emph {Phys. Rev. D}}\ }\textbf {\bibinfo {volume} {85}},\
  \bibinfo {pages} {024023} (\bibinfo {year} {2012})},\ \Eprint
  {http://arxiv.org/abs/1111.6749} {arXiv:1111.6749 [astro-ph.CO]} \BibitemShut
  {NoStop}%
\bibitem [{\citenamefont {Koyama}\ \emph {et~al.}(2013)\citenamefont {Koyama},
  \citenamefont {Niz},\ and\ \citenamefont {Tasinato}}]{Koyama:2013paa}%
  \BibitemOpen
  \bibfield  {author} {\bibinfo {author} {\bibfnamefont {K.}~\bibnamefont
  {Koyama}}, \bibinfo {author} {\bibfnamefont {G.}~\bibnamefont {Niz}},  and
  \bibinfo {author} {\bibfnamefont {G.}~\bibnamefont {Tasinato}},\ }\href
  {\doibase 10.1103/PhysRevD.88.021502} {\bibfield  {journal} {\bibinfo
  {journal} {\emph {Phys. Rev. D}}\ }\textbf {\bibinfo {volume} {88}},\
  \bibinfo {pages} {021502} (\bibinfo {year} {2013})},\ \Eprint
  {http://arxiv.org/abs/1305.0279} {arXiv:1305.0279 [hep-th]} \BibitemShut
  {NoStop}%
\bibitem [{\citenamefont {Kase}\ and\ \citenamefont
  {Tsujikawa}(2013)}]{Kase:2013uja}%
  \BibitemOpen
  \bibfield  {author} {\bibinfo {author} {\bibfnamefont {R.}~\bibnamefont
  {Kase}} and \bibinfo {author} {\bibfnamefont {S.}~\bibnamefont {Tsujikawa}},\
  }\href {\doibase 10.1088/1475-7516/2013/08/054} {\bibfield  {journal}
  {\bibinfo  {journal} {\emph {JCAP}}\ }\textbf {\bibinfo {volume} {08}},\
  \bibinfo {pages} {054} (\bibinfo {year} {2013})},\ \Eprint
  {http://arxiv.org/abs/1306.6401} {arXiv:1306.6401 [gr-qc]} \BibitemShut
  {NoStop}%
\bibitem [{\citenamefont {Vainshtein}(1972)}]{Vainshtein:1972sx}%
  \BibitemOpen
  \bibfield  {author} {\bibinfo {author} {\bibfnamefont {A.~I.}\ \bibnamefont
  {Vainshtein}},\ }\href {\doibase 10.1016/0370-2693(72)90147-5} {\bibfield
  {journal} {\bibinfo  {journal} {\emph {Phys. Lett. B}}\ }\textbf {\bibinfo
  {volume} {39}},\ \bibinfo {pages} {393} (\bibinfo {year} {1972})}\BibitemShut
  {NoStop}%
\bibitem [{\citenamefont {Khoury}\ and\ \citenamefont
  {Weltman}(2004)}]{Khoury:2003aq}%
  \BibitemOpen
  \bibfield  {author} {\bibinfo {author} {\bibfnamefont {J.}~\bibnamefont
  {Khoury}} and \bibinfo {author} {\bibfnamefont {A.}~\bibnamefont {Weltman}},\
  }\href {\doibase 10.1103/PhysRevLett.93.171104} {\bibfield  {journal}
  {\bibinfo  {journal} {\emph {Phys. Rev. Lett.}}\ }\textbf {\bibinfo {volume}
  {93}},\ \bibinfo {pages} {171104} (\bibinfo {year} {2004})},\ \Eprint
  {http://arxiv.org/abs/astro-ph/0309300} {arXiv:astro-ph/0309300} \BibitemShut
  {NoStop}%
\bibitem [{\citenamefont {Anabalon}\ \emph {et~al.}(2014)\citenamefont
  {Anabalon}, \citenamefont {Cisterna},\ and\ \citenamefont
  {Oliva}}]{Anabalon:2013oea}%
  \BibitemOpen
  \bibfield  {author} {\bibinfo {author} {\bibfnamefont {A.}~\bibnamefont
  {Anabalon}}, \bibinfo {author} {\bibfnamefont {A.}~\bibnamefont {Cisterna}},
  and \bibinfo {author} {\bibfnamefont {J.}~\bibnamefont {Oliva}},\ }\href
  {\doibase 10.1103/PhysRevD.89.084050} {\bibfield  {journal} {\bibinfo
  {journal} {\emph {Phys. Rev. D}}\ }\textbf {\bibinfo {volume} {89}},\
  \bibinfo {pages} {084050} (\bibinfo {year} {2014})},\ \Eprint
  {http://arxiv.org/abs/1312.3597} {arXiv:1312.3597 [gr-qc]} \BibitemShut
  {NoStop}%
\bibitem [{\citenamefont {Hawking}(1972{\natexlab{a}})}]{Hawking:1971vc}%
  \BibitemOpen
  \bibfield  {author} {\bibinfo {author} {\bibfnamefont {S.~W.}\ \bibnamefont
  {Hawking}},\ }\href {\doibase 10.1007/BF01877517} {\bibfield  {journal}
  {\bibinfo  {journal} {\emph {Commun. Math. Phys.}}\ }\textbf {\bibinfo
  {volume} {25}},\ \bibinfo {pages} {152} (\bibinfo {year}
  {1972}{\natexlab{a}})}\BibitemShut {NoStop}%
\bibitem [{\citenamefont {Bekenstein}(1972)}]{Bekenstein:1972ny}%
  \BibitemOpen
  \bibfield  {author} {\bibinfo {author} {\bibfnamefont {J.~D.}\ \bibnamefont
  {Bekenstein}},\ }\href {\doibase 10.1103/PhysRevLett.28.452} {\bibfield
  {journal} {\bibinfo  {journal} {\emph {Phys. Rev. Lett.}}\ }\textbf {\bibinfo
  {volume} {28}},\ \bibinfo {pages} {452} (\bibinfo {year} {1972})}\BibitemShut
  {NoStop}%
\bibitem [{\citenamefont {Graham}\ and\ \citenamefont
  {Jha}(2014)}]{Graham:2014mda}%
  \BibitemOpen
  \bibfield  {author} {\bibinfo {author} {\bibfnamefont {A.~A.~H.}\
  \bibnamefont {Graham}} and \bibinfo {author} {\bibfnamefont {R.}~\bibnamefont
  {Jha}},\ }\href {\doibase 10.1103/PhysRevD.89.084056} {\bibfield  {journal}
  {\bibinfo  {journal} {\emph {Phys. Rev. D}}\ }\textbf {\bibinfo {volume}
  {89}},\ \bibinfo {pages} {084056} (\bibinfo {year} {2014})},\ \bibinfo {note}
  {[Erratum: {\it Phys. Rev. D} {\bf 92}, 069901 (2015)]},\ \Eprint
  {http://arxiv.org/abs/1401.8203} {arXiv:1401.8203 [gr-qc]} \BibitemShut
  {NoStop}%
\bibitem [{\citenamefont {Hawking}(1972{\natexlab{b}})}]{Hawking:1972qk}%
  \BibitemOpen
  \bibfield  {author} {\bibinfo {author} {\bibfnamefont {S.~W.}\ \bibnamefont
  {Hawking}},\ }\href {\doibase 10.1007/BF01877518} {\bibfield  {journal}
  {\bibinfo  {journal} {\emph {Commun. Math. Phys.}}\ }\textbf {\bibinfo
  {volume} {25}},\ \bibinfo {pages} {167} (\bibinfo {year}
  {1972}{\natexlab{b}})}\BibitemShut {NoStop}%
\bibitem [{\citenamefont {Bekenstein}(1995)}]{Bekenstein:1995un}%
  \BibitemOpen
  \bibfield  {author} {\bibinfo {author} {\bibfnamefont {J.~D.}\ \bibnamefont
  {Bekenstein}},\ }\href {\doibase 10.1103/PhysRevD.51.R6608} {\bibfield
  {journal} {\bibinfo  {journal} {\emph {Phys. Rev. D}}\ }\textbf {\bibinfo
  {volume} {51}},\ \bibinfo {pages} {R6608} (\bibinfo {year}
  {1995})}\BibitemShut {NoStop}%
\bibitem [{\citenamefont {Sotiriou}\ and\ \citenamefont
  {Faraoni}(2012)}]{Sotiriou:2011dz}%
  \BibitemOpen
  \bibfield  {author} {\bibinfo {author} {\bibfnamefont {T.~P.}\ \bibnamefont
  {Sotiriou}} and \bibinfo {author} {\bibfnamefont {V.}~\bibnamefont
  {Faraoni}},\ }\href {\doibase 10.1103/PhysRevLett.108.081103} {\bibfield
  {journal} {\bibinfo  {journal} {\emph {Phys. Rev. Lett.}}\ }\textbf {\bibinfo
  {volume} {108}},\ \bibinfo {pages} {081103} (\bibinfo {year} {2012})},\
  \Eprint {http://arxiv.org/abs/1109.6324} {arXiv:1109.6324 [gr-qc]}
  \BibitemShut {NoStop}%
\bibitem [{\citenamefont {Faraoni}(2017)}]{Faraoni:2017ock}%
  \BibitemOpen
  \bibfield  {author} {\bibinfo {author} {\bibfnamefont {V.}~\bibnamefont
  {Faraoni}},\ }\href {\doibase 10.1103/PhysRevD.95.124013} {\bibfield
  {journal} {\bibinfo  {journal} {\emph {Phys. Rev. D}}\ }\textbf {\bibinfo
  {volume} {95}},\ \bibinfo {pages} {124013} (\bibinfo {year} {2017})},\
  \Eprint {http://arxiv.org/abs/1705.07134} {arXiv:1705.07134 [gr-qc]}
  \BibitemShut {NoStop}%
\bibitem [{\citenamefont {Zwiebach}(1985)}]{Zwiebach:1985uq}%
  \BibitemOpen
  \bibfield  {author} {\bibinfo {author} {\bibfnamefont {B.}~\bibnamefont
  {Zwiebach}},\ }\href {\doibase 10.1016/0370-2693(85)91616-8} {\bibfield
  {journal} {\bibinfo  {journal} {\emph {Phys. Lett. B}}\ }\textbf {\bibinfo
  {volume} {156}},\ \bibinfo {pages} {315} (\bibinfo {year}
  {1985})}\BibitemShut {NoStop}%
\bibitem [{\citenamefont {Antoniadis}\ \emph {et~al.}(1994)\citenamefont
  {Antoniadis}, \citenamefont {Rizos},\ and\ \citenamefont
  {Tamvakis}}]{Antoniadis:1993jc}%
  \BibitemOpen
  \bibfield  {author} {\bibinfo {author} {\bibfnamefont {I.}~\bibnamefont
  {Antoniadis}}, \bibinfo {author} {\bibfnamefont {J.}~\bibnamefont {Rizos}},
  and \bibinfo {author} {\bibfnamefont {K.}~\bibnamefont {Tamvakis}},\ }\href
  {\doibase 10.1016/0550-3213(94)90120-1} {\bibfield  {journal} {\bibinfo
  {journal} {\emph {Nucl. Phys. B}}\ }\textbf {\bibinfo {volume} {415}},\
  \bibinfo {pages} {497} (\bibinfo {year} {1994})},\ \Eprint
  {http://arxiv.org/abs/hep-th/9305025} {arXiv:hep-th/9305025} \BibitemShut
  {NoStop}%
\bibitem [{\citenamefont {Gasperini}\ \emph {et~al.}(1997)\citenamefont
  {Gasperini}, \citenamefont {Maggiore},\ and\ \citenamefont
  {Veneziano}}]{Gasperini:1996fu}%
  \BibitemOpen
  \bibfield  {author} {\bibinfo {author} {\bibfnamefont {M.}~\bibnamefont
  {Gasperini}}, \bibinfo {author} {\bibfnamefont {M.}~\bibnamefont {Maggiore}},
   and \bibinfo {author} {\bibfnamefont {G.}~\bibnamefont {Veneziano}},\ }\href
  {\doibase 10.1016/S0550-3213(97)00149-1} {\bibfield  {journal} {\bibinfo
  {journal} {\emph {Nucl. Phys. B}}\ }\textbf {\bibinfo {volume} {494}},\
  \bibinfo {pages} {315} (\bibinfo {year} {1997})},\ \Eprint
  {http://arxiv.org/abs/hep-th/9611039} {arXiv:hep-th/9611039} \BibitemShut
  {NoStop}%
\bibitem [{\citenamefont {Kanti}\ \emph {et~al.}(1996)\citenamefont {Kanti},
  \citenamefont {Mavromatos}, \citenamefont {Rizos}, \citenamefont {Tamvakis},\
  and\ \citenamefont {Winstanley}}]{Kanti:1995vq}%
  \BibitemOpen
  \bibfield  {author} {\bibinfo {author} {\bibfnamefont {P.}~\bibnamefont
  {Kanti}}, \bibinfo {author} {\bibfnamefont {N.~E.}\ \bibnamefont
  {Mavromatos}}, \bibinfo {author} {\bibfnamefont {J.}~\bibnamefont {Rizos}},
  \bibinfo {author} {\bibfnamefont {K.}~\bibnamefont {Tamvakis}},  and \bibinfo
  {author} {\bibfnamefont {E.}~\bibnamefont {Winstanley}},\ }\href {\doibase
  10.1103/PhysRevD.54.5049} {\bibfield  {journal} {\bibinfo  {journal} {\emph
  {Phys. Rev. D}}\ }\textbf {\bibinfo {volume} {54}},\ \bibinfo {pages} {5049}
  (\bibinfo {year} {1996})},\ \Eprint {http://arxiv.org/abs/hep-th/9511071}
  {arXiv:hep-th/9511071} \BibitemShut {NoStop}%
\bibitem [{\citenamefont {Torii}\ \emph {et~al.}(1997)\citenamefont {Torii},
  \citenamefont {Yajima},\ and\ \citenamefont {Maeda}}]{Torii:1996yi}%
  \BibitemOpen
  \bibfield  {author} {\bibinfo {author} {\bibfnamefont {T.}~\bibnamefont
  {Torii}}, \bibinfo {author} {\bibfnamefont {H.}~\bibnamefont {Yajima}},  and
  \bibinfo {author} {\bibfnamefont {K.-i.}\ \bibnamefont {Maeda}},\ }\href
  {\doibase 10.1103/PhysRevD.55.739} {\bibfield  {journal} {\bibinfo  {journal}
  {\emph {Phys. Rev. D}}\ }\textbf {\bibinfo {volume} {55}},\ \bibinfo {pages}
  {739} (\bibinfo {year} {1997})},\ \Eprint
  {http://arxiv.org/abs/gr-qc/9606034} {arXiv:gr-qc/9606034} \BibitemShut
  {NoStop}%
\bibitem [{\citenamefont {Kanti}\ \emph {et~al.}(1998)\citenamefont {Kanti},
  \citenamefont {Mavromatos}, \citenamefont {Rizos}, \citenamefont {Tamvakis},\
  and\ \citenamefont {Winstanley}}]{Kanti:1997br}%
  \BibitemOpen
  \bibfield  {author} {\bibinfo {author} {\bibfnamefont {P.}~\bibnamefont
  {Kanti}}, \bibinfo {author} {\bibfnamefont {N.~E.}\ \bibnamefont
  {Mavromatos}}, \bibinfo {author} {\bibfnamefont {J.}~\bibnamefont {Rizos}},
  \bibinfo {author} {\bibfnamefont {K.}~\bibnamefont {Tamvakis}},  and \bibinfo
  {author} {\bibfnamefont {E.}~\bibnamefont {Winstanley}},\ }\href {\doibase
  10.1103/PhysRevD.57.6255} {\bibfield  {journal} {\bibinfo  {journal} {\emph
  {Phys. Rev. D}}\ }\textbf {\bibinfo {volume} {57}},\ \bibinfo {pages} {6255}
  (\bibinfo {year} {1998})},\ \Eprint {http://arxiv.org/abs/hep-th/9703192}
  {arXiv:hep-th/9703192} \BibitemShut {NoStop}%
\bibitem [{\citenamefont {Chen}\ \emph {et~al.}(2007)\citenamefont {Chen},
  \citenamefont {Gal'tsov},\ and\ \citenamefont {Orlov}}]{Chen:2006ge}%
  \BibitemOpen
  \bibfield  {author} {\bibinfo {author} {\bibfnamefont {C.-M.}\ \bibnamefont
  {Chen}}, \bibinfo {author} {\bibfnamefont {D.~V.}\ \bibnamefont {Gal'tsov}},
  and \bibinfo {author} {\bibfnamefont {D.~G.}\ \bibnamefont {Orlov}},\ }\href
  {\doibase 10.1103/PhysRevD.75.084030} {\bibfield  {journal} {\bibinfo
  {journal} {\emph {Phys. Rev. D}}\ }\textbf {\bibinfo {volume} {75}},\
  \bibinfo {pages} {084030} (\bibinfo {year} {2007})},\ \Eprint
  {http://arxiv.org/abs/hep-th/0701004} {arXiv:hep-th/0701004} \BibitemShut
  {NoStop}%
\bibitem [{\citenamefont {Guo}\ \emph {et~al.}(2008)\citenamefont {Guo},
  \citenamefont {Ohta},\ and\ \citenamefont {Torii}}]{Guo:2008hf}%
  \BibitemOpen
  \bibfield  {author} {\bibinfo {author} {\bibfnamefont {Z.-K.}\ \bibnamefont
  {Guo}}, \bibinfo {author} {\bibfnamefont {N.}~\bibnamefont {Ohta}},  and
  \bibinfo {author} {\bibfnamefont {T.}~\bibnamefont {Torii}},\ }\href
  {\doibase 10.1143/PTP.120.581} {\bibfield  {journal} {\bibinfo  {journal}
  {\emph {Prog. Theor. Phys.}}\ }\textbf {\bibinfo {volume} {120}},\ \bibinfo
  {pages} {581} (\bibinfo {year} {2008})},\ \Eprint
  {http://arxiv.org/abs/0806.2481} {arXiv:0806.2481 [gr-qc]} \BibitemShut
  {NoStop}%
\bibitem [{\citenamefont {Guo}\ \emph {et~al.}(2009)\citenamefont {Guo},
  \citenamefont {Ohta},\ and\ \citenamefont {Torii}}]{Guo:2008eq}%
  \BibitemOpen
  \bibfield  {author} {\bibinfo {author} {\bibfnamefont {Z.-K.}\ \bibnamefont
  {Guo}}, \bibinfo {author} {\bibfnamefont {N.}~\bibnamefont {Ohta}},  and
  \bibinfo {author} {\bibfnamefont {T.}~\bibnamefont {Torii}},\ }\href
  {\doibase 10.1143/PTP.121.253} {\bibfield  {journal} {\bibinfo  {journal}
  {\emph {Prog. Theor. Phys.}}\ }\textbf {\bibinfo {volume} {121}},\ \bibinfo
  {pages} {253} (\bibinfo {year} {2009})},\ \Eprint
  {http://arxiv.org/abs/0811.3068} {arXiv:0811.3068 [gr-qc]} \BibitemShut
  {NoStop}%
\bibitem [{\citenamefont {Pani}\ and\ \citenamefont
  {Cardoso}(2009)}]{Pani:2009wy}%
  \BibitemOpen
  \bibfield  {author} {\bibinfo {author} {\bibfnamefont {P.}~\bibnamefont
  {Pani}} and \bibinfo {author} {\bibfnamefont {V.}~\bibnamefont {Cardoso}},\
  }\href {\doibase 10.1103/PhysRevD.79.084031} {\bibfield  {journal} {\bibinfo
  {journal} {\emph {Phys. Rev. D}}\ }\textbf {\bibinfo {volume} {79}},\
  \bibinfo {pages} {084031} (\bibinfo {year} {2009})},\ \Eprint
  {http://arxiv.org/abs/0902.1569} {arXiv:0902.1569 [gr-qc]} \BibitemShut
  {NoStop}%
\bibitem [{\citenamefont {Ayzenberg}\ and\ \citenamefont
  {Yunes}(2014)}]{Ayzenberg:2014aka}%
  \BibitemOpen
  \bibfield  {author} {\bibinfo {author} {\bibfnamefont {D.}~\bibnamefont
  {Ayzenberg}} and \bibinfo {author} {\bibfnamefont {N.}~\bibnamefont
  {Yunes}},\ }\href {\doibase 10.1103/PhysRevD.90.044066} {\bibfield  {journal}
  {\bibinfo  {journal} {\emph {Phys. Rev. D}}\ }\textbf {\bibinfo {volume}
  {90}},\ \bibinfo {pages} {044066} (\bibinfo {year} {2014})},\ \bibinfo {note}
  {[Erratum: Phys.Rev.D 91, 069905 (2015)]},\ \Eprint
  {http://arxiv.org/abs/1405.2133} {arXiv:1405.2133 [gr-qc]} \BibitemShut
  {NoStop}%
\bibitem [{\citenamefont {Maselli}\ \emph {et~al.}(2015)\citenamefont
  {Maselli}, \citenamefont {Pani}, \citenamefont {Gualtieri},\ and\
  \citenamefont {Ferrari}}]{Maselli:2015tta}%
  \BibitemOpen
  \bibfield  {author} {\bibinfo {author} {\bibfnamefont {A.}~\bibnamefont
  {Maselli}}, \bibinfo {author} {\bibfnamefont {P.}~\bibnamefont {Pani}},
  \bibinfo {author} {\bibfnamefont {L.}~\bibnamefont {Gualtieri}},  and
  \bibinfo {author} {\bibfnamefont {V.}~\bibnamefont {Ferrari}},\ }\href
  {\doibase 10.1103/PhysRevD.92.083014} {\bibfield  {journal} {\bibinfo
  {journal} {\emph {Phys. Rev. D}}\ }\textbf {\bibinfo {volume} {92}},\
  \bibinfo {pages} {083014} (\bibinfo {year} {2015})},\ \Eprint
  {http://arxiv.org/abs/1507.00680} {arXiv:1507.00680 [gr-qc]} \BibitemShut
  {NoStop}%
\bibitem [{\citenamefont {Kleihaus}\ \emph {et~al.}(2011)\citenamefont
  {Kleihaus}, \citenamefont {Kunz},\ and\ \citenamefont
  {Radu}}]{Kleihaus:2011tg}%
  \BibitemOpen
  \bibfield  {author} {\bibinfo {author} {\bibfnamefont {B.}~\bibnamefont
  {Kleihaus}}, \bibinfo {author} {\bibfnamefont {J.}~\bibnamefont {Kunz}},  and
  \bibinfo {author} {\bibfnamefont {E.}~\bibnamefont {Radu}},\ }\href {\doibase
  10.1103/PhysRevLett.106.151104} {\bibfield  {journal} {\bibinfo  {journal}
  {\emph {Phys. Rev. Lett.}}\ }\textbf {\bibinfo {volume} {106}},\ \bibinfo
  {pages} {151104} (\bibinfo {year} {2011})},\ \Eprint
  {http://arxiv.org/abs/1101.2868} {arXiv:1101.2868 [gr-qc]} \BibitemShut
  {NoStop}%
\bibitem [{\citenamefont {Kleihaus}\ \emph {et~al.}(2016)\citenamefont
  {Kleihaus}, \citenamefont {Kunz}, \citenamefont {Mojica},\ and\ \citenamefont
  {Radu}}]{Kleihaus:2015aje}%
  \BibitemOpen
  \bibfield  {author} {\bibinfo {author} {\bibfnamefont {B.}~\bibnamefont
  {Kleihaus}}, \bibinfo {author} {\bibfnamefont {J.}~\bibnamefont {Kunz}},
  \bibinfo {author} {\bibfnamefont {S.}~\bibnamefont {Mojica}},  and \bibinfo
  {author} {\bibfnamefont {E.}~\bibnamefont {Radu}},\ }\href {\doibase
  10.1103/PhysRevD.93.044047} {\bibfield  {journal} {\bibinfo  {journal} {\emph
  {Phys. Rev. D}}\ }\textbf {\bibinfo {volume} {93}},\ \bibinfo {pages}
  {044047} (\bibinfo {year} {2016})},\ \Eprint
  {http://arxiv.org/abs/1511.05513} {arXiv:1511.05513 [gr-qc]} \BibitemShut
  {NoStop}%
\bibitem [{\citenamefont {Sotiriou}\ and\ \citenamefont
  {Zhou}(2014{\natexlab{a}})}]{Sotiriou:2013qea}%
  \BibitemOpen
  \bibfield  {author} {\bibinfo {author} {\bibfnamefont {T.~P.}\ \bibnamefont
  {Sotiriou}} and \bibinfo {author} {\bibfnamefont {S.-Y.}\ \bibnamefont
  {Zhou}},\ }\href {\doibase 10.1103/PhysRevLett.112.251102} {\bibfield
  {journal} {\bibinfo  {journal} {\emph {Phys. Rev. Lett.}}\ }\textbf {\bibinfo
  {volume} {112}},\ \bibinfo {pages} {251102} (\bibinfo {year}
  {2014}{\natexlab{a}})},\ \Eprint {http://arxiv.org/abs/1312.3622}
  {arXiv:1312.3622 [gr-qc]} \BibitemShut {NoStop}%
\bibitem [{\citenamefont {Sotiriou}\ and\ \citenamefont
  {Zhou}(2014{\natexlab{b}})}]{Sotiriou:2014pfa}%
  \BibitemOpen
  \bibfield  {author} {\bibinfo {author} {\bibfnamefont {T.~P.}\ \bibnamefont
  {Sotiriou}} and \bibinfo {author} {\bibfnamefont {S.-Y.}\ \bibnamefont
  {Zhou}},\ }\href {\doibase 10.1103/PhysRevD.90.124063} {\bibfield  {journal}
  {\bibinfo  {journal} {\emph {Phys. Rev. D}}\ }\textbf {\bibinfo {volume}
  {90}},\ \bibinfo {pages} {124063} (\bibinfo {year} {2014}{\natexlab{b}})},\
  \Eprint {http://arxiv.org/abs/1408.1698} {arXiv:1408.1698 [gr-qc]}
  \BibitemShut {NoStop}%
\bibitem [{\citenamefont {Doneva}\ and\ \citenamefont
  {Yazadjiev}(2018)}]{Doneva:2017bvd}%
  \BibitemOpen
  \bibfield  {author} {\bibinfo {author} {\bibfnamefont {D.~D.}\ \bibnamefont
  {Doneva}} and \bibinfo {author} {\bibfnamefont {S.~S.}\ \bibnamefont
  {Yazadjiev}},\ }\href {\doibase 10.1103/PhysRevLett.120.131103} {\bibfield
  {journal} {\bibinfo  {journal} {\emph {Phys. Rev. Lett.}}\ }\textbf {\bibinfo
  {volume} {120}},\ \bibinfo {pages} {131103} (\bibinfo {year} {2018})},\
  \Eprint {http://arxiv.org/abs/1711.01187} {arXiv:1711.01187 [gr-qc]}
  \BibitemShut {NoStop}%
\bibitem [{\citenamefont {Silva}\ \emph {et~al.}(2018)\citenamefont {Silva},
  \citenamefont {Sakstein}, \citenamefont {Gualtieri}, \citenamefont
  {Sotiriou},\ and\ \citenamefont {Berti}}]{Silva:2017uqg}%
  \BibitemOpen
  \bibfield  {author} {\bibinfo {author} {\bibfnamefont {H.~O.}\ \bibnamefont
  {Silva}}, \bibinfo {author} {\bibfnamefont {J.}~\bibnamefont {Sakstein}},
  \bibinfo {author} {\bibfnamefont {L.}~\bibnamefont {Gualtieri}}, \bibinfo
  {author} {\bibfnamefont {T.~P.}\ \bibnamefont {Sotiriou}},  and \bibinfo
  {author} {\bibfnamefont {E.}~\bibnamefont {Berti}},\ }\href {\doibase
  10.1103/PhysRevLett.120.131104} {\bibfield  {journal} {\bibinfo  {journal}
  {\emph {Phys. Rev. Lett.}}\ }\textbf {\bibinfo {volume} {120}},\ \bibinfo
  {pages} {131104} (\bibinfo {year} {2018})},\ \Eprint
  {http://arxiv.org/abs/1711.02080} {arXiv:1711.02080 [gr-qc]} \BibitemShut
  {NoStop}%
\bibitem [{\citenamefont {Antoniou}\ \emph {et~al.}(2018)\citenamefont
  {Antoniou}, \citenamefont {Bakopoulos},\ and\ \citenamefont
  {Kanti}}]{Antoniou:2017acq}%
  \BibitemOpen
  \bibfield  {author} {\bibinfo {author} {\bibfnamefont {G.}~\bibnamefont
  {Antoniou}}, \bibinfo {author} {\bibfnamefont {A.}~\bibnamefont
  {Bakopoulos}},  and \bibinfo {author} {\bibfnamefont {P.}~\bibnamefont
  {Kanti}},\ }\href {\doibase 10.1103/PhysRevLett.120.131102} {\bibfield
  {journal} {\bibinfo  {journal} {\emph {Phys. Rev. Lett.}}\ }\textbf {\bibinfo
  {volume} {120}},\ \bibinfo {pages} {131102} (\bibinfo {year} {2018})},\
  \Eprint {http://arxiv.org/abs/1711.03390} {arXiv:1711.03390 [hep-th]}
  \BibitemShut {NoStop}%
\bibitem [{\citenamefont {Bl\'azquez-Salcedo}\ \emph
  {et~al.}(2018)\citenamefont {Bl\'azquez-Salcedo}, \citenamefont {Doneva},
  \citenamefont {Kunz},\ and\ \citenamefont
  {Yazadjiev}}]{Blazquez-Salcedo:2018jnn}%
  \BibitemOpen
  \bibfield  {author} {\bibinfo {author} {\bibfnamefont {J.~L.}\ \bibnamefont
  {Bl\'azquez-Salcedo}}, \bibinfo {author} {\bibfnamefont {D.~D.}\ \bibnamefont
  {Doneva}}, \bibinfo {author} {\bibfnamefont {J.}~\bibnamefont {Kunz}},  and
  \bibinfo {author} {\bibfnamefont {S.~S.}\ \bibnamefont {Yazadjiev}},\ }\href
  {\doibase 10.1103/PhysRevD.98.084011} {\bibfield  {journal} {\bibinfo
  {journal} {\emph {Phys. Rev. D}}\ }\textbf {\bibinfo {volume} {98}},\
  \bibinfo {pages} {084011} (\bibinfo {year} {2018})},\ \Eprint
  {http://arxiv.org/abs/1805.05755} {arXiv:1805.05755 [gr-qc]} \BibitemShut
  {NoStop}%
\bibitem [{\citenamefont {Minamitsuji}\ and\ \citenamefont
  {Ikeda}(2019{\natexlab{a}})}]{Minamitsuji:2018xde}%
  \BibitemOpen
  \bibfield  {author} {\bibinfo {author} {\bibfnamefont {M.}~\bibnamefont
  {Minamitsuji}} and \bibinfo {author} {\bibfnamefont {T.}~\bibnamefont
  {Ikeda}},\ }\href {\doibase 10.1103/PhysRevD.99.044017} {\bibfield  {journal}
  {\bibinfo  {journal} {\emph {Phys. Rev. D}}\ }\textbf {\bibinfo {volume}
  {99}},\ \bibinfo {pages} {044017} (\bibinfo {year} {2019}{\natexlab{a}})},\
  \Eprint {http://arxiv.org/abs/1812.03551} {arXiv:1812.03551 [gr-qc]}
  \BibitemShut {NoStop}%
\bibitem [{\citenamefont {Silva}\ \emph {et~al.}(2019)\citenamefont {Silva},
  \citenamefont {Macedo}, \citenamefont {Sotiriou}, \citenamefont {Gualtieri},
  \citenamefont {Sakstein},\ and\ \citenamefont {Berti}}]{Silva:2018qhn}%
  \BibitemOpen
  \bibfield  {author} {\bibinfo {author} {\bibfnamefont {H.~O.}\ \bibnamefont
  {Silva}}, \bibinfo {author} {\bibfnamefont {C.~F.~B.}\ \bibnamefont
  {Macedo}}, \bibinfo {author} {\bibfnamefont {T.~P.}\ \bibnamefont
  {Sotiriou}}, \bibinfo {author} {\bibfnamefont {L.}~\bibnamefont {Gualtieri}},
  \bibinfo {author} {\bibfnamefont {J.}~\bibnamefont {Sakstein}},  and \bibinfo
  {author} {\bibfnamefont {E.}~\bibnamefont {Berti}},\ }\href {\doibase
  10.1103/PhysRevD.99.064011} {\bibfield  {journal} {\bibinfo  {journal} {\emph
  {Phys. Rev. D}}\ }\textbf {\bibinfo {volume} {99}},\ \bibinfo {pages}
  {064011} (\bibinfo {year} {2019})},\ \Eprint
  {http://arxiv.org/abs/1812.05590} {arXiv:1812.05590 [gr-qc]} \BibitemShut
  {NoStop}%
\bibitem [{\citenamefont {Macedo}\ \emph {et~al.}(2019)\citenamefont {Macedo},
  \citenamefont {Sakstein}, \citenamefont {Berti}, \citenamefont {Gualtieri},
  \citenamefont {Silva},\ and\ \citenamefont {Sotiriou}}]{Macedo:2019sem}%
  \BibitemOpen
  \bibfield  {author} {\bibinfo {author} {\bibfnamefont {C.~F.~B.}\
  \bibnamefont {Macedo}}, \bibinfo {author} {\bibfnamefont {J.}~\bibnamefont
  {Sakstein}}, \bibinfo {author} {\bibfnamefont {E.}~\bibnamefont {Berti}},
  \bibinfo {author} {\bibfnamefont {L.}~\bibnamefont {Gualtieri}}, \bibinfo
  {author} {\bibfnamefont {H.~O.}\ \bibnamefont {Silva}},  and \bibinfo
  {author} {\bibfnamefont {T.~P.}\ \bibnamefont {Sotiriou}},\ }\href {\doibase
  10.1103/PhysRevD.99.104041} {\bibfield  {journal} {\bibinfo  {journal} {\emph
  {Phys. Rev. D}}\ }\textbf {\bibinfo {volume} {99}},\ \bibinfo {pages}
  {104041} (\bibinfo {year} {2019})},\ \Eprint
  {http://arxiv.org/abs/1903.06784} {arXiv:1903.06784 [gr-qc]} \BibitemShut
  {NoStop}%
\bibitem [{\citenamefont {Doneva}\ and\ \citenamefont
  {Yazadjiev}(2022)}]{Doneva:2021tvn}%
  \BibitemOpen
  \bibfield  {author} {\bibinfo {author} {\bibfnamefont {D.~D.}\ \bibnamefont
  {Doneva}} and \bibinfo {author} {\bibfnamefont {S.~S.}\ \bibnamefont
  {Yazadjiev}},\ }\href {\doibase 10.1103/PhysRevD.105.L041502} {\bibfield
  {journal} {\bibinfo  {journal} {\emph {Phys. Rev. D}}\ }\textbf {\bibinfo
  {volume} {105}},\ \bibinfo {pages} {L041502} (\bibinfo {year} {2022})},\
  \Eprint {http://arxiv.org/abs/2107.01738} {arXiv:2107.01738 [gr-qc]}
  \BibitemShut {NoStop}%
\bibitem [{\citenamefont {Rinaldi}(2012)}]{Rinaldi:2012vy}%
  \BibitemOpen
  \bibfield  {author} {\bibinfo {author} {\bibfnamefont {M.}~\bibnamefont
  {Rinaldi}},\ }\href {\doibase 10.1103/PhysRevD.86.084048} {\bibfield
  {journal} {\bibinfo  {journal} {\emph {Phys. Rev. D}}\ }\textbf {\bibinfo
  {volume} {86}},\ \bibinfo {pages} {084048} (\bibinfo {year} {2012})},\
  \Eprint {http://arxiv.org/abs/1208.0103} {arXiv:1208.0103 [gr-qc]}
  \BibitemShut {NoStop}%
\bibitem [{\citenamefont
  {Minamitsuji}(2014{\natexlab{a}})}]{Minamitsuji:2013ura}%
  \BibitemOpen
  \bibfield  {author} {\bibinfo {author} {\bibfnamefont {M.}~\bibnamefont
  {Minamitsuji}},\ }\href {\doibase 10.1103/PhysRevD.89.064017} {\bibfield
  {journal} {\bibinfo  {journal} {\emph {Phys. Rev. D}}\ }\textbf {\bibinfo
  {volume} {89}},\ \bibinfo {pages} {064017} (\bibinfo {year}
  {2014}{\natexlab{a}})},\ \Eprint {http://arxiv.org/abs/1312.3759}
  {arXiv:1312.3759 [gr-qc]} \BibitemShut {NoStop}%
\bibitem [{\citenamefont {Cisterna}\ and\ \citenamefont
  {Erices}(2014)}]{Cisterna:2014nua}%
  \BibitemOpen
  \bibfield  {author} {\bibinfo {author} {\bibfnamefont {A.}~\bibnamefont
  {Cisterna}} and \bibinfo {author} {\bibfnamefont {C.}~\bibnamefont
  {Erices}},\ }\href {\doibase 10.1103/PhysRevD.89.084038} {\bibfield
  {journal} {\bibinfo  {journal} {\emph {Phys. Rev. D}}\ }\textbf {\bibinfo
  {volume} {89}},\ \bibinfo {pages} {084038} (\bibinfo {year} {2014})},\
  \Eprint {http://arxiv.org/abs/1401.4479} {arXiv:1401.4479 [gr-qc]}
  \BibitemShut {NoStop}%
\bibitem [{\citenamefont {Kolyvaris}\ \emph {et~al.}(2012)\citenamefont
  {Kolyvaris}, \citenamefont {Koutsoumbas}, \citenamefont {Papantonopoulos},\
  and\ \citenamefont {Siopsis}}]{Kolyvaris:2011fk}%
  \BibitemOpen
  \bibfield  {author} {\bibinfo {author} {\bibfnamefont {T.}~\bibnamefont
  {Kolyvaris}}, \bibinfo {author} {\bibfnamefont {G.}~\bibnamefont
  {Koutsoumbas}}, \bibinfo {author} {\bibfnamefont {E.}~\bibnamefont
  {Papantonopoulos}},  and \bibinfo {author} {\bibfnamefont {G.}~\bibnamefont
  {Siopsis}},\ }\href {\doibase 10.1088/0264-9381/29/20/205011} {\bibfield
  {journal} {\bibinfo  {journal} {\emph {Class. Quant. Grav.}}\ }\textbf
  {\bibinfo {volume} {29}},\ \bibinfo {pages} {205011} (\bibinfo {year}
  {2012})},\ \Eprint {http://arxiv.org/abs/1111.0263} {arXiv:1111.0263 [gr-qc]}
  \BibitemShut {NoStop}%
\bibitem [{\citenamefont
  {Minamitsuji}(2014{\natexlab{b}})}]{Minamitsuji:2014hha}%
  \BibitemOpen
  \bibfield  {author} {\bibinfo {author} {\bibfnamefont {M.}~\bibnamefont
  {Minamitsuji}},\ }\href {\doibase 10.1007/s10714-014-1785-0} {\bibfield
  {journal} {\bibinfo  {journal} {\emph {Gen. Rel. Grav.}}\ }\textbf {\bibinfo
  {volume} {46}},\ \bibinfo {pages} {1785} (\bibinfo {year}
  {2014}{\natexlab{b}})},\ \Eprint {http://arxiv.org/abs/1407.4901}
  {arXiv:1407.4901 [gr-qc]} \BibitemShut {NoStop}%
\bibitem [{\citenamefont {Minamitsuji}\ \emph {et~al.}(2022)\citenamefont
  {Minamitsuji}, \citenamefont {Takahashi},\ and\ \citenamefont
  {Tsujikawa}}]{Minamitsuji:2022mlv}%
  \BibitemOpen
  \bibfield  {author} {\bibinfo {author} {\bibfnamefont {M.}~\bibnamefont
  {Minamitsuji}}, \bibinfo {author} {\bibfnamefont {K.}~\bibnamefont
  {Takahashi}},  and \bibinfo {author} {\bibfnamefont {S.}~\bibnamefont
  {Tsujikawa}},\ }\href {\doibase 10.1103/PhysRevD.105.104001} {\bibfield
  {journal} {\bibinfo  {journal} {\emph {Phys. Rev. D}}\ }\textbf {\bibinfo
  {volume} {105}},\ \bibinfo {pages} {104001} (\bibinfo {year} {2022})},\
  \Eprint {http://arxiv.org/abs/2201.09687} {arXiv:2201.09687 [gr-qc]}
  \BibitemShut {NoStop}%
\bibitem [{\citenamefont {Kobayashi}\ \emph {et~al.}(2012)\citenamefont
  {Kobayashi}, \citenamefont {Motohashi},\ and\ \citenamefont
  {Suyama}}]{Kobayashi:2012kh}%
  \BibitemOpen
  \bibfield  {author} {\bibinfo {author} {\bibfnamefont {T.}~\bibnamefont
  {Kobayashi}}, \bibinfo {author} {\bibfnamefont {H.}~\bibnamefont
  {Motohashi}},  and \bibinfo {author} {\bibfnamefont {T.}~\bibnamefont
  {Suyama}},\ }\href {\doibase 10.1103/PhysRevD.85.084025} {\bibfield
  {journal} {\bibinfo  {journal} {\emph {Phys. Rev. D}}\ }\textbf {\bibinfo
  {volume} {85}},\ \bibinfo {pages} {084025} (\bibinfo {year} {2012})},\
  \bibinfo {note} {[Erratum:
  \href{https://doi.org/10.1103/PhysRevD.96.109903}{{\it Phys. Rev. D} {\bf
  96}, 109903 (2017)}]},\ \Eprint {http://arxiv.org/abs/1202.4893}
  {arXiv:1202.4893 [gr-qc]} \BibitemShut {NoStop}%
\bibitem [{\citenamefont {Kobayashi}\ \emph {et~al.}(2014)\citenamefont
  {Kobayashi}, \citenamefont {Motohashi},\ and\ \citenamefont
  {Suyama}}]{Kobayashi:2014wsa}%
  \BibitemOpen
  \bibfield  {author} {\bibinfo {author} {\bibfnamefont {T.}~\bibnamefont
  {Kobayashi}}, \bibinfo {author} {\bibfnamefont {H.}~\bibnamefont
  {Motohashi}},  and \bibinfo {author} {\bibfnamefont {T.}~\bibnamefont
  {Suyama}},\ }\href {\doibase 10.1103/PhysRevD.89.084042} {\bibfield
  {journal} {\bibinfo  {journal} {\emph {Phys. Rev. D}}\ }\textbf {\bibinfo
  {volume} {89}},\ \bibinfo {pages} {084042} (\bibinfo {year} {2014})},\
  \Eprint {http://arxiv.org/abs/1402.6740} {arXiv:1402.6740 [gr-qc]}
  \BibitemShut {NoStop}%
\bibitem [{\citenamefont {Kase}\ and\ \citenamefont
  {Tsujikawa}(2022)}]{Kase:2021mix}%
  \BibitemOpen
  \bibfield  {author} {\bibinfo {author} {\bibfnamefont {R.}~\bibnamefont
  {Kase}} and \bibinfo {author} {\bibfnamefont {S.}~\bibnamefont {Tsujikawa}},\
  }\href {\doibase 10.1103/PhysRevD.105.024059} {\bibfield  {journal} {\bibinfo
   {journal} {\emph {Phys. Rev. D}}\ }\textbf {\bibinfo {volume} {105}},\
  \bibinfo {pages} {024059} (\bibinfo {year} {2022})},\ \Eprint
  {http://arxiv.org/abs/2110.12728} {arXiv:2110.12728 [gr-qc]} \BibitemShut
  {NoStop}%
\bibitem [{\citenamefont {Hui}\ and\ \citenamefont
  {Nicolis}(2013)}]{Hui:2012qt}%
  \BibitemOpen
  \bibfield  {author} {\bibinfo {author} {\bibfnamefont {L.}~\bibnamefont
  {Hui}} and \bibinfo {author} {\bibfnamefont {A.}~\bibnamefont {Nicolis}},\
  }\href {\doibase 10.1103/PhysRevLett.110.241104} {\bibfield  {journal}
  {\bibinfo  {journal} {\emph {Phys. Rev. Lett.}}\ }\textbf {\bibinfo {volume}
  {110}},\ \bibinfo {pages} {241104} (\bibinfo {year} {2013})},\ \Eprint
  {http://arxiv.org/abs/1202.1296} {arXiv:1202.1296 [hep-th]} \BibitemShut
  {NoStop}%
\bibitem [{\citenamefont {Babichev}\ and\ \citenamefont
  {Charmousis}(2014)}]{Babichev:2013cya}%
  \BibitemOpen
  \bibfield  {author} {\bibinfo {author} {\bibfnamefont {E.}~\bibnamefont
  {Babichev}} and \bibinfo {author} {\bibfnamefont {C.}~\bibnamefont
  {Charmousis}},\ }\href {\doibase 10.1007/JHEP08(2014)106} {\bibfield
  {journal} {\bibinfo  {journal} {\emph {JHEP}}\ }\textbf {\bibinfo {volume}
  {08}},\ \bibinfo {pages} {106} (\bibinfo {year} {2014})},\ \Eprint
  {http://arxiv.org/abs/1312.3204} {arXiv:1312.3204 [gr-qc]} \BibitemShut
  {NoStop}%
\bibitem [{\citenamefont {Kobayashi}\ and\ \citenamefont
  {Tanahashi}(2014)}]{Kobayashi:2014eva}%
  \BibitemOpen
  \bibfield  {author} {\bibinfo {author} {\bibfnamefont {T.}~\bibnamefont
  {Kobayashi}} and \bibinfo {author} {\bibfnamefont {N.}~\bibnamefont
  {Tanahashi}},\ }\href {\doibase 10.1093/ptep/ptu096} {\bibfield  {journal}
  {\bibinfo  {journal} {\emph {PTEP}}\ }\textbf {\bibinfo {volume} {2014}},\
  \bibinfo {pages} {073E02} (\bibinfo {year} {2014})},\ \Eprint
  {http://arxiv.org/abs/1403.4364} {arXiv:1403.4364 [gr-qc]} \BibitemShut
  {NoStop}%
\bibitem [{\citenamefont {Ogawa}\ \emph {et~al.}(2016)\citenamefont {Ogawa},
  \citenamefont {Kobayashi},\ and\ \citenamefont {Suyama}}]{Ogawa:2015pea}%
  \BibitemOpen
  \bibfield  {author} {\bibinfo {author} {\bibfnamefont {H.}~\bibnamefont
  {Ogawa}}, \bibinfo {author} {\bibfnamefont {T.}~\bibnamefont {Kobayashi}},
  and \bibinfo {author} {\bibfnamefont {T.}~\bibnamefont {Suyama}},\ }\href
  {\doibase 10.1103/PhysRevD.93.064078} {\bibfield  {journal} {\bibinfo
  {journal} {\emph {Phys. Rev. D}}\ }\textbf {\bibinfo {volume} {93}},\
  \bibinfo {pages} {064078} (\bibinfo {year} {2016})},\ \Eprint
  {http://arxiv.org/abs/1510.07400} {arXiv:1510.07400 [gr-qc]} \BibitemShut
  {NoStop}%
\bibitem [{\citenamefont {Takahashi}\ \emph {et~al.}(2016)\citenamefont
  {Takahashi}, \citenamefont {Suyama},\ and\ \citenamefont
  {Kobayashi}}]{Takahashi:2015pad}%
  \BibitemOpen
  \bibfield  {author} {\bibinfo {author} {\bibfnamefont {K.}~\bibnamefont
  {Takahashi}}, \bibinfo {author} {\bibfnamefont {T.}~\bibnamefont {Suyama}},
  and \bibinfo {author} {\bibfnamefont {T.}~\bibnamefont {Kobayashi}},\ }\href
  {\doibase 10.1103/PhysRevD.93.064068} {\bibfield  {journal} {\bibinfo
  {journal} {\emph {Phys. Rev. D}}\ }\textbf {\bibinfo {volume} {93}},\
  \bibinfo {pages} {064068} (\bibinfo {year} {2016})},\ \Eprint
  {http://arxiv.org/abs/1511.06083} {arXiv:1511.06083 [gr-qc]} \BibitemShut
  {NoStop}%
\bibitem [{\citenamefont {Takahashi}\ and\ \citenamefont
  {Suyama}(2017)}]{Takahashi:2016dnv}%
  \BibitemOpen
  \bibfield  {author} {\bibinfo {author} {\bibfnamefont {K.}~\bibnamefont
  {Takahashi}} and \bibinfo {author} {\bibfnamefont {T.}~\bibnamefont
  {Suyama}},\ }\href {\doibase 10.1103/PhysRevD.95.024034} {\bibfield
  {journal} {\bibinfo  {journal} {\emph {Phys. Rev. D}}\ }\textbf {\bibinfo
  {volume} {95}},\ \bibinfo {pages} {024034} (\bibinfo {year} {2017})},\
  \Eprint {http://arxiv.org/abs/1610.00432} {arXiv:1610.00432 [gr-qc]}
  \BibitemShut {NoStop}%
\bibitem [{\citenamefont {Babichev}\ \emph {et~al.}(2018)\citenamefont
  {Babichev}, \citenamefont {Charmousis}, \citenamefont {Esposito-Far\`ese},\
  and\ \citenamefont {Leh\'ebel}}]{Babichev:2017lmw}%
  \BibitemOpen
  \bibfield  {author} {\bibinfo {author} {\bibfnamefont {E.}~\bibnamefont
  {Babichev}}, \bibinfo {author} {\bibfnamefont {C.}~\bibnamefont
  {Charmousis}}, \bibinfo {author} {\bibfnamefont {G.}~\bibnamefont
  {Esposito-Far\`ese}},  and \bibinfo {author} {\bibfnamefont {A.}~\bibnamefont
  {Leh\'ebel}},\ }\href {\doibase 10.1103/PhysRevLett.120.241101} {\bibfield
  {journal} {\bibinfo  {journal} {\emph {Phys. Rev. Lett.}}\ }\textbf {\bibinfo
  {volume} {120}},\ \bibinfo {pages} {241101} (\bibinfo {year} {2018})},\
  \Eprint {http://arxiv.org/abs/1712.04398} {arXiv:1712.04398 [gr-qc]}
  \BibitemShut {NoStop}%
\bibitem [{\citenamefont {Tretyakova}\ and\ \citenamefont
  {Takahashi}(2017)}]{Tretyakova:2017lyg}%
  \BibitemOpen
  \bibfield  {author} {\bibinfo {author} {\bibfnamefont {D.~A.}\ \bibnamefont
  {Tretyakova}} and \bibinfo {author} {\bibfnamefont {K.}~\bibnamefont
  {Takahashi}},\ }\href {\doibase 10.1088/1361-6382/aa8057} {\bibfield
  {journal} {\bibinfo  {journal} {\emph {Class. Quant. Grav.}}\ }\textbf
  {\bibinfo {volume} {34}},\ \bibinfo {pages} {175007} (\bibinfo {year}
  {2017})},\ \Eprint {http://arxiv.org/abs/1702.03502} {arXiv:1702.03502
  [gr-qc]} \BibitemShut {NoStop}%
\bibitem [{\citenamefont {Takahashi}\ \emph {et~al.}(2019)\citenamefont
  {Takahashi}, \citenamefont {Motohashi},\ and\ \citenamefont
  {Minamitsuji}}]{Takahashi:2019oxz}%
  \BibitemOpen
  \bibfield  {author} {\bibinfo {author} {\bibfnamefont {K.}~\bibnamefont
  {Takahashi}}, \bibinfo {author} {\bibfnamefont {H.}~\bibnamefont
  {Motohashi}},  and \bibinfo {author} {\bibfnamefont {M.}~\bibnamefont
  {Minamitsuji}},\ }\href {\doibase 10.1103/PhysRevD.100.024041} {\bibfield
  {journal} {\bibinfo  {journal} {\emph {Phys. Rev. D}}\ }\textbf {\bibinfo
  {volume} {100}},\ \bibinfo {pages} {024041} (\bibinfo {year} {2019})},\
  \Eprint {http://arxiv.org/abs/1904.03554} {arXiv:1904.03554 [gr-qc]}
  \BibitemShut {NoStop}%
\bibitem [{\citenamefont {Motohashi}\ and\ \citenamefont
  {Minamitsuji}(2019)}]{Motohashi:2019sen}%
  \BibitemOpen
  \bibfield  {author} {\bibinfo {author} {\bibfnamefont {H.}~\bibnamefont
  {Motohashi}} and \bibinfo {author} {\bibfnamefont {M.}~\bibnamefont
  {Minamitsuji}},\ }\href {\doibase 10.1103/PhysRevD.99.064040} {\bibfield
  {journal} {\bibinfo  {journal} {\emph {Phys. Rev. D}}\ }\textbf {\bibinfo
  {volume} {99}},\ \bibinfo {pages} {064040} (\bibinfo {year} {2019})},\
  \Eprint {http://arxiv.org/abs/1901.04658} {arXiv:1901.04658 [gr-qc]}
  \BibitemShut {NoStop}%
\bibitem [{\citenamefont {Takahashi}\ and\ \citenamefont
  {Motohashi}(2020)}]{Takahashi:2020hso}%
  \BibitemOpen
  \bibfield  {author} {\bibinfo {author} {\bibfnamefont {K.}~\bibnamefont
  {Takahashi}} and \bibinfo {author} {\bibfnamefont {H.}~\bibnamefont
  {Motohashi}},\ }\href {\doibase 10.1088/1475-7516/2020/06/034} {\bibfield
  {journal} {\bibinfo  {journal} {\emph {JCAP}}\ }\textbf {\bibinfo {volume}
  {06}},\ \bibinfo {pages} {034} (\bibinfo {year} {2020})},\ \Eprint
  {http://arxiv.org/abs/2004.03883} {arXiv:2004.03883 [gr-qc]} \BibitemShut
  {NoStop}%
\bibitem [{\citenamefont {Khoury}\ \emph {et~al.}(2020)\citenamefont {Khoury},
  \citenamefont {Trodden},\ and\ \citenamefont {Wong}}]{Khoury:2020aya}%
  \BibitemOpen
  \bibfield  {author} {\bibinfo {author} {\bibfnamefont {J.}~\bibnamefont
  {Khoury}}, \bibinfo {author} {\bibfnamefont {M.}~\bibnamefont {Trodden}},
  and \bibinfo {author} {\bibfnamefont {S.~S.~C.}\ \bibnamefont {Wong}},\
  }\href {\doibase 10.1088/1475-7516/2020/11/044} {\bibfield  {journal}
  {\bibinfo  {journal} {\emph {JCAP}}\ }\textbf {\bibinfo {volume} {11}},\
  \bibinfo {pages} {044} (\bibinfo {year} {2020})},\ \Eprint
  {http://arxiv.org/abs/2007.01320} {arXiv:2007.01320 [astro-ph.CO]}
  \BibitemShut {NoStop}%
\bibitem [{\citenamefont {Takahashi}\ and\ \citenamefont
  {Motohashi}(2021)}]{Takahashi:2021bml}%
  \BibitemOpen
  \bibfield  {author} {\bibinfo {author} {\bibfnamefont {K.}~\bibnamefont
  {Takahashi}} and \bibinfo {author} {\bibfnamefont {H.}~\bibnamefont
  {Motohashi}},\ }\href {\doibase 10.1088/1475-7516/2021/08/013} {\bibfield
  {journal} {\bibinfo  {journal} {\emph {JCAP}}\ }\textbf {\bibinfo {volume}
  {08}},\ \bibinfo {pages} {013} (\bibinfo {year} {2021})},\ \Eprint
  {http://arxiv.org/abs/2106.07128} {arXiv:2106.07128 [gr-qc]} \BibitemShut
  {NoStop}%
\bibitem [{\citenamefont {Nakashi}\ \emph {et~al.}(2022)\citenamefont
  {Nakashi}, \citenamefont {Kimura}, \citenamefont {Motohashi},\ and\
  \citenamefont {Takahashi}}]{Nakashi:2022wdg}%
  \BibitemOpen
  \bibfield  {author} {\bibinfo {author} {\bibfnamefont {K.}~\bibnamefont
  {Nakashi}}, \bibinfo {author} {\bibfnamefont {M.}~\bibnamefont {Kimura}},
  \bibinfo {author} {\bibfnamefont {H.}~\bibnamefont {Motohashi}},  and
  \bibinfo {author} {\bibfnamefont {K.}~\bibnamefont {Takahashi}},\ }\Eprint
  {http://arxiv.org/abs/2204.05054} {arXiv:2204.05054 [gr-qc]} \BibitemShut
  {NoStop}%
\bibitem [{\citenamefont {Babichev}\ \emph {et~al.}(2017)\citenamefont
  {Babichev}, \citenamefont {Charmousis},\ and\ \citenamefont
  {Leh\'ebel}}]{Babichev:2017guv}%
  \BibitemOpen
  \bibfield  {author} {\bibinfo {author} {\bibfnamefont {E.}~\bibnamefont
  {Babichev}}, \bibinfo {author} {\bibfnamefont {C.}~\bibnamefont
  {Charmousis}},  and \bibinfo {author} {\bibfnamefont {A.}~\bibnamefont
  {Leh\'ebel}},\ }\href {\doibase 10.1088/1475-7516/2017/04/027} {\bibfield
  {journal} {\bibinfo  {journal} {\emph {JCAP}}\ }\textbf {\bibinfo {volume}
  {04}},\ \bibinfo {pages} {027} (\bibinfo {year} {2017})},\ \Eprint
  {http://arxiv.org/abs/1702.01938} {arXiv:1702.01938 [gr-qc]} \BibitemShut
  {NoStop}%
\bibitem [{\citenamefont {Torii}\ and\ \citenamefont
  {Maeda}(1998)}]{Torii:1998gm}%
  \BibitemOpen
  \bibfield  {author} {\bibinfo {author} {\bibfnamefont {T.}~\bibnamefont
  {Torii}} and \bibinfo {author} {\bibfnamefont {K.-i.}\ \bibnamefont
  {Maeda}},\ }\href {\doibase 10.1103/PhysRevD.58.084004} {\bibfield  {journal}
  {\bibinfo  {journal} {\emph {Phys. Rev. D}}\ }\textbf {\bibinfo {volume}
  {58}},\ \bibinfo {pages} {084004} (\bibinfo {year} {1998})}\BibitemShut
  {NoStop}%
\bibitem [{\citenamefont {De~Felice}\ \emph {et~al.}(2011)\citenamefont
  {De~Felice}, \citenamefont {Suyama},\ and\ \citenamefont
  {Tanaka}}]{DeFelice:2011ka}%
  \BibitemOpen
  \bibfield  {author} {\bibinfo {author} {\bibfnamefont {A.}~\bibnamefont
  {De~Felice}}, \bibinfo {author} {\bibfnamefont {T.}~\bibnamefont {Suyama}},
  and \bibinfo {author} {\bibfnamefont {T.}~\bibnamefont {Tanaka}},\ }\href
  {\doibase 10.1103/PhysRevD.83.104035} {\bibfield  {journal} {\bibinfo
  {journal} {\emph {Phys. Rev. D}}\ }\textbf {\bibinfo {volume} {83}},\
  \bibinfo {pages} {104035} (\bibinfo {year} {2011})},\ \Eprint
  {http://arxiv.org/abs/1102.1521} {arXiv:1102.1521 [gr-qc]} \BibitemShut
  {NoStop}%
\bibitem [{\citenamefont {Bl\'azquez-Salcedo}\ \emph
  {et~al.}(2020{\natexlab{a}})\citenamefont {Bl\'azquez-Salcedo}, \citenamefont
  {Doneva}, \citenamefont {Kahlen}, \citenamefont {Kunz}, \citenamefont
  {Nedkova},\ and\ \citenamefont {Yazadjiev}}]{Blazquez-Salcedo:2020rhf}%
  \BibitemOpen
  \bibfield  {author} {\bibinfo {author} {\bibfnamefont {J.~L.}\ \bibnamefont
  {Bl\'azquez-Salcedo}}, \bibinfo {author} {\bibfnamefont {D.~D.}\ \bibnamefont
  {Doneva}}, \bibinfo {author} {\bibfnamefont {S.}~\bibnamefont {Kahlen}},
  \bibinfo {author} {\bibfnamefont {J.}~\bibnamefont {Kunz}}, \bibinfo {author}
  {\bibfnamefont {P.}~\bibnamefont {Nedkova}},  and \bibinfo {author}
  {\bibfnamefont {S.~S.}\ \bibnamefont {Yazadjiev}},\ }\href {\doibase
  10.1103/PhysRevD.101.104006} {\bibfield  {journal} {\bibinfo  {journal}
  {\emph {Phys. Rev. D}}\ }\textbf {\bibinfo {volume} {101}},\ \bibinfo {pages}
  {104006} (\bibinfo {year} {2020}{\natexlab{a}})},\ \Eprint
  {http://arxiv.org/abs/2003.02862} {arXiv:2003.02862 [gr-qc]} \BibitemShut
  {NoStop}%
\bibitem [{\citenamefont {Bl\'azquez-Salcedo}\ \emph
  {et~al.}(2020{\natexlab{b}})\citenamefont {Bl\'azquez-Salcedo}, \citenamefont
  {Doneva}, \citenamefont {Kahlen}, \citenamefont {Kunz}, \citenamefont
  {Nedkova},\ and\ \citenamefont {Yazadjiev}}]{Blazquez-Salcedo:2020caw}%
  \BibitemOpen
  \bibfield  {author} {\bibinfo {author} {\bibfnamefont {J.~L.}\ \bibnamefont
  {Bl\'azquez-Salcedo}}, \bibinfo {author} {\bibfnamefont {D.~D.}\ \bibnamefont
  {Doneva}}, \bibinfo {author} {\bibfnamefont {S.}~\bibnamefont {Kahlen}},
  \bibinfo {author} {\bibfnamefont {J.}~\bibnamefont {Kunz}}, \bibinfo {author}
  {\bibfnamefont {P.}~\bibnamefont {Nedkova}},  and \bibinfo {author}
  {\bibfnamefont {S.~S.}\ \bibnamefont {Yazadjiev}},\ }\href {\doibase
  10.1103/PhysRevD.102.024086} {\bibfield  {journal} {\bibinfo  {journal}
  {\emph {Phys. Rev. D}}\ }\textbf {\bibinfo {volume} {102}},\ \bibinfo {pages}
  {024086} (\bibinfo {year} {2020}{\natexlab{b}})},\ \Eprint
  {http://arxiv.org/abs/2006.06006} {arXiv:2006.06006 [gr-qc]} \BibitemShut
  {NoStop}%
\bibitem [{\citenamefont {Bl\'azquez-Salcedo}\ \emph
  {et~al.}(2022)\citenamefont {Bl\'azquez-Salcedo}, \citenamefont {Doneva},
  \citenamefont {Kunz},\ and\ \citenamefont
  {Yazadjiev}}]{Blazquez-Salcedo:2022omw}%
  \BibitemOpen
  \bibfield  {author} {\bibinfo {author} {\bibfnamefont {J.~L.}\ \bibnamefont
  {Bl\'azquez-Salcedo}}, \bibinfo {author} {\bibfnamefont {D.~D.}\ \bibnamefont
  {Doneva}}, \bibinfo {author} {\bibfnamefont {J.}~\bibnamefont {Kunz}},  and
  \bibinfo {author} {\bibfnamefont {S.~S.}\ \bibnamefont {Yazadjiev}},\
  }\Eprint {http://arxiv.org/abs/2203.00709} {arXiv:2203.00709 [gr-qc]}
  \BibitemShut {NoStop}%
\bibitem [{\citenamefont {Langlois}\ \emph {et~al.}(2022)\citenamefont
  {Langlois}, \citenamefont {Noui},\ and\ \citenamefont
  {Roussille}}]{Langlois:2022eta}%
  \BibitemOpen
  \bibfield  {author} {\bibinfo {author} {\bibfnamefont {D.}~\bibnamefont
  {Langlois}}, \bibinfo {author} {\bibfnamefont {K.}~\bibnamefont {Noui}},  and
  \bibinfo {author} {\bibfnamefont {H.}~\bibnamefont {Roussille}},\ }\Eprint
  {http://arxiv.org/abs/2204.04107} {arXiv:2204.04107 [gr-qc]} \BibitemShut
  {NoStop}%
\bibitem [{\citenamefont {Nojiri}\ and\ \citenamefont
  {Odintsov}(2005)}]{Nojiri:2005jg}%
  \BibitemOpen
  \bibfield  {author} {\bibinfo {author} {\bibfnamefont {S.}~\bibnamefont
  {Nojiri}} and \bibinfo {author} {\bibfnamefont {S.~D.}\ \bibnamefont
  {Odintsov}},\ }\href {\doibase 10.1016/j.physletb.2005.10.010} {\bibfield
  {journal} {\bibinfo  {journal} {\emph {Phys. Lett. B}}\ }\textbf {\bibinfo
  {volume} {631}},\ \bibinfo {pages} {1} (\bibinfo {year} {2005})},\ \Eprint
  {http://arxiv.org/abs/hep-th/0508049} {arXiv:hep-th/0508049} \BibitemShut
  {NoStop}%
\bibitem [{\citenamefont {De~Felice}\ and\ \citenamefont
  {Hindmarsh}(2007)}]{DeFelice:2007zq}%
  \BibitemOpen
  \bibfield  {author} {\bibinfo {author} {\bibfnamefont {A.}~\bibnamefont
  {De~Felice}} and \bibinfo {author} {\bibfnamefont {M.}~\bibnamefont
  {Hindmarsh}},\ }\href {\doibase 10.1088/1475-7516/2007/06/028} {\bibfield
  {journal} {\bibinfo  {journal} {\emph {JCAP}}\ }\textbf {\bibinfo {volume}
  {06}},\ \bibinfo {pages} {028} (\bibinfo {year} {2007})},\ \Eprint
  {http://arxiv.org/abs/0705.3375} {arXiv:0705.3375 [astro-ph]} \BibitemShut
  {NoStop}%
\bibitem [{\citenamefont {Li}\ \emph {et~al.}(2007)\citenamefont {Li},
  \citenamefont {Barrow},\ and\ \citenamefont {Mota}}]{Li:2007jm}%
  \BibitemOpen
  \bibfield  {author} {\bibinfo {author} {\bibfnamefont {B.}~\bibnamefont
  {Li}}, \bibinfo {author} {\bibfnamefont {J.~D.}\ \bibnamefont {Barrow}},  and
  \bibinfo {author} {\bibfnamefont {D.~F.}\ \bibnamefont {Mota}},\ }\href
  {\doibase 10.1103/PhysRevD.76.044027} {\bibfield  {journal} {\bibinfo
  {journal} {\emph {Phys. Rev. D}}\ }\textbf {\bibinfo {volume} {76}},\
  \bibinfo {pages} {044027} (\bibinfo {year} {2007})},\ \Eprint
  {http://arxiv.org/abs/0705.3795} {arXiv:0705.3795 [gr-qc]} \BibitemShut
  {NoStop}%
\bibitem [{\citenamefont {De~Felice}\ and\ \citenamefont
  {Tsujikawa}(2009{\natexlab{a}})}]{DeFelice:2008wz}%
  \BibitemOpen
  \bibfield  {author} {\bibinfo {author} {\bibfnamefont {A.}~\bibnamefont
  {De~Felice}} and \bibinfo {author} {\bibfnamefont {S.}~\bibnamefont
  {Tsujikawa}},\ }\href {\doibase 10.1016/j.physletb.2009.03.060} {\bibfield
  {journal} {\bibinfo  {journal} {\emph {Phys. Lett. B}}\ }\textbf {\bibinfo
  {volume} {675}},\ \bibinfo {pages} {1} (\bibinfo {year}
  {2009}{\natexlab{a}})},\ \Eprint {http://arxiv.org/abs/0810.5712}
  {arXiv:0810.5712 [hep-th]} \BibitemShut {NoStop}%
\bibitem [{\citenamefont {De~Felice}\ and\ \citenamefont
  {Tsujikawa}(2009{\natexlab{b}})}]{DeFelice:2009aj}%
  \BibitemOpen
  \bibfield  {author} {\bibinfo {author} {\bibfnamefont {A.}~\bibnamefont
  {De~Felice}} and \bibinfo {author} {\bibfnamefont {S.}~\bibnamefont
  {Tsujikawa}},\ }\href {\doibase 10.1103/PhysRevD.80.063516} {\bibfield
  {journal} {\bibinfo  {journal} {\emph {Phys. Rev. D}}\ }\textbf {\bibinfo
  {volume} {80}},\ \bibinfo {pages} {063516} (\bibinfo {year}
  {2009}{\natexlab{b}})},\ \Eprint {http://arxiv.org/abs/0907.1830}
  {arXiv:0907.1830 [hep-th]} \BibitemShut {NoStop}%
\bibitem [{\citenamefont {Motohashi}\ \emph {et~al.}(2016)\citenamefont
  {Motohashi}, \citenamefont {Suyama},\ and\ \citenamefont
  {Takahashi}}]{Motohashi:2016prk}%
  \BibitemOpen
  \bibfield  {author} {\bibinfo {author} {\bibfnamefont {H.}~\bibnamefont
  {Motohashi}}, \bibinfo {author} {\bibfnamefont {T.}~\bibnamefont {Suyama}},
  and \bibinfo {author} {\bibfnamefont {K.}~\bibnamefont {Takahashi}},\ }\href
  {\doibase 10.1103/PhysRevD.94.124021} {\bibfield  {journal} {\bibinfo
  {journal} {\emph {Phys. Rev. D}}\ }\textbf {\bibinfo {volume} {94}},\
  \bibinfo {pages} {124021} (\bibinfo {year} {2016})},\ \Eprint
  {http://arxiv.org/abs/1608.00071} {arXiv:1608.00071 [gr-qc]} \BibitemShut
  {NoStop}%
\bibitem [{\citenamefont {Kase}\ \emph {et~al.}(2020)\citenamefont {Kase},
  \citenamefont {Kimura}, \citenamefont {Sato},\ and\ \citenamefont
  {Tsujikawa}}]{Kase:2020qvz}%
  \BibitemOpen
  \bibfield  {author} {\bibinfo {author} {\bibfnamefont {R.}~\bibnamefont
  {Kase}}, \bibinfo {author} {\bibfnamefont {R.}~\bibnamefont {Kimura}},
  \bibinfo {author} {\bibfnamefont {S.}~\bibnamefont {Sato}},  and \bibinfo
  {author} {\bibfnamefont {S.}~\bibnamefont {Tsujikawa}},\ }\href {\doibase
  10.1103/PhysRevD.102.084037} {\bibfield  {journal} {\bibinfo  {journal}
  {\emph {Phys. Rev. D}}\ }\textbf {\bibinfo {volume} {102}},\ \bibinfo {pages}
  {084037} (\bibinfo {year} {2020})},\ \Eprint
  {http://arxiv.org/abs/2007.09864} {arXiv:2007.09864 [gr-qc]} \BibitemShut
  {NoStop}%
\bibitem [{\citenamefont {Chagoya}\ and\ \citenamefont
  {Tasinato}(2019)}]{Chagoya:2018yna}%
  \BibitemOpen
  \bibfield  {author} {\bibinfo {author} {\bibfnamefont {J.}~\bibnamefont
  {Chagoya}} and \bibinfo {author} {\bibfnamefont {G.}~\bibnamefont
  {Tasinato}},\ }\href {\doibase 10.1088/1361-6382/ab0a4b} {\bibfield
  {journal} {\bibinfo  {journal} {\emph {Class. Quant. Grav.}}\ }\textbf
  {\bibinfo {volume} {36}},\ \bibinfo {pages} {075014} (\bibinfo {year}
  {2019})},\ \Eprint {http://arxiv.org/abs/1805.12010} {arXiv:1805.12010
  [hep-th]} \BibitemShut {NoStop}%
\bibitem [{\citenamefont {Creminelli}\ \emph {et~al.}(2020)\citenamefont
  {Creminelli}, \citenamefont {Loayza}, \citenamefont {Serra}, \citenamefont
  {Trincherini},\ and\ \citenamefont {Trombetta}}]{Creminelli:2020lxn}%
  \BibitemOpen
  \bibfield  {author} {\bibinfo {author} {\bibfnamefont {P.}~\bibnamefont
  {Creminelli}}, \bibinfo {author} {\bibfnamefont {N.}~\bibnamefont {Loayza}},
  \bibinfo {author} {\bibfnamefont {F.}~\bibnamefont {Serra}}, \bibinfo
  {author} {\bibfnamefont {E.}~\bibnamefont {Trincherini}},  and \bibinfo
  {author} {\bibfnamefont {L.~G.}\ \bibnamefont {Trombetta}},\ }\href {\doibase
  10.1007/JHEP08(2020)045} {\bibfield  {journal} {\bibinfo  {journal} {\emph
  {JHEP}}\ }\textbf {\bibinfo {volume} {08}},\ \bibinfo {pages} {045} (\bibinfo
  {year} {2020})},\ \Eprint {http://arxiv.org/abs/2004.02893} {arXiv:2004.02893
  [hep-th]} \BibitemShut {NoStop}%
\bibitem [{\citenamefont {Abbott}\ \emph
  {et~al.}(2017{\natexlab{a}})\citenamefont {Abbott} \emph
  {et~al.}}]{LIGOScientific:2017vwq}%
  \BibitemOpen
  \bibfield  {author} {\bibinfo {author} {\bibfnamefont {B.~P.}\ \bibnamefont
  {Abbott}} \emph {et~al.},\ }\href {\doibase 10.1103/PhysRevLett.119.161101}
  {\bibfield  {journal} {\bibinfo  {journal} {\emph {Phys. Rev. Lett.}}\
  }\textbf {\bibinfo {volume} {119}},\ \bibinfo {pages} {161101} (\bibinfo
  {year} {2017}{\natexlab{a}})},\ \Eprint {http://arxiv.org/abs/1710.05832}
  {arXiv:1710.05832 [gr-qc]} \BibitemShut {NoStop}%
\bibitem [{\citenamefont {Abbott}\ \emph
  {et~al.}(2017{\natexlab{b}})\citenamefont {Abbott} \emph
  {et~al.}}]{LIGOScientific:2017ync}%
  \BibitemOpen
  \bibfield  {author} {\bibinfo {author} {\bibfnamefont {B.~P.}\ \bibnamefont
  {Abbott}} \emph {et~al.},\ }\href {\doibase 10.3847/2041-8213/aa91c9}
  {\bibfield  {journal} {\bibinfo  {journal} {\emph {Astrophys. J. Lett.}}\
  }\textbf {\bibinfo {volume} {848}},\ \bibinfo {pages} {L12} (\bibinfo {year}
  {2017}{\natexlab{b}})},\ \Eprint {http://arxiv.org/abs/1710.05833}
  {arXiv:1710.05833 [astro-ph.HE]} \BibitemShut {NoStop}%
\bibitem [{\citenamefont {Abbott}\ \emph
  {et~al.}(2017{\natexlab{c}})\citenamefont {Abbott} \emph
  {et~al.}}]{LIGOScientific:2017zic}%
  \BibitemOpen
  \bibfield  {author} {\bibinfo {author} {\bibfnamefont {B.~P.}\ \bibnamefont
  {Abbott}} \emph {et~al.},\ }\href {\doibase 10.3847/2041-8213/aa920c}
  {\bibfield  {journal} {\bibinfo  {journal} {\emph {Astrophys. J. Lett.}}\
  }\textbf {\bibinfo {volume} {848}},\ \bibinfo {pages} {L13} (\bibinfo {year}
  {2017}{\natexlab{c}})},\ \Eprint {http://arxiv.org/abs/1710.05834}
  {arXiv:1710.05834 [astro-ph.HE]} \BibitemShut {NoStop}%
\bibitem [{\citenamefont {Doneva}\ \emph {et~al.}(2019)\citenamefont {Doneva},
  \citenamefont {Staykov},\ and\ \citenamefont {Yazadjiev}}]{Doneva:2019vuh}%
  \BibitemOpen
  \bibfield  {author} {\bibinfo {author} {\bibfnamefont {D.~D.}\ \bibnamefont
  {Doneva}}, \bibinfo {author} {\bibfnamefont {K.~V.}\ \bibnamefont {Staykov}},
   and \bibinfo {author} {\bibfnamefont {S.~S.}\ \bibnamefont {Yazadjiev}},\
  }\href {\doibase 10.1103/PhysRevD.99.104045} {\bibfield  {journal} {\bibinfo
  {journal} {\emph {Phys. Rev. D}}\ }\textbf {\bibinfo {volume} {99}},\
  \bibinfo {pages} {104045} (\bibinfo {year} {2019})},\ \Eprint
  {http://arxiv.org/abs/1903.08119} {arXiv:1903.08119 [gr-qc]} \BibitemShut
  {NoStop}%
\bibitem [{\citenamefont {Antoniou}\ \emph {et~al.}(2021)\citenamefont
  {Antoniou}, \citenamefont {Leh\'ebel}, \citenamefont {Ventagli},\ and\
  \citenamefont {Sotiriou}}]{Antoniou:2021zoy}%
  \BibitemOpen
  \bibfield  {author} {\bibinfo {author} {\bibfnamefont {G.}~\bibnamefont
  {Antoniou}}, \bibinfo {author} {\bibfnamefont {A.}~\bibnamefont {Leh\'ebel}},
  \bibinfo {author} {\bibfnamefont {G.}~\bibnamefont {Ventagli}},  and \bibinfo
  {author} {\bibfnamefont {T.~P.}\ \bibnamefont {Sotiriou}},\ }\href {\doibase
  10.1103/PhysRevD.104.044002} {\bibfield  {journal} {\bibinfo  {journal}
  {\emph {Phys. Rev. D}}\ }\textbf {\bibinfo {volume} {104}},\ \bibinfo {pages}
  {044002} (\bibinfo {year} {2021})},\ \Eprint
  {http://arxiv.org/abs/2105.04479} {arXiv:2105.04479 [gr-qc]} \BibitemShut
  {NoStop}%
\bibitem [{\citenamefont {Minamitsuji}\ and\ \citenamefont
  {Ikeda}(2019{\natexlab{b}})}]{Minamitsuji:2019iwp}%
  \BibitemOpen
  \bibfield  {author} {\bibinfo {author} {\bibfnamefont {M.}~\bibnamefont
  {Minamitsuji}} and \bibinfo {author} {\bibfnamefont {T.}~\bibnamefont
  {Ikeda}},\ }\href {\doibase 10.1103/PhysRevD.99.104069} {\bibfield  {journal}
  {\bibinfo  {journal} {\emph {Phys. Rev. D}}\ }\textbf {\bibinfo {volume}
  {99}},\ \bibinfo {pages} {104069} (\bibinfo {year} {2019}{\natexlab{b}})},\
  \Eprint {http://arxiv.org/abs/1904.06572} {arXiv:1904.06572 [gr-qc]}
  \BibitemShut {NoStop}%
\end{thebibliography}%

\end{document}